\documentclass[12pt]{article}

\usepackage[utf8]{inputenc}
\usepackage[T1]{fontenc}

\usepackage{setspace} 
\usepackage[vmargin = 1.25in, hmargin = 1.25in]{geometry}

\usepackage[usenames,dvipsnames,svgnames]{xcolor}
\usepackage{amsmath, amssymb, amsfonts, graphicx, tikz,pdflscape, mathtools, amsthm, upgreek, bm,pgfplots,csvsimple,multirow, multicol, booktabs}
\usepackage{verbatim}
\usepackage{natbib}
\usepackage{accents}

\usepackage{needspace}
\def\citeapos#1{\citeauthor{#1}'s (\citeyear{#1})}
\usepackage{comment}
\usepackage[inline]{enumitem}
\usepackage{tocvsec2}
\usepackage{threeparttable}
\usetikzlibrary{calc, cd, arrows.meta}
\usepackage[skip = 10pt]{caption}
\allowdisplaybreaks[1]
\allowdisplaybreaks[1]

\definecolor{Blue}{RGB}{86,180,233}
\definecolor{Orange}{RGB}{230,159,0}
\definecolor{Green}{RGB}{0,158,115}
\definecolor{GmailBlue}{RGB}{42, 93, 176} 
\usepackage[
	pagebackref,
	colorlinks=true,
	citecolor= GmailBlue,
	linkcolor=GmailBlue,
	urlcolor = GmailBlue
]{hyperref}

\newcommand{\bibtexorder}[1]{}

\usepackage{pgfplots}
\usepgfplotslibrary{groupplots,colorbrewer}
\pgfplotsset{compat=newest}
\pgfplotsset{cycle list/Set1}
\usepackage{tikz}
\usetikzlibrary{matrix,calc,shapes,arrows.meta,positioning}
\tikzset{
    vertex/.style = {shape=circle,draw, minimum size = 1.8em, inner sep = 0pt},
    edge/.style = {->,> = latex}
}

\usepackage[capitalize,noabbrev]{cleveref}

\newtheoremstyle{break}
{}
{}
{\itshape}
{}
{\bfseries}
{}
{\newline}
{}

\theoremstyle{break}
\newtheorem{thm}{Theorem}
\newtheorem*{theorem*}{Theorem}
\newtheorem*{cor*}{Corollary}

\newtheorem{lem}{Lemma}

\crefname{prop}{Proposition}{Propositions}
\crefname{thm}{Theorem}{Theorems}
\crefname{lem}{Lemma}{Lemmas}
\crefname{blem}{Lemma}{Lemmas}

\theoremstyle{definition}

\newtheorem{exmp}{Example}

\newtheorem{rem}{Remark}
\newtheorem*{rem*}{Remark}

\def\a{\alpha}
\def\b{\beta}
\def\g{\gamma}
\def\d{\delta}
\def\e{\varepsilon}
\def\z{\zeta}

\def\th{\theta}
\def\i{\iota}

\def\l{\lambda}

\def\s{\sigma}

\def\D{\Delta}
\def\Th{\Theta}

\def\R{\mathbf{R}}

\def\XX{\mathcal{X}}

\def\P{\mathbf{P}}

\DeclareMathOperator{\E}{\mathbf{E}}
\DeclareMathOperator{\supp}{supp} 
\DeclareMathOperator*{\argmax}{argmax}

\DeclareMathOperator{\var}{var}

\DeclareMathOperator{\marg}{marg} 

\DeclareMathOperator{\oprod}{\otimes}

\newcommand{\Abs}[1]{\left\lvert #1 \right\rvert}

\newcommand{\Paren}[1]{\left( #1 \right)}

\newcommand{\Brac}[1]{\left[ #1 \right]}

\newcommand{\Set}[1]{\left\{ #1 \right\}}

\newcommand{\de}{\mathop{}\!\mathrm{d}}

\usepackage{datetime}
\newdateformat{specialdate}{\THEDAY~\monthname[\THEMONTH] \THEYEAR}

\title{Quota Mechanisms: Finite-Sample \\ Optimality and Robustness\thanks{This paper was previously circulated under the title ``Linking Mechanisms: Limits and Robustness.'' The paper was initially written while Ian Ball was visiting Northwestern University. For helpful discussions, we thank Sandeep Baliga,  Alex Frankel, Drew Fudenberg, Alkis Georgiadis-Harris, Johannes H\"{o}rner, Yuhta Ishii, Matt Jackson, Elliot Lipnowski, Stephen Morris, Ludvig Sinander, and Juuso Toikka. For useful feedback, we also thank the seminar audiences at the ASSA Annual Meeting, Bonn Junior Workshop in Economic Theory, Warwick, CIREQ, Northwestern, Oxford, Toronto, and Toulouse.}}
\date{\specialdate\today}

\author{Ian Ball\thanks{MIT, \texttt{ianball@mit.edu}.} \and Deniz Kattwinkel\thanks{UCL, \texttt{d.kattwinkel@ucl.ac.uk}.}}

\begin{document}


\maketitle \thispagestyle{empty} 

\begin{abstract}
A quota mechanism, such as a mandatory grading curve, links together multiple decisions. We analyze the performance of quota mechanisms when the number of linked decisions is finite and the designer has imperfect knowledge of the type distribution. Using a new optimal transport approach, we derive an ex-post decision error guarantee for quota mechanisms. This guarantee cannot be improved by any mechanisms without transfers. We quantify the sensitivity of quota mechanisms to errors in the designer's estimate of the type distribution. Finally, we show that quotas are robust to a range of agents' beliefs about each other.
\end{abstract}



\newpage

\section{Introduction}

When stakeholders have private information but transfers are restricted, a common approach is to link together multiple similar decisions by imposing an aggregate quota.
For example, a mandatory grading curve constrains the distribution of grades that an instructor can assign to his students. Prescription drug monitoring programs are used by regulators to identify and punish doctors who prescribe opioids to an unusually large share of their patients.\footnote{See the CDC's \href{https://www.cdc.gov/drugoverdose/pdf/policyimpact-prescriptionpainkillerod-a.pdf}{Policy Impact: Prescription Painkiller Overdoses}.} In multi-issue voting, the storable votes procedure \citep{Casella2005} allows each voter to distribute a fixed number of votes across different issues. Under  Temporary Assistance for Needy Families (TANF), a family can only collect cash assistance for up to $60$ months over their lifetime.\footnote{See the CBPP's \href{https://www.cbpp.org/research/family-income-support/policy-basics-an-introduction-to-tanf}{Policy Basics: Temporary Assistance for Needy Families}. When allocating money, side-payments cannot serve as an \emph{additional} instrument. Thus, allocating money fits within the framework of allocation without transfers.}

Even when preferences are misaligned, quotas can motivate agents to represent their private information truthfully. The reason is that quotas force agents to make tradeoffs across different decisions. Under a grading curve, an instructor who prefers to grade leniently cannot choose in isolation \emph{whether} a student gets an $A$, but must instead choose \emph{which} students get the $A$s. Similarly, under a prescription quota, a doctor who is inclined to overprescribe painkillers must choose \emph{which} patients receive opioids.

But what if most of the class really deserves an $A$? Or if many of a doctor's patients genuinely need painkillers? The difficulty with quota mechanisms is that the empirical distribution of cases may differ significantly from the quota. This happens in small samples due to standard sampling variation and even in large samples if the quota designer does not know the exact population distribution.

In practice, these two sources of error---sampling variation and misspecified quotas---can almost never be ruled out. The existing literature, however,  sidesteps these two sources of error. Under the assumption that the designer knows the exact population distribution, this literature proves asymptotic results as the number of decisions grows large (and hence sampling variation vanishes). Thus, the existing theoretical rationale for using quota mechanisms in practice is incomplete. 

Our paper is the first to analyze the performance of quota mechanisms outside of these idealized conditions.
We quantify how the distance between an agent's quota and his realized type frequencies translates into decision errors. Moreover, we show that quota mechanisms satisfy a robust optimality property: the decision error guarantee under quota mechanisms cannot be improved by any other mechanisms without transfers. Using this guarantee, we bound the decision error that results from sampling variation and misspecification of the type distribution. These results complete the rationale for using quota mechanisms in practice. 

Concretely, our results yield guarantees such as the following. \emph{If a doctor is bound by a correctly specified quota when prescribing one of three treatments to each of his 200 patients, then the expected share of patients who will receive the wrong treatment is at most 10\%} (\cref{res:convergence_rate}). We also quantify how using the wrong quota translates into decision errors, even when many decisions are linked. 
\emph{If a doctor has many patients and is regulated by a quota that underestimates the need for one (of three) treatments in the population by $1$ percentage point, then at most $2$ percent of patients will be prescribed the wrong treatment} (\cref{res:error_bound}).\footnote{Here, we assume that the need for each of the other two treatments is not strictly underestimated.}


We work in the decision setting of \cite{jackson2007overcoming}. Consider a principal (she) and one or more agents (he). There are multiple independent copies of a primitive collective decision problem with independent, private values. Each agent knows his type in each problem copy. Utilities are additively separable across the problems. A \emph{linking mechanism} elicits a message from each agent and then selects a decision in every problem simultaneously. A \emph{quota mechanism} is a special kind of linking mechanism. Each agent's reports across the problem copies must satisfy an aggregate quota. In each problem copy, the desired social choice function is applied to the submitted reports.

In the grading curve example, there is a single agent---the instructor. The university is the principal. Each problem copy corresponds to a student in the instructor's class. The instructor privately observes the performance of each student in the class and reports to the university registrar a grade recommendation for each student. These recommended grades must satisfy the curve set by the university. Based on the instructor's reports, the university registrar assigns a grade to each student.\footnote{Here, the quota is effectively imposed on the agent's \emph{decisions} rather than on his \emph{type reports}. These two formulations are equivalent with a single agent (and a deterministic social choice function); see \cref{rem:decisions}.}

We introduce a new class of quota mechanisms, in which each agent reports a \emph{type distribution} on each problem, rather than a single \emph{type} as in \cite{jackson2007overcoming}. Technically, this is convenient because it allows us to directly apply tools from optimal transport theory.  But all of our results extend to  \citeapos{jackson2007overcoming} more familiar quota mechanisms, up to a small error in our finite-sample bounds; see \cref{rem:JS_quota}. 

Consider a quota mechanism with a social choice function that is cyclically monotone. \cref{res:optimal_transport_bound} gives a tight bound on the decision error for each realization of the agents' private information. The bound depends on the distance between each agent's quota and the empirical distribution of that agent's realized type vector. The challenge in proving this bound is that when an agent's realized type frequencies differ from the quota, that agent may find it optimal to respond with a ``cascade of lies.'' We quantify the size of this cascade by reformulating each agent's choice under the quota mechanism as an optimal transport problem. Our decision error bound in \cref{res:optimal_transport_bound} cannot be improved, even if the principal uses more complicated linking mechanisms. Next, we set each quota equal to the prior.  Taking expectations in our ex-post bound, we bound the expected decision error (\cref{res:convergence_rate}). This bound is decreasing in the number of problem copies and increasing in the number of types of each agent.

Building on our finite-sample bounds, we unify and strengthen the asymptotic results from the literature. \cref{res:equivalence} says that a social choice function is asymptotically implementable by quota mechanisms if and only if it is cyclically monotone. Recall from \cite{Rochet1987} that a social choice function is cyclically monotone if and only if it is one-shot implementable with transfers. Thus, cyclical monotonicity characterizes both forms of implementability. Moreover, if a social choice function is not cyclically monotone, then it cannot be asymptotically implemented by any linking mechanisms, even with transfers. This justifies our focus on quota mechanisms. 

Next, we turn to the robustness properties of quota mechanisms. We first consider the sensitivity of quota mechanisms to the true type distribution.  In our motivating example of opioid prescription regulation, this sensitivity is relevant if the regulator can only imperfectly estimate the share of a particular patient population that would benefit from opioids. We establish in \cref{res:error_bound} a tight bound on the equilibrium decision error when the quota is set equal to an incorrect estimate of the  type distribution. 

Finally, we study the robustness of quota mechanisms to agents' beliefs about each other. This kind of robustness is relevant, for example, to the design of multi-issue voting in a committee, where the designer is unlikely to know what each committee member believes about the other members' preferences.  Whatever each agent believes about the other agents, he knows that their \emph{reports} must obey the quota. As a result, quota mechanisms satisfy a belief-robustness property, which we formalize in the rich type-space framework of \cite{BergemannMorris2005}. In \cref{res:robust}, we show that for a class of type spaces satisfying certain independence and exchangeability properties, a quota mechanism admits a special equilibrium that approximates the desired social choice function. In this equilibrium, each agent's reports depend only on his payoff type, not his belief type.

The rest of the paper is organized as follows. \cref{sec:literature} discusses related literature. \cref{sec:model} introduces the model, which is then illustrated in a simple example in \cref{sec:simple_example}. In \cref{sec:optimal_transport_bound}, we bound the decision error under quota mechanisms when there are finitely many problem copies. In \cref{sec:Bayesian}, we characterize which social choice functions can be asymptotically implemented by quota mechanisms.  In \cref{sec:robustness_single}, we analyze robustness to the type distributions. In  \cref{sec:belief_robustness}, we analyze robustness to the agents' beliefs about each other. \cref{sec:discussion} discusses extensions to interdependent values and dynamics. The conclusion is in \cref{sec:conclusion}. The main proofs are in \cref{sec:proofs}. \cref{sec:comparison} compares our quota mechanisms with those in \cite{jackson2007overcoming}. Additional proofs are in \cref{sec:additional}.

\subsection{Related literature} \label{sec:literature}

We depart from the previous literature by analyzing quota mechanisms under more realistic conditions---with finitely many problem copies and uncertainty about the population distribution. Quota mechanisms were introduced by \cite{jackson2007overcoming}.\footnote{A special case of a quota mechanism appears in \cite{Townsend1982}.  In a setting with transferable utility, \cite{FangNorman2006} analyze the power of quota mechanisms to overcome participation constraints, rather than incentive constraints.} They show that every ex-ante Pareto efficient social choice function can be asymptotically implemented by quota mechanisms, as the number of linked problems grows large.\footnote{\cite{jackson2007overcoming} also show how quota mechanisms can be modified to punish collusion. They augment their quota mechanisms with statistical tests of correlation between the agents' reports. With this modification, they show that under all equilibria, the agents' payoffs converge to the desired payoff profile as the number of problem copies grows large. See \cite{CsokaEtal2023} for an alternative approach to collusion-proofness.} \cite{MatsushimaEtal2010} extend this result to cyclically monotone social choice functions, but they use a weaker notion of implementation in $\e$-equilibrium. Both papers study asymptotic implementation in 
the common-prior setting. 

A few papers study the robustness of quota-like mechanisms in special environments. \cite{Hortala-Vallve2010} proves that with finitely many copies of a binary decision problem, no nontrivial social choice function can be implemented in ex-post equilibrium without transfers. \cite{Frankel2014} considers finitely many copies of a delegation problem in which the principal and the agent both prefer higher actions in higher states.%
\footnote{\cite{Frankel2016} considers a Bayesian version of the multi-task delegation model. With quadratic losses, constant bias, and normally distributed states,  it is optimal for the principal to cap a weighted average of the agent's decisions.} He gives conditions under which a quota-like mechanism maximizes the principal's worst-case expected payoff, where the worst-case is taken over a set of state-dependent utility functions for the agent. Relative to \cite{Frankel2014}, we consider a dual sense of robustness. Whereas he considers robustness to the agent's utility function given a fixed prior over states,  we analyze robustness to the distribution of states  given a fixed utility function for the agents. 


Our results also relate to the literature on belief-robust implementation. \cite{BergemannMorris2005} study robustness to all possible information structures. We adopt their type-space framework, but we consider robustness to a subclass of information structures. In settings with transfers, some work has studied robustness to restricted subclasses of beliefs. \cite{PeiStrulovici2024} consider robustness to large state perturbations that occur with small probability. \cite{Lopomoetal2022} and \cite{OllarPenta2023} study robust full-surplus extraction under belief restrictions. By contrast, our results are about quota mechanisms, which do not use transfers. 

Another strand of the literature generalizes quota mechanisms to dynamic environments in which agents’ types follow a Markov chain. These mechanisms rely on precise knowledge of the underlying state process. \cite{EscobarToikka2013} propose a \emph{credible reporting mechanism} that uses statistical tests to reject histories of reports that are unlikely if agents are truthful. For any payoff profile that can be achieved by a convex combination of an efficient decision rule and a constant decision rule, they construct an associated mechanism. In every equilibrium of this mechanism, payoffs converge to  
the desired payoff profile in the patient limit.\footnote{\cite{EscobarToikka2013} build upon this mechanism to show that these payoff vectors can also be approximated by equilibria in the associated game in which each player controls his own actions. \cite{RenaultSolanVieille2013} provide a similar characterization of the limit set of equilibrium payoffs in a dynamic sender--receiver game.}
Their proof establishes bounds on the agents' payoffs, without solving for equilibrium strategies. Thus, little can be concluded about the implemented \emph{decisions}, which determine the \emph{principal's} payoffs.\footnote{\cite{GorokhEtal2021} use a payoff approach to show that artificial currency mechanisms can approximate static monetary mechanisms, up to incentive and welfare errors that depend on the length of the horizon. In a discounted, infinite-horizon repeated allocation problem, 
\cite{BalseiroEtal2019} present a mechanism that asymptotically implements an efficient allocation. They show that the welfare loss converges to zero, as the discount factor tends to $1$, at a rate that is faster than under quota mechanisms. Their focus is on the agents' welfare, rather than on the implemented decisions.}  
\cite{RenouTomala2015} construct a similar mechanism. They show that a given ``undetectable efficient'' social choice function is approximately implemented in every communication equilibrium of their mechanism. 

\cite{GuoHorner2018} analyze the case of a fixed discount factor. They consider a repeated single-good allocation problem in which the agent's valuation is binary and follows a Markov chain. They solve for the welfare-maximizing linking mechanism; it is not a discounted quota mechanism because the state is persistent. By contrast, \cite{Frankel2016discounted} shows that a discounted quota mechanism is exactly optimal in a repeated delegation problem (with transfers) in which the state is distributed independently across periods and the agent has state-independent preferences. 


Methodologically, we introduce a new form of quota mechanism in which each agent's choice of message can be formulated as an optimal transport problem. We analyze this optimal transport problem in order to establish a tight bound on the ex-post decision error under these quota mechanisms. We believe that our paper is the first to explicitly apply optimal transport techniques to quota mechanisms. Our approach is related to \cite{Rahman2011}, which uses linear duality to give an alternative proof of  \citeapos{Rochet1987} characterization of implementable allocation rules. \cite{Rahman2011} does not mention optimal transport or quota mechanisms, but the linear duality in that paper shows that transfers correspond to Lagrange multipliers on ``detectable deviations.''  
We interpret essentially the same duality in our statement of the equivalence between quota and transfer implementation. Finally, \cite{LinLiu2022} use optimal transport to relate their notion of credibility in Bayesian persuasion with a form of cyclical monotonicity. 


\section{Model} \label{sec:model}

\subsection{Setting} \label{sec:setting}

There is a principal  and there are $n$ agents, labeled $i = 1, \ldots, n$. Consider a Bayesian collective decision problem with independent, private values.  This problem is denoted by $(\XX, \Th, u, \pi)$, where $\XX$ is a measurable space of decisions; $\Th = \prod_{i=1}^{n} \Th_i$ is a finite set of type profiles; $u = (u_1, \ldots, u_n)$ specifies each agent $i$'s bounded von Neumann--Morgenstern utility function $u_i \colon \XX \times \Th_i \to \R$; and $\pi = (\pi_1, \ldots, \pi_n) \in \prod_{i=1}^{n} \D (\Th_i)$ is a profile of full-support priors. A social choice function is a map $x \colon \Th \to \D(\XX)$, which assigns to each type profile $\th = (\th_1, \ldots, \th_n)$ a decision lottery $x(\th)$ in $\D(\XX)$. We linearly extend each utility function $u_i$ from the domain $\XX$ to the domain $\D (\XX)$. 

As in \citet[hereafter JS]{jackson2007overcoming}, there are $K$ independent copies of the primitive problem, labeled $k = 1, \ldots, K$. In this $K$-composite problem, each agent $i$ knows his type vector $\bm{\th}_i = (\th_i^1, \ldots, \th_i^K) \in \Th_i^K$.\footnote{In \cref{sec:dynamics}, we analyze a setting in which each agent's information arrives over time.} Agent $i$'s utility is additively separable across problems: his utility from a decision vector $\bm{x} = (x^1, \ldots, x^K) \in \XX^K$ is the average $\frac{1}{K} \sum_{k=1}^{K} u_i ( x^k, \th_i^k)$.
Types are drawn independently across agents and problems according to the priors in $\pi$.\footnote{For some of our robustness results, the assumption of independence across problems can either be dropped (as in \cref{res:optimal_transport_bound}, with $n =1$) or relaxed to exchangeability (as in \cref{res:robust}).} All private information in the $K$-composite problem can be collected in a single vector $\bm{\th} = (\bm{\th}_1, \ldots, \bm{\th}_n) = (\th^1, \ldots, \th^K)$. Here and below, agents are indicated by subscripts, problem copies by superscripts. We bold vectors that range over problems $k = 1,\ldots, K$. If there is a single agent ($n=1$), we drop agent subscripts. 

\subsection{Linking mechanisms and quota mechanisms}

In the $K$-composite problem, a \emph{linking mechanism} is a pair $(M ,g)$ consisting of a measurable message space $M = \prod_{i=1}^{n} M_i$ and an outcome rule\footnote{Here and below, all maps are assumed measurable; any product of measurable spaces is endowed with the product $\s$-algebra. We also make the technical assumption that for each measurable space we consider, the associated $\s$-algebra separates points. That is, for any pair of distinct points, there exists a measurable set that contains the first point but not the second.}
\[
    g = (g^1, \ldots, g^K) \colon M \to [\D ( \XX)]^{K}.
\]
The outcome rule specifies only the marginal distribution over decisions in each problem, rather than the joint distribution over $K$-vectors of decisions. These marginals determine payoffs because utilities are additively separable across problem copies. In the $K$-composite problem, a linking mechanism $(M,g)$ induces a game between the agents. In this game, a (behavior) strategy for agent $i$ is a map $\s_{i} \colon \Th_i^K \to \D ( M_{i})$. 

We now define a special class of linking mechanisms, called quota mechanisms. Consider a social choice function $x \colon \Th \to \D (\XX)$ and a quota profile $q = (q_1, \ldots, q_n) \in \prod_{i=1}^{n} \D (\Th_i)$. We first describe the $(x,q)$-quota mechanism informally. Agent $i$ is asked to report on each problem a type \emph{distribution}, subject to the constraint that the $K$ reported distributions average to his quota $q_i$. On each problem, the principal independently samples a type from each agent's reported distribution and then applies the social choice function $x$ to the sampled type profile.

Formally, in the $K$-composite problem, the \emph{$(x,q)$-quota mechanism} is the linking mechanism $(M,g)$ defined as follows.\footnote{Technically, there is a distinct $(x,q)$-quota mechanism for each $K$. Below, when we allow $K$ to vary, we refer to the $(x,q)$-quota mechanisms.} Let $M = \prod_{i=1}^{n} M_i$, where
\[
    M_i = \Set{ \bm{r}_i = (r_i^1, \ldots, r_i^K) \in [\D (\Th_i)]^K: \frac{1}{K} \sum_{k=1}^{K} r_i^k = q_i}.
\]
The quota $q_i$ links together agent $i$'s reports across the $K$ problems; agent $i$'s quota is not affected by the reports of the other agents. For each $\bm{r} = (\bm{r}_1, \ldots, \bm{r}_n) \in M$, let
\[
    g ( \bm{r} ) =  \bigl( x( \oprod_{i=1}^{n} r_i^1), \ldots, x(\oprod_{i=1}^{n} r_i^K)  \bigr), 
\]
where $\otimes$ denotes the product of probability measures, and the map $x \colon \Th \to \D (\XX)$ is extended linearly to the domain $\D ( \Th)$.\footnote{For each problem $k$ and each type profile $\th' = (\th_1', \ldots, \th_n') \in \Th$ we have $(\oprod_{i=1}^{n} r_i^k) (\th') = \prod_{i=1}^{n} r_i^k ( \th_i')$, so
\[
    x ( \oprod_{i=1}^{n} r_i^k) =
    \sum_{\th' \in \Th} x (\th') \prod_{i=1}^{n} r_i^k(\th_i') \in \D (\XX). 
\]}

To illustrate this definition, consider the following simple example with two players. Let $\Th_1 = \Th_2 = \{A, B, C\}$. There are $K = 3$ problem copies. Fix a social choice function $x \colon \{A,B,C\}^2 \to \D( \XX)$. The principal uses the $(x,q)$-quota mechanism with $q_1 = q_2 = (1/3,1/3,1/3) \in \D(\{A,B,C\})$. Suppose that agent $1$'s realized type vector is $(A,B,C)$ and agent $2$'s realized type vector is $(A,A,B)$. Suppose that the agents report, respectively,
\[
    \bm{r}_1 = (\d_A, \d_B, \d_C), 
    \qquad
    \bm{r}_2 = ( \d_A, (1/2) \d_B + (1/2) \d_C, (1/2) \d_B + (1/2) \d_C).
\]
Note that each agent's report vector satisfies his quota. Informally, we say that agent $1$ reports type $A$ (respectively $B$, $C$) on problem $1$ (respectively $2$, $3$). So agent $1$ is truthful on each problem, and agent $2$ is truthful on problem $1$. Note, however, that a quota mechanism is not a direct mechanism because $M_i \neq \Th_i^K$.\footnote{By the revelation principle, any equilibrium of a quota mechanism could be implemented as a truthful equilibrium of an associated direct mechanism. We find it clearer to work with quota mechanisms.} Given these reports, the principal selects $x(A,A)$ on problem $1$; on problem $2$, she selects $x(B,B)$ with probability $1/2$ and $x(B,C)$ with probability $1/2$; on problem $3$, she selects $x(C,B)$ with probability $1/2$ and $x(C,C)$ with probability $1/2$. A more substantive example (with a single agent) is given in \cref{sec:simple_example}.

\begin{rem}[Implementation via decision restrictions] \label{rem:decisions} If there is a single agent ($n=1$) and the social choice function $x$ is deterministic, then the $(x,q)$-quota mechanism can be implemented by letting the agent choose directly from the following menu of decision vectors:
\[
    \Set{ (\bar{x}^1, \ldots, \bar{x}^K) \in  [\D(\XX)]^K : \frac{1}{K} \sum_{k=1}^{K} \bar{x}^k = x(q) },
\]
where $x$ is the linear extension that maps $\D(\Th)$ to $\D(\XX)$. See \cref{sec:proof_decisions} for the proof. Like a grading curve or prescription quota, this implementation restricts decisions rather than reports. 
\end{rem}
 
Our definition of a quota mechanism is slightly different from that in JS. In JS's quota mechanism, agent $i$ reports a $K$-vector of \emph{types}, subject to the constraint that the frequencies of the reported types match the quota $q_i$. In order for this constraint to be feasible, the components of the quota $q_i$ must be integer multiples of $1/K$. To accommodate general quotas that are not divisible by $1/K$, JS's mechanism involves further modifications. These modifications can introduce additional decision errors, as we discuss after \cref{res:optimal_transport_bound}.

\section{Simple example of a quota mechanism} \label{sec:simple_example}

In this section, we illustrate our quota mechanisms in a simple example with a single agent. In the primitive problem, the principal chooses the probability of allocating a good to the agent. The agent's valuation for the good is high ($\th_H$) with probability $\pi$ and low ($\th_L$) with probability $1 - \pi$, where $0 < \th_L < \th_H$ and $0 < \pi < 1$. Let $x$ be the social choice function that allocates the good if and only if the agent's valuation is high:  $x(\th_H) = 1$ and $x (\th_L) = 0$. 

Consider the $K$-composite problem. On each problem $k$, there is a copy of the good that the principal can allocate. For example, the principal could be a manager who chooses whether to allocate some resource (such as computing power or the support of an intern) to each of $K$ projects that an employee is working on. The agent's type vector $\bm{\th} = (\th^1, \ldots, \th^K) \in \{ \th_L, \th_H \}^{K}$ specifies his valuation for the good in each of the $K$ problems. The valuations are drawn from $\pi$, independently across problem copies; here  we identify a distribution over $\{\th_L, \th_H\}$ with the probability assigned to $\th_H$. The agent's utility from a decision vector $(x^1, \ldots, x^K) \in [0,1]^K$ is $\frac{1}{K} \sum_{k=1}^{K} \th^k x^k$. 

Suppose that the principal seeks to implement this social choice function $x$ on each problem copy. Consider the $(x,q)$-quota mechanism, where the quota $q$ is in $(0,1)$. The agent is asked to report a vector $(r^1, \ldots, r^K) \in [0,1]^K$ satisfying $\frac{1}{K} \sum_{k=1}^{K} r^k = q$. On problem $k$, the agent gets the allocation $x(\th_H)$ with probability $r^k$ and the allocation $x(\th_L)$ with probability $1 - r^k$. Thus, the agent is allocated the good with probability $r^k$. Equivalently, the principal allocates the agent an aggregate probability mass $q K$ of receiving the good. The agent chooses how to distribute this mass across the $K$ problems. (This implementation is a special case of the construction in \cref{rem:decisions}.)

To maximize expected utility, the agent chooses his report vector to maximize his aggregate probability of receiving the good \emph{on problems in which his valuation is high}. Denote by $K_H = K_H (\bm{\th})$ the number of high-valuation problems. If $K_H > q K$, then it is not feasible for the agent to receive the good on every high-valuation problem. In this case, it is optimal to report $0$ on every low-valuation problem; his reports on the high-valuation problems must then average to $q K / K_H$. If $K_H \leq q K$, then it is feasible for the agent to receive the good on every high-valuation problem. In this case, it is optimal to report $1$ on every high-valuation problem; his reports on the low-valuation problems must then average to $(q K - K_H)/(K - K_H)$. If $K_H / K$ is close to $q$, then the agent's average probability of getting the good on high-valuation (respectively, low-valuation) problems is close to $1$ (respectively, $0$). By the law of large numbers, $K_H / K$ is likely to be close to the prior $\pi$ if $K$ is large. Thus, if the quota $q$ is set equal to $\pi$, then this quota mechanism approximately
implements the social choice function $x$.

\section{Finite-sample decision error under quota mechanisms} \label{sec:optimal_transport_bound}

The fundamental challenge for quota mechanisms is that the empirical distribution of each agent $i$'s realized type vector $\bm{\th}_i$ may differ significantly from his quota $q_i$, particularly when the number $K$ of problems (the sample size) is small. In this section, we bound the decision error that results from such a discrepancy between each agent's realized type frequencies  and his quota. Moreover, we show that this error guarantee cannot be improved by any other linking mechanisms.


To state our bound, we need a few definitions. We begin with cyclical monotonicity. First suppose that there is a single agent ($n=1$). In this case, a social choice function $x \colon \Th \to \D(\XX)$ is \emph{cyclically monotone} if for all integers $J \geq 2$ and all types $\th_1, \ldots, \th_J \in \Th$, 
we have
\[
    \sum_{j=1}^{J} u (x(\th_j), \th_j) \geq \sum_{j=1}^{J} u (x(\th_{j+1}), \th_j),
\]
where we use the convention that $\th_{J+1} = \th_1$. In words, cyclical monotonicity requires that there is no cycle of types that would strictly gain, on average, if each type received the allocation intended for the next type instead of his own type. For example, if $\XX$ and $\Th$ are totally ordered, and $u \colon \XX \times \Th \to \R$ is supermodular,\footnote{That is, for all $x,x' \in \XX$ and $\th, \th' \in \Th$, if 
$x < x'$ and $\th < \th'$, then $u(x,\th) + u(x', \th') \geq u ( x, \th') + u(x', \th)$.} then every weakly increasing deterministic function $x \colon \Th \to \XX$ is cyclically monotone.

To extend this definition to the case of multiple agents, we must take expectations over the types of the other agents. Suppose that $n > 1$. Given any distribution profile $p = (p_1, \ldots, p_n) \in \prod_{i=1}^{n} \D(\Th_i)$, let $p_{-i} = \oprod_{j \neq i} p_j$. A social choice function $x \colon \Th \to \D(\XX)$ is \emph{$p$-cyclically monotone} if for each agent $i$ the following holds: for all integers $J \geq 2$ and all types $\th_{i,1}, \ldots, \th_{i,J} \in \Th_i$, we have
\begin{equation} \label{eq:CM}
    \sum_{j=1}^{J}  \E_{\th_{-i} \sim p_{-i}} \Brac{u_i ( x (\th_{i,j}, \th_{-i}), \th_{i,j})}
 \geq \sum_{j=1}^{J}  \E_{\th_{-i} \sim p_{-i}} \Brac{ u_i ( x ( \th_{i,j+1}, \th_{-i}), \th_{i,j})},
\end{equation}
where $\th_{i,J+1} = \th_{i,1}$. If there is a single agent, we adopt the convention that for any type distribution $p$, the term $p$-cyclical monotonicity means cyclical monotonicity. 

We need a few more definitions. Consider a fixed type vector $\bm{\th}_i$ in $\Th_i^K$. The empirical (or marginal) distribution of the realized vector $\bm{\th}_i$, denoted 
$\marg \bm{\th}_i$ or $\marg ( \cdot | \bm{\th}_i)$, is the probability measure on $\Th_i$ defined by
\[
    \marg ( \th_i | \bm{\th}_i) = \frac{| \{ k : \th_i^k = \th_i \}|}{K}, \qquad \th_i \in \Th_i.
\]
For example, if $\Th_i = \{ A, B, C\}$ and $K = 4$, then $\marg ( A, C, B, A)$ assigns probability $1/2$ to $A$ and probability $1/4$ each to $B$ and $C$.

Finally, in the spaces $\D(\Th_i)$ and $\D(\XX)$, we measure the distance between distributions using the total variation norm, denoted by ${\| \cdot \|}$.\footnote{Given a measurable space $Z$, for any $\mu, \nu \in \D (Z)$, let $\| \mu - \nu \| = \sup_{A} | \mu (A) - \nu (A) |$, where the supremum is over all measurable subsets $A$ of $Z$. This norm does not require a topology on the space $Z$. The total variation norm is convenient because of its optimal transport foundation: $\| \mu - \nu\|$ is the minimum probability that is moved when transporting $\mu$ to $\nu$.} On $\D(\XX)$, we use this norm to measure the frequency of incorrect decisions.

\subsection{Ex-post error bound}

We first bound the decision error under a quota mechanism, for each realization of the agents' private information. This foundational bound will be used to prove many of our subsequent results. 

\begin{thm}[Optimal ex-post error bound] \label{res:optimal_transport_bound} 
Fix $q \in \prod_{i=1}^{n} \D(\Th_i)$. Let $x \colon \Th \to \D ( \XX)$ be $q$-cyclically monotone. In the $K$-composite problem, the $(x,q)$-quota mechanism $(M,g)$ has a Bayes--Nash equilibrium $\s$ satisfying, for all realizations $\bm{\th}$ in $\Th^K$,\footnote{This equilibrium $\s$ is pure, so $\s( \bm{\th}) \coloneqq (\s_i (\bm{\th}_i))_{i=1}^{n} \in M$. By the definition of the $(x,q)$-quota mechanism $(M,g)$, we have $g^k (\s ( \bm{\th})) = x ( \oprod_{i=1}^{n} \s_i^k ( \bm{\th}_i))$. This decision lottery assigns to each measurable subset $A$ of $\XX$ the probability
\[
    \sum_{(\th_1', \ldots, \th_n') \in \Th} x(A | \th_1', \ldots, \th_n')  \prod_{i=1}^{n} \s_i^k ( \th_i' | \bm{\th}_i),
\]
where $x ( A | \th_1', \ldots, \th_n')$ is the probability that $x (\th_1', \ldots, \th_n')$ assigns to the set $A$. More generally, if the strategy $\s$ is mixed, then $\s ( \bm{\th})$ denotes the product measure $\oprod_{i=1}^{n} \s_{i} (\bm{\th}_i)$, and each map $g^k \colon M \to \D (\XX)$ is extended linearly to the domain $\D (M)$. Under a quota mechanism, however, every mixed strategy is outcome-equivalent to some pure strategy.} 
\begin{equation} \label{eq:EP_bound}
\frac{1}{K} \sum_{k=1}^{K} \| g^k ( \s(\bm{\theta})) - x( \theta^k) \| \leq \sum_{i=1}^{n} ( | \Theta_i| - 1) \| q_i - \marg \bm{\theta}_i \|.
\end{equation}
Moreover, the constants $|\Th_i| - 1$ cannot be reduced, even using arbitrary linking mechanisms.
\end{thm}

For each realization $\bm{\th}$ of the agents' private information,  the inequality in \eqref{eq:EP_bound} bounds the frequency with which the decision is incorrect (relative to the decision specified by the social choice function $x$).\footnote{The average on the left side of \eqref{eq:EP_bound} could alternatively be defined by first averaging the decisions over problems with the same realized preference profile. This alternative average is weakly smaller than our average, with equality if $x$ is deterministic. \cref{res:optimal_transport_bound} holds with this alternative definition.}
The bound depends on the distance, for each agent $i$, between the quota $q_i$ and the empirical distribution $\marg \bm{\th}_i$ of agent $i$'s realized type vector $\bm{\th}_i$. Moreover, this bound is optimal in the following sense. If the constants $|\Th_i| - 1$ are reduced, then the result no longer holds, even if the principal can use arbitrary linking mechanisms. To show this, we consider the case $n = 1$. We drop agent subscripts. For any type space $\Th$ with $|\Th| \geq 2$ and any integer $K$ satisfying $K \geq |\Th|$, we construct a decision environment $(\XX, u)$, a cyclically monotone social choice function $x$, and a quota $q \in \D(\Th)$ such that the following holds: if the coefficient $|\Th| - 1$ in \eqref{eq:EP_bound} is strictly reduced, then for any linking mechanism $(M,g)$ in the $K$-composite problem, the modified version of \eqref{eq:EP_bound} fails for some realization $\bm{\th}$ in $\Th^K$.\footnote{In the special setting of bilateral trade with transfers, \cite{Cohn2010} proposes an alternative to JS's quota mechanism. Under this mechanism, the share of problems in which the induced allocation is inefficient converges to $0$ exponentially in $K$. By contrast, our bound in \cref{res:optimal_transport_bound} holds for every $q$-cyclically monotone social choice function.}

The bound on the average decision error in \eqref{eq:EP_bound} implies a bound on the principal's utility loss. If the principal's utility function is normalized to have range $[0,1]$, then at realization $\bm{\th}$, the principal's utility loss (relative to implementing the social choice function $x$) is bounded by the right side of \eqref{eq:EP_bound}.

\begin{rem}[JS's quota mechanisms] \label{rem:JS_quota} \cref{res:optimal_transport_bound} would not hold if we used JS's definition of a quota mechanism in place of our definition; see \cref{sec:comparison} for a counterexample. In fact, every type of every agent gets weakly higher expected utility under the equilibrium $\s$ from \cref{res:optimal_transport_bound} than under the equilibrium constructed by JS in their associated quota mechanism; \cref{sec:comparison} gives the proof and a numerical example.  

On the other hand, we show in \cref{sec:comparison} that under JS's definition of a quota mechanism, a weaker version of \cref{res:optimal_transport_bound} holds, where \eqref{eq:EP_bound} is relaxed to
\begin{equation} \label{eq:EP_bound_JS}
   \frac{1}{K} \sum_{k=1}^{K} \| g^k ( \s(\bm{\theta})) - x( \theta^k) \| \leq  \sum_{i=1}^{n} ( |\Th_i| - 1) \Paren{ \| q_i - \marg \bm{\th}_i \| + \frac{ 2 |\Th_i| -1}{K} }.
\end{equation}
Under JS's quota mechanism, each quota $q_i$ is approximated by a quota $q_{K,i}$ whose components are integer multiples of $1/K$. Their mechanism elicits type reports satisfying these new quotas. The submitted reports are then modified randomly so that the original quotas are satisfied in expectation. Finally, the desired social choice function is applied to the modified reports. These modifications can introduce additional errors into the decision. By bounding these errors, we obtain \eqref{eq:EP_bound_JS}.
\end{rem}


\begin{rem}[Lower bound] \label{rem:lower_bound}
\cref{res:optimal_transport_bound} gives an upper bound on the ex-post decision error. For certain social choice functions, we can obtain an accompanying lower bound. A social choice function $x \colon \Th \to \D (\XX)$ is \emph{injective} if for any distinct type profiles $\th, \th' \in \Th$, the lotteries $x ( \th)$ and $x(\th')$ concentrate on disjoint measurable sets.\footnote{If $x$ is deterministic, then this reduces to the usual definition of injectivity of a function; if $x$ is stochastic, then this definition is stronger than the usual definition.} If $x$ is injective, then in the $K$-composite problem, \emph{every} strategy profile $\s$ in the $(x,q)$-quota mechanism $(M,g)$ satisfies, for each realization $\bm{\th}$ in $\Th^K$, 
\begin{equation} \label{eq:lower_bound}
\frac{1}{K} \sum_{k=1}^{K} \| g^k ( \s(\bm{\theta})) - x( \theta^k) \| \geq \max_{i=1,\ldots, n} \|q_i -  \marg \bm{\theta}_i \|. 
\end{equation} 
See \cref{sec:proof_optimal_transport_bound} for the proof. In the special case of a single agent with two possible types, this lower bound agrees with the upper bound in \eqref{eq:EP_bound}, so there exists a best response for the agent that minimizes the decision error over all strategies.\end{rem}

\begin{rem}[Decision-space upper bound] \label{rem:refined_upper_bound} Suppose that there is a single agent ($n =1$) and the social choice function $x$ is deterministic. In this case, the $(x,q)$-quota mechanism can be implemented by restricting decisions rather than reports; see \cref{rem:decisions}. Moreover, if $x$ is cyclically monotone, then the bound in \eqref{eq:EP_bound} can be strengthened to
\begin{equation} \label{eq:EP_bound_decision}
   \frac{1}{K} \sum_{k=1}^{K} \| g^k ( \s(\bm{\theta})) - x( \theta^k) \| \leq ( | x(\Th)|  - 1) \| x(q) - x(\marg \bm{\th}) \|.
\end{equation}
See \cref{sec:proof_optimal_transport_bound} for the proof. This inequality \eqref{eq:EP_bound_decision} coincides with \eqref{eq:EP_bound} if $x$ is injective, but \eqref{eq:EP_bound_decision} is strictly stronger if $x$ is not injective.\footnote{We thank Drew Fudenberg for suggesting such a result.} In the context of a mandatory grading curve, $|x(\Th)|$ is the number of grades and $\| x(q) - x (\marg \bm{\th}) \|$ is the difference between the mandatory grade distribution and the
empirical distribution of grades that the realized class would receive under $x$. The guarantee is independent of $|\Th|$, the number of possible raw scores.
\end{rem}

\begin{rem}[Partial versus full implementation] \label{rem:full}
\cref{res:optimal_transport_bound} is about partial implementation---it says that there exists \emph{some} equilibrium satisfying the bound \eqref{eq:EP_bound}. Without further assumptions, \eqref{eq:EP_bound} need not hold for \emph{every} equilibrium. In particular, if each agent is indifferent between all decisions, then every strategy profile is an equilibrium.  We can sharpen the conclusion, in the case of a single agent, under a condition that rules out such indifference.

Suppose $n = 1$.  A social choice function $x \colon \Th \to \D(\XX)$ is \emph{strictly cyclically monotone} if for all integers $J \geq 2$ and all types $\th_1, \ldots, \th_J \in \Th$ such that $x(\th_1), \ldots, x(\th_{J})$ are distinct, we have
\begin{equation} \label{eq:strict_CM}
    \sum_{j=1}^{J} u ( x (\th_j), \th_j) > \sum_{j=1}^{J} u ( x (\th_{j+1}), \th_j),
\end{equation}
where $\th_{J +1} = \th_1$. If $x$ is strictly cyclically monotone, then \eqref{eq:EP_bound} holds for \emph{every} best-response $\s$ to the $(x, q)$-quota mechanism;\footnote{With a single agent, \citeapos{RenouTomala2015} notion of ``undetectable efficiency'' reduces to our notion of strict cyclical monotonicity.} see \cref{sec:proof_optimal_transport_bound} for the proof. 
\end{rem}


To prove \cref{res:optimal_transport_bound}, we first observe that each agent effectively faces an optimal transport problem.\footnote{For background on optimal transport, see  \cref{sec:OT}.}  To see this, first consider the case of a single agent. In the $K$-composite problem, under the $(x,q)$-quota mechanism, suppose that the agent has type vector $\bm{\th} = (\th^1, \ldots, \th^K)$ and the agent reports $\bm{r} = (r^1, \ldots, r^K)$. The agent's expected payoff is 
\[
    \frac{1}{K} \sum_{k=1}^{K} \sum_{\th' \in \Th} u ( x( \th'), \th^k) r^k (\th').
\]
Grouping the outer summation according to the values of $\th^1, \ldots, \th^K$, we get
\begin{equation}\label{eq:OT_obj}
     \sum_{\th, \th' \in \Th} u (x(\th'), \th) \g(\th, \th'),
\end{equation}
where $\g$ is the probability distribution on $\Th \times \Th$ defined by
\[
    \g( \th, \th')   = \frac{1}{K} \sum_{k: \th^k = \th} r^k ( \th'), \qquad \th, \th' \in \Th.
\]
The first marginal of $\g$ is $\marg \bm{\th}$. Since the report vector $\bm{r}$ satisfies the quota $q$, the second marginal of $\g$ is $q$. Thus, $\g$ is a \emph{coupling} of $\marg \bm{\th}$ and $q$. Conversely, any coupling  of $\marg \bm{\th}$ and $q$ can be induced in this way by some report vector $\bm{r}$ that satisfies the quota $q$.\footnote{\label{ft:plan}The coupling $\g$ can be interpreted as a \emph{transport plan} where each type $\th$ in $\supp (\marg \bm{\th})$ is mapped to the distribution $r(\th)$ defined by 
$r(\cdot |\th) = \frac{\g(\th, \cdot)}{\sum_{\th'} \g(\th, \th')}$. For each $k$, set $r^k = r(\th^k)$.} Thus, the agent equivalently chooses a coupling of $\marg \bm{\th}$ and $q$ to maximize the expression in \eqref{eq:OT_obj}.

We illustrate this optimal transport construction in a simple example with $\Th = \{ A, B, C, D \}$ and $K = 4$. The principal uses the $(x,q)$-quota mechanism with the uniform quota $q = (1/4, 1/4, 1/4, 1/4)$. The agent has type vector $\bm{\th}$, and he considers two different report vectors, $\bm{r}$ and $\bm{r}'$, where
\begin{equation*}
\begin{aligned}
    \bm{\th} &= ( A, A, B,C), \\
    \bm{r} &= ( \d_A, \d_D,\d_B, \d_C), \\
    \bm{r}' &= (\d_A, \d_B, \d_C, \d_D).
\end{aligned}
\end{equation*}
Here, $\marg \bm{\th} = (1/2, 1/4, 1/4, 0)$. \cref{fig:transport} illustrates the couplings induced by $\bm{r}$ (left) and $\bm{r}'$ (right). Under each coupling, every pair $(\th, \th')$ is assigned probability $0$ or $1/4$; we shade the corresponding square if it is assigned probability $1/4$. Summing each column yields the initial distribution $\marg \bm{\th}$, shown on the horizontal axis. Summing each row yields the final distribution $q$, shown on the vertical axis. 
In each grid,  we highlight the diagonal. The total probability that is moved (i.e., the probability off the diagonal) represents the share of problems on which the agent is untruthful. 

\begin{figure}
\centering
\begin{minipage}{0.95\textwidth}
\begin{multicols}{2}

\centering

\begin{tikzpicture}[scale=0.8]

\draw[thick] (0,0) grid (4,4);

\fill[Blue] (1,-2) rectangle (2,-1);
\fill[Blue] (2,-2) rectangle (3,-1);
\fill[Blue] (0,-2) rectangle (1,-1);
\fill[Blue] (0,-3) rectangle (1,-2);

\fill[Blue] (-2,1) rectangle (-1,2);
\fill[Blue] (-2,2) rectangle (-1,3);
\fill[Blue] (-2,3) rectangle (-1,4);
\fill[Blue] (-2,0) rectangle (-1,1);

\fill[Blue] (1,1) rectangle (2,2);
\fill[Blue] (2,2) rectangle (3,3);
\fill[Blue] (0,0) rectangle (1,1);
\fill[Blue] (0,3) rectangle (1,4);

\draw[thick] (0,0) grid (4,4);
\draw[thick] (0,-2) grid (4,-1);
\draw[thick] (0,-3) grid (1,-1);
\draw[thick] (-2,0) grid (-1,4);

\path (-1,0) rectangle (0,1) node[midway]{\large A};
\path (-1,1) rectangle (0,2) node[midway]{\large B};
\path (-1,2) rectangle (0,3) node[midway]{\large C};
\path (-1,3) rectangle (0,4) node[midway]{\large D};

\path (0,-2) rectangle (1,1) node[midway]{\large A};
\path (1,-2) rectangle (2,1) node[midway]{\large B};
\path (2,-2) rectangle (3,1) node[midway]{\large C};
\path (3,-2) rectangle (4,1) node[midway]{\large D};

\foreach \i in {0,...,3} {
    \draw[Red, ultra thick] (\i,\i) rectangle ++(1,1);
}

\end{tikzpicture}

\columnbreak

\begin{tikzpicture}[scale=0.8]

\draw[thick] (0,0) grid (4,4);

\fill[Blue] (1,-2) rectangle (2,-1);
\fill[Blue] (2,-2) rectangle (3,-1);
\fill[Blue] (0,-2) rectangle (1,-1);
\fill[Blue] (0,-3) rectangle (1,-2);

\fill[Blue] (-2,1) rectangle (-1,2);
\fill[Blue] (-2,2) rectangle (-1,3);
\fill[Blue] (-2,3) rectangle (-1,4);
\fill[Blue] (-2,0) rectangle (-1,1);

\fill[Blue] (1,2) rectangle (2,3);
\fill[Blue] (2,3) rectangle (3,4);
\fill[Blue] (0,0) rectangle (1,1);
\fill[Blue] (0,1) rectangle (1,2);

\draw[thick] (0,0) grid (4,4);
\draw[thick] (0,-2) grid (4,-1);
\draw[thick] (0,-3) grid (1,-1);
\draw[thick] (-2,0) grid (-1,4);

\path (-1,0) rectangle (0,1) node[midway]{\large A};
\path (-1,1) rectangle (0,2) node[midway]{\large B};
\path (-1,2) rectangle (0,3) node[midway]{\large C};
\path (-1,3) rectangle (0,4) node[midway]{\large D};

\path (0,-2) rectangle (1,1) node[midway]{\large A};
\path (1,-2) rectangle (2,1) node[midway]{\large B};
\path (2,-2) rectangle (3,1) node[midway]{\large C};
\path (3,-2) rectangle (4,1) node[midway]{\large D};

\foreach \i in {0,...,3} {
    \draw[Red, ultra thick] (\i,\i) rectangle ++(1,1);
}

\end{tikzpicture}
\end{multicols}
\end{minipage}
\caption{Couplings under a quota mechanism}
\label{fig:transport}
\end{figure}

Now consider the case of multiple agents. Under the $(x,q)$-quota mechanism, agent $i$ knows that his opponents must submit reports that satisfy their quotas. Say that an agent $j$'s strategy is \emph{label-free} if any permutation of agent $j$'s type vector results in the same permutation of agent $j$'s reports. As long as agent $j$'s strategy is label-free, then agent $j$'s report on every problem $k$ has the same expectation, namely $q_j$. Facing label-free strategies by his opponents, type $\bm{\th}_i$ equivalently chooses a coupling $\g_i$ of $\marg \bm{\th}_i$ and $q_i$ to maximize
\[
   \sum_{\th_i, \th_i' \in \Th_i} u_i (\th_i' | \th_i) \g_i (\th_i, \th_i'), 
    \quad
    \text{where}
    \quad
    u_i (\th_i' | \th_i ) = \E_{ \th_{-i} \sim q_{-i}} \Brac{ u_i (x(\th_i', \th_{-i}), \th_i)}.
\]
To prove \cref{res:optimal_transport_bound}, we analyze this optimal transport problem.

First, suppose that $\marg \bm{\th}_i = q_i$, so the initial and final distributions agree. Since $x$ is $q$-cyclically monotone, it follows from a standard result in optimal transport theory \citep[e.g.,][Theorem 5.10, pp.~57--59]{Villani2009}  that keeping all mass fixed is an optimal coupling.

Next, suppose that $\marg \bm{\th}_i \neq q_i$.  By a standard property of the total variation norm $\| \cdot\|$, there is a \emph{feasible} coupling of $\marg \bm{\th}_i$ and $q_i$ that moves probability $\| q_i - \marg \bm{\th}_i \|$ and keeps the remaining probability fixed. But this coupling may not be optimal. Instead, the discrepancy between $\marg \bm{\th}_i$ and $q_i$ can trigger a ``cascade of lies.'' We bound the size of this cascade. Since $x$ is  $q$-cyclical monotone, we can apply \cref{res:DZ_bound} (\cref{sec:OT}) to conclude that there is an \emph{optimal} coupling that moves at most probability $( | \Th_i| - 1) \| q_i - \marg \bm{\th}_i \|$. To prove \cref{res:DZ_bound}, we show that there exists an optimal coupling whose support contains no nontrivial cycles. Therefore, the probability moved under this coupling moves along paths in $\Th_i$ such that no path visits the same type twice. Thus, each path has length at most $|\Th_i| - 1$. The net probability moved (summing over paths) is exactly $\| q_i - \marg \bm{\th}_i\|$. Since each path has at most $|\Th_i| - 1$ edges, the total probability moved (summing over edges) is at most $(|\Th_i| -1 ) \| q_i - \marg \bm{\th}_i\|$.

We illustrate this argument with the two couplings in \cref{fig:transport}. Let
$\Th_i = \{A,B,C,D\}$. Note that neither coupling contains a nontrivial cycle. The coupling on the left moves probability $1/4 = \| q_i - \marg \bm{\th}_i \|$.  The coupling on the right moves probability $3/4 = ( | \Th_i| - 1) \| q_i - \marg \bm{\th}_i \|$. \cref{fig:cascade} shows the initial distribution $\marg \bm{\th}_i$ in blue and the final distribution $q_i$ in orange. In each coupling, the net probability of $1/4$ is moved along one path. For the coupling on the left, this path has a single edge, shown as a dotted arrow. For the coupling on the right, this path has three edges, shown as solid arrows.



\begin{figure}
\centering
\begin{tikzpicture}
  \begin{axis}[
    ybar, axis on top,
    height=8cm, width=12cm,
    bar width=0.7cm,
    ymajorgrids, tick align=inside,
    major grid style={draw=white},
    enlarge y limits={value=.1,upper},
    ymin=0, ymax=0.6,
    axis x line*=bottom,
    axis y line*=left,
    ytick={0,0.25,0.5},
    y axis line style={opacity=1},
    tickwidth=0.8pt,
    enlarge x limits=0.1,
    symbolic x coords={A,B,C,D},
    xtick=data,
  ]

    \addplot [draw=none, fill=Orange] coordinates {
      (A,0.25) (B,0.25) (C,0.25) (D,0.25)
    };

    \addplot [draw=none, fill=Blue] coordinates {
      (A,0.5) (B,0.25) (C,0.25) (D,0)
    };

    
    \coordinate (Ablue75Right) at
      ($(axis cs:A,0.375) + (0.9cm,0cm)$);

    \coordinate (Ablue75RightU) at
      ($(axis cs:A,0.375) + (0.9cm,0.2cm)$);

    \coordinate (BblueTopMid) at
      ($(axis cs:B,0.25) + (0.35cm,0.15cm)$);
    \coordinate (CblueTopMid) at
      ($(axis cs:C,0.25) + (0.35cm,0.15cm)$);

    \coordinate (BorangeTopMid) at
      ($(axis cs:B,0.25) - (0.35cm,-0.15cm)$);
    \coordinate (CorangeTopMid) at
      ($(axis cs:C,0.25) - (0.35cm,-0.15cm)$);
    \coordinate (DorangeTopMid) at
      ($(axis cs:D,0.25) - (0.35cm,-0.15cm)$);
    \coordinate (DorangeTopMidU) at
      ($(axis cs:D,0.25) - (0.35cm,-0.5cm)$);

  \end{axis}


  \draw[->, very thick, Blue, bend left=30]
    (Ablue75Right) to (BorangeTopMid);

  \draw[->, very thick, Blue, bend left=40]
    (BblueTopMid) to (CorangeTopMid);

  \draw[->, very thick, Blue, bend left=40]
    (CblueTopMid) to (DorangeTopMid);

  \draw[->, very thick, Blue, dashed, bend left=40]
    (Ablue75RightU) to (DorangeTopMidU);

\end{tikzpicture}

\caption{Cascade of lies}
\label{fig:cascade}
\end{figure}

The bound in \cref{res:DZ_bound} is a special case of a more general continuity property of optimal transport solutions, which we establish in \cref{res:Lip}. For optimal transport problems between two fixed finite sets, we show that the solution set is Lipschitz continuous as a function of the marginals (with respect to the total variation norm), and we identify the sharp Lipschitz constant. Previous work \citep[e.g.,][]{MangasarianShiau1987,Li1993} shows that the solution set of a general linear program is Lipschitz continuous as a function of the right-side data. With the special optimal transport structure, we obtain a smaller and simpler Lipschitz constant; for details,  see \cref{sec:OT}.


\subsection{Expected error bound} \label{sec:rate_of_convergence}

We now bound the expected decision error under a quota mechanism. Consider the ex-post decision error bound in \cref{res:optimal_transport_bound}. Taking expectations over $\bm{\th}$ (with respect to the profile $\pi$ of priors), we get
\begin{equation}  \label{eq:bound_in_q}
\E_{\bm{\th}} \Brac{ \frac{1}{K} \sum_{k=1}^{K}  \| g^k ( \s(\bm{\theta})) - x( \theta^k) \|} \leq \sum_{i=1}^{n} ( | \Theta_i| - 1) \E_{\bm{\th}_i} \| q_i - \marg \bm{\theta}_i \|.
\end{equation}
The right side depends on the quotas $q_1, \ldots, q_n$. If the components of the probability vectors $\pi_1, \ldots, \pi_n$ are all integer multiples of $1/K$, then the right side is minimized by setting each agent $i$'s quota $q_i$ equal to the prior $\pi_i$.\footnote{For each $\th_i$ in $\Th_i$, the random variable $K \marg (\th_i | \bm{\th}_i)$ follows a binomial distribution. If $K \pi_i(\th_i)$ is an integer, then  $K \pi_i(\th_i)$ is a median of this distribution; see \cite{KaasBuhrman1980}. Thus, $\E_{\bm{\th}_i} | q_i (\th_i) - \marg (\th_i | \bm{\th}_i)|$ is minimized by setting $q_i (\th_i) = \pi_i (\th_i)$.} We plug the quota profile $q = \pi$ into  \eqref{eq:bound_in_q} and then bound each expectation $\E_{ \bm{\theta}_i} \| \pi_i - \marg \bm{\theta}_i  \|$ to get the following. 

\begin{thm}[Expected error bound] \label{res:convergence_rate}
Let $x \colon \Th \to \D ( \XX)$ be $\pi$-cyclically monotone. In the $K$-composite problem, the $(x,\pi)$-quota mechanism $(M,g)$ has a Bayes--Nash equilibrium $\s$ that satisfies
\begin{equation} \label{eq:bound_expected}
    \E_{\bm{\th}} \Brac{ \frac{1}{K} \sum_{k=1}^{K}  \| g^k ( \s(\bm{\theta})) - x( \theta^k) \|}  \leq  \frac{1}{2 \sqrt{K}}  \sum_{i=1}^{n}  ( | \Theta_i | - 1)^{3/2}.
\end{equation}
\end{thm}

\cref{res:convergence_rate} provides a simple guarantee on the expected frequency of incorrect decisions. This guarantee depends only on the number $K$ of problem copies and the number of types of each agent in the primitive problem. The bound in \eqref{eq:bound_expected} cannot be improved by more than a factor of approximately $1.25$, as we show in the proof. This bound is of order $1/\sqrt{K}$. By  \eqref{eq:EP_bound_JS}, the  additional approximation error in JS's quota mechanism is of order $1/K$, so the relative size of the approximation error vanishes in the limit as $K$ grows large.

If there is a single agent and the social choice function $x$ is deterministic, then we can use the refined bound in \cref{rem:refined_upper_bound} to reduce the right side of \eqref{eq:bound_expected} to $\frac{1}{2 \sqrt{K}} ( |x(\Th)| - 1)^{3/2}$. In the grading curve example, $|x(\Th)|$ is the number of grades and $K$ is the number of students in the class. The refined bound is independent of $|\Th|$, the number of raw scores. Thus, the expected share of students receiving the wrong grade is controlled by the square root of the ratio between the number of grades cubed and the size of the class. 

\section{Asymptotic implementation} \label{sec:Bayesian}

In this section, we characterize the social choice functions that can be asymptotically implemented by quota mechanisms, as the number of problem copies grows large. 


\subsection{Implementation equivalence} \label{sec:implementation_equivalence}

We begin by defining asymptotic implementation. For each $K$, let $(M_K, g_K)$ be a linking mechanism in the $K$-composite problem.  Let $x \colon \Th \to \D (\XX)$ be a social choice function. The  sequence $(M_K, g_K)_{K \geq 1}$ \emph{asymptotically implements} $x$ if there is an associated sequence $(\s_K)_{K \geq 1}$ of Bayes--Nash equilibria of $(M_K, g_K)_{K \geq 1}$ such that 
\begin{equation} \label{eq:convergence}
 \lim_{K \to \infty}  \E_{\bm{\th}} \Brac{ \frac{1}{K} \sum_{k=1}^{K} \| g_K^k ( \s_K ( \bm{\th})) -  x(\th^k) \|} = 0.
\end{equation}
Condition \eqref{eq:convergence} requires that the expected average decision error in the $K$-composite problem converges to $0$ as $K$ tends to $\infty$. 


To state our implementation equivalence result, we need a few more definitions. A social choice function $x \colon \Th \to \D(\XX)$ is  \emph{one-shot implementable with transfers} if for each agent $i$ there exists a transfer function $T_i \colon \Th_i \to \R$ such that for all $\th_i, \th_i' \in \Th_i$, we have
\begin{equation} \label{eq:one_shot_def}
    \E_{\th_{-i}} \Brac{ u_i (x (\th_i, \th_{-i}), \th_i)} - T_i (\th_i) \geq \E_{\th_{-i}} \Brac{ u_i ( x (\th_i', \th_{-i}), \th_i)} - T_i (\th_i').
\end{equation}
If there is a single agent, then no expectations are needed in \eqref{eq:one_shot_def}. Next, a \emph{linking mechanism with transfers} is a tuple $(M,g,t)$, where $(M,g)$ is a linking mechanism and $t = (t_1, \ldots, t_n)  \colon M \to \R^n$ specifies a transfer payment from each agent. Assuming quasilinear utility, our definition of asymptotic implementation can be naturally extended to linking  mechanisms with transfers. The term \emph{linking mechanism}, by itself, always refers to a mechanism without transfers. 
\begin{thm}[Implementation equivalence] \label{res:equivalence}
For any social choice function $x \colon \Th \to \D (\XX)$, the following are equivalent:
\begin{enumerate}[label = (\roman*), ref = \roman*]
    \item \label{it:oneshot} $x$ is one-shot implementable with transfers;
    \item \label{it:CM} $x$ is $\pi$-cyclically monotone;
    \item \label{it:quota} $x$ is asymptotically implemented by the $(x, \pi)$-quota mechanisms;
    \item \label{it:transfers} $x$ is asymptotically implementable by linking mechanisms with transfers.
\end{enumerate}
\end{thm}


\cref{res:equivalence} says that $\pi$-cyclical monotonicity characterizes three different forms of implementability. The equivalence between \eqref{it:oneshot} and \eqref{it:CM} is due to
\cite{Rochet1987}. The equivalence between \eqref{it:CM}, \eqref{it:quota}, and \eqref{it:transfers} can be interpreted as follows. Consider a social choice function $x$. If $x$ is $\pi$-cyclically monotone, then $x$ can be asymptotically implemented by the $(x, \pi)$-quota mechanisms.  If $x$ is not $\pi$-cyclically monotone, then $x$ cannot be asymptotically implemented by the $(x, \pi)$-quota mechanisms, nor by any sequence of linking mechanisms, even with transfers.\footnote{Note an important difference from the finite-sample case. In the $K$-composite problem, the decision error guarantee under quota mechanisms cannot be improved by linking mechanisms (\cref{res:optimal_transport_bound}), but it can be improved with transfers (see \cref{ex:binary} below). Transfers are useful when there is uncertainty about the empirical distribution of the type vector, but this uncertainty vanishes in the limit as the number of problem copies grows large.} This result justifies our focus on $\pi$-cyclically monotone social choice functions $x$ and the associated $(x, \pi)$-quota mechanisms. More complicated linking mechanisms, even with transfers, cannot asymptotically implement any social choice functions that quota mechanisms cannot.

Weaker versions of the implication from \eqref{it:CM} to \eqref{it:quota} appear in JS and \cite{MatsushimaEtal2010}.\footnote{Both proofs contain errors. For corrections, see \cite{BJK2022} and \cite{BallKattwinkel2023}, respectively.} Specifically, JS prove that ex-ante Pareto efficient social choice functions---a proper subset of $\pi$-cyclically monotone social choice functions---are asymptotically implementable by quota mechanisms. Efficiency is measured with respect to the agents' preferences only, so efficient social choice functions may be unattractive for the principal. \cite{MatsushimaEtal2010} prove that $\pi$-cyclically monotone social choice functions are asymptotically implementable in $\e$-equilibrium by quota mechanisms.\footnote{With this notion of asymptotic implementation, they also prove that \eqref{it:transfers} implies \eqref{it:CM}.} 

We now sketch the proof of \cref{res:equivalence}. By \cref{res:convergence_rate}, $\pi$-cyclical monotonicity is sufficient for asymptotic implementation by quota mechanisms; the expected average decision error is of order $1/\sqrt{K}$, so it converges to $0$ as $K \to \infty$. To prove that $\pi$-cyclical monotonicity is necessary for asymptotic implementation, we follow the proof in \cite{MatsushimaEtal2010}. Suppose for a contradiction that some social choice function $x$, which is not $\pi$-cyclically monotone, can be asymptotically implemented by a  sequence of linking mechanisms with transfers. By the revelation principle, we may assume that these mechanisms are direct and that each agent is truthful in each equilibrium. Since $x$ is not $\pi$-cyclically monotone, there is some agent $i$ and some cycle of types in $\Th_i$ that violates \eqref{eq:CM}. We construct a deviation for agent $i$ in which each type in this cycle misreports as the next type in the cycle with positive probability in such a way that 
the ex-ante distribution of agent $i$'s reported type vector does not change. This deviation is \emph{undetectable}, in the language of \cite{Rahman2011}. 
Therefore, agent $i$'s ex-ante expected transfer payment is also unchanged. For $K$ sufficiently large, this deviation strictly increases agent $i$'s ex-ante expected decision utility, giving a contradiction. 

\subsection{Quota--transfer duality and robustness}

The implementation equivalence between transfers and quotas in \cref{res:equivalence} reflects a formal duality: each transfer $T_i(\th_i')$ in the one-shot problem corresponds to the Lagrange multiplier attached to the quota on reporting type $\th_i'$.\footnote{\cite{Rahman2011} identifies a very similar duality, outside the context of quota mechanisms.} These dual forms of implementation also require dual information. In the quota implementation, the quota $q_i$ is set equal to the prior $\pi_i$. Thus, the quota $q_i$ does not depend on  agent $i$'s utility function $u_i$ or agent $i$'s interim belief $\pi_{-i}$, provided that the $\pi$-cyclical monotonicity condition for agent $i$ is satisfied. In the one-shot implementation with transfers, the transfer function $T_i$ for agent $i$ must be tailored to agent $i$'s utility function $u_i$ and interim belief $\pi_{-i}$ in order to ensure incentive compatibility. But $T_i$ does not depend on the distribution $\pi_i$ of agent $i$'s type, as we illustrate in the next example.

\begin{exmp}[Quotas v.~prices] \label{ex:binary} Recall the example in \cref{sec:simple_example}. Let $x$ denote the social choice function that allocates the good if and only if the agent's valuation is high. The $(x,\pi)$-quota mechanisms asymptotically implement $x$, but this implementation requires the principal to know the true type distribution $\pi$. By contrast, if transfers are available, then the principal can implement $x$ without any knowledge of the type distribution, as follows.  In each problem, post a price between $\th_L$ and $\th_H$. The agent will buy the good exactly in those problems in which his valuation is high, even if his valuation is high more (or less) often than the principal anticipates.\footnote{ \cite{Weitzman1974} famously compares price and quantity controls. In that model, the agent is privately informed of costs, not benefits. Translated into that language, our example corresponds to the case of flat marginal benefits. In that case, \cite{Weitzman1974} shows that price controls yield higher expected social welfare than quantity controls, consistent with our finding here.} 

\end{exmp}


The reasoning in \cref{ex:binary} extends to any single-agent problem, as follows. Consider a cyclically monotone social choice function $x \colon \Th \to \D(\XX)$. By \cref{res:equivalence}, this social choice function $x$ is one-shot implementable using some transfer function $T$. In the $K$-composite problem, if transfers are allowed, then in each problem the principal can separately apply the one-shot mechanism with transfer function $T$.  This mechanism implements $x$ exactly, for every realized type vector. Thus, with a single agent, implementation with transfers does not depend on the type distribution. In the rest of the paper, we explore the robustness of quota mechanisms to the distribution of the agents' types and to agents' beliefs about each other.

\section{Robustness to type distributions} \label{sec:robustness_single}

We now analyze the robustness of quota mechanisms to the distribution of the agents' types.  Throughout \cref{sec:robustness_single}, we maintain the assumption that the environment, including the profile $\pi$ of type distributions, is common knowledge. We relax this assumption in \cref{sec:belief_robustness}, where we model interim beliefs using rich type spaces.

In practice, the principal can only imperfectly estimate the distribution of each agent's type. Suppose that the principal sets each quota $q_i$ equal to her estimate of agent $i$'s type distribution. Since these estimates are imperfect, the principal may be concerned with the performance of the mechanism for other type distributions near these estimates. Here, we bound the decision error that results from the principal's estimation errors.

We use the expectation notation $\E^{\pi}$ to emphasize the profile $\pi$ of distributions from which types are independently drawn. 

\begin{thm}[Approximate distributional robustness] \label{res:error_bound}
Fix $q \in \prod_{i=1}^{n} \D(\Th_i)$. Let $x \colon \Th \to \D(\XX)$ be $q$-cyclically monotone. For each $\pi \in \prod_{i=1}^{n} \D ( \Th_i)$, the $(x,q)$-quota mechanisms asymptotically implement, under the distribution profile $\pi$, some social choice function $x_\pi \colon \Th \to \D (\XX)$ satisfying
\begin{equation} \label{eq:distributional_robustness_bound}
    \E_{\th}^{\pi} \| x_\pi (\th) - x(\th) \| \leq \sum_{i=1}^{n}( | \Th_i| - 1) \| q_i - \pi_i \|.
\end{equation}
Moreover, the constants $|\Th_i| - 1$ cannot be reduced, even using arbitrary linking mechanisms.
\end{thm}

Suppose that the principal uses the quota profile $q$, but the true profile of type distributions is $\pi$. \cref{res:error_bound} says that the $(x,q)$-quota mechanisms asymptotically implement some social choice function $x_{\pi}$ that approximates $x$. The expected decision error (under $\pi$) from this approximation $x_\pi$ is bounded by an expression depending on the estimation error $\| q_i - \pi_i \|$ for each $i$. In the proof, we construct the social choice function $x_\pi$ for each $\pi \in \prod_{i=1}^{n} \D(\Th_i)$ as follows. By \cref{res:DZ_bound}, for each agent $i$ there exists an optimal transport plan\footnote{See \cref{ft:plan} for the definition.} $r_i \colon \Th_i \to \D(\Th_i)$ from $\pi_i$ to $q_i$ that moves at most probability $(|\Th_i| - 1) \| q_i - \pi_i \|$. For each $\th = (\th_1, \ldots, \th_n) \in \Th$, let $x_\pi (\th) = x ( \oprod_{i=1}^{n} r_i ( \th_i))$. In the proof, we check that $x_\pi$ satisfies \eqref{eq:distributional_robustness_bound}. Then we show that under the distribution profile $\pi$, the $(x,q)$-quota mechanisms asymptotically implement $x_\pi$. Here is a sketch of the argument. For each agent $i$, the realized empirical distribution $\marg \bm{\th}_i$ is likely to be close to $\pi_i$ when $K$ is large, by the law of large numbers. By \cref{res:Lip}, the solution set of an optimal transport problem is continuous in the marginals. Therefore, if $\marg \bm{\th}_i$ is close to $\pi_i$, then there is an optimal transport plan from $\marg \bm{\th}_i$ to $q_i$ that is close to $r_i$ for each agent $i$. We use these transport plans to construct an equilibrium that  approximates $x_\pi$.

The bound on the expected decision error in 
\eqref{eq:distributional_robustness_bound} implies a bound on the principal's expected utility loss (under $\pi$).  If the principal's utility function is normalized to have range $[0,1]$, then  the right side of \eqref{eq:distributional_robustness_bound} is an upper bound on the principal's expected utility loss from implementing $x_\pi$ rather than $x$. This loss can be interpreted as the principal's regret from incorrectly estimating the distribution profile $\pi$ to be $q$.

\cref{res:error_bound} provides a guarantee on the expected decision error when a single quota is applied in different local conditions. For example, different doctors face different patient populations, and different courses attract different kinds of students. However, for reasons of fairness and simplicity, it is common to apply the same quota to every doctor or to every class. The guarantee in \cref{res:error_bound} depends on the distance between the local population distribution and the quota.

\section{Robustness to agents' beliefs} \label{sec:belief_robustness}

In this section, we show that quota mechanisms are robust to a range of agents' beliefs about each other. Crucially, the quota $q_i$ imposed on agent $i$ assures agent $i$'s opponents that agent $i$'s reports average to $q_i$ over the $K$ problems. We illustrate this property in a simple voting example before turning to the general result. 

\subsection{Voting example} \label{sec:voting}

Two agents are voting on $K$ issues. On each issue, there are three possible policies: left ($L$), center ($C$), and right ($R$). Let $\XX = \{ L, C, R\}$. On each issue $k$, agent $i$ has single-peaked preferences determined by his type $\th_i^k \in \Th_i = \{-1, 0, +1\}$. Type $-1$ strictly prefers $L$; type $0$ strictly prefers $C$; and type $+1$ strictly prefers $R$. Type $0$ is indifferent between $L$ and $R$. 

The principal seeks to implement, on each issue, the deterministic social choice function $x \colon \Th \to \XX$ defined by
\[
    x(\th_1, \th_2) 
    =
    \begin{cases}
    L &\text{if}~ \th_1 + \th_2 < 0, \\
    C &\text{if}~\th_1 + \th_2 = 0, \\
    R &\text{if}~\th_1 + \th_2 > 0.
    \end{cases}
\]
Here, $x$ is one social choice function that respects unanimous preferences of the agents (but there are others). 

There is a common prior that types are uniformly distributed, independently across agents and issues. Define agent $i$'s interim social choice function $X_i \colon \Th_i \to \D(\XX)$ by $X_i (\th_i') = \E_{\th_{-i}} \Brac{ x( \th_i', \th_{-i})}$. We have 
\begin{equation*}
\begin{aligned}
 X_i (-1) &= (2/3) L + (1/3) C, \\
    X_i (0) &= (1/3) L + (1/3) C + (1/3) R,\\
    X_i (+1) &= (1/3) C + (2/3) R.
\end{aligned}
\end{equation*}
Note that $u_i (X_i (\th_i), \th_i) \geq u_i (X_i (\th_i'), \th_i)$ for all types $\th_i$ and $\th_i'$. Thus, $x$ is one-shot implementable without transfers. 

Consider two different mechanisms that asymptotically implement $x$: (a) unconstrained voting and (b) voting with quotas, namely, the $(x,q)$-quota mechanism with the uniform quotas $q_1 = q_2 = (1/3, 1/3,1/3)$. On each issue $k$, we interpret the report $-1$ (respectively~$0$, $+1$) as a vote for policy $L$ (respectively~$C$, $R$). Under unconstrained voting, each agent submits a vote (i.e., report) on each issue. On each issue, the votes are aggregated, and the policy is selected according to the social choice function $x$. With quotas, the votes are aggregated in the same way, but each agent is required to allocate exactly $1/3$ of his total votes to each of the three policies $L$, $C$, and $R$.  Each agent is free to distribute these votes across the $K$ issues however he wishes. On each issue, an agent can split his vote by reporting a probability distribution over votes. The principal samples a realized vote from the reported distribution. 

In the $K$-composite problem, unconstrained voting exactly implements the social choice function $x$, even if an agent turns out to prefer $R$ (or $L$ or $C$) on more issues than the principal expects. Crucially, in equilibrium, each agent believes that on each issue, his opponent is equally likely to vote for $L$, $C$, or $R$. Suppose instead that agent $1$ believes that agent $2$ is more likely to vote for $R$ than for $C$ or for $L$. If agent $1$ prefers $C$ on an issue, then it is uniquely optimal for him to vote for $L$ in order to offset the expected vote of agent $2$. Indeed, an arbitrarily small change in agent $1$'s belief about agent $2$'s vote can dramatically change agent $1$'s best response. 

By contrast, voting with quotas is robust to a range of agents' beliefs. Suppose that agent $1$ believes that agent $2$ tends to prefer $R$. Since agent $1$ knows that agent $2$'s votes must satisfy the quota, agent $1$ expects that on some issues, agent $2$ will vote for $L$ or $C$ when he actually prefers $R$. In this private-values setting, agent $1$'s optimal reporting strategy depends only on his belief about agent $2$'s votes, not on his belief about agent $2$'s true preferences. As long as agent $1$ believes that on each issue agent $2$ is equally likely to vote for $L$, $C$, and $R$, then agent $1$'s reporting incentives are the same.

This voting example illustrates a general point. The common prior assumption specifies the distribution of each agent's preferences, and it also pins down each agent's belief about his opponents' preferences. As we move away from the common prior idealization, the optimal choice of mechanism depends on the principal's relative concern for different uncertainties. If the principal is primarily concerned that she has incorrectly estimated the distribution of the agents' preferences, then unconstrained voting is more appealing. If the principal is primarily concerned that she has incorrectly estimated the agents' beliefs about others' preferences, then voting with quotas may be more appealing.

\subsection{Quota implementation on rich type spaces}

Motivated by the voting example, we now formalize a general belief-robustness property of quota mechanisms. To analyze beliefs, we adopt the robust private-values framework of \cite{BergemannMorris2005}. The environment $(\XX, \Th, u; K)$ is common knowledge. Assume $n \geq 2$ so that beliefs are nontrivial. Given this environment, quota mechanisms and social choice functions are defined as before. A type space consists of a measurable product space $T  = T_1 \times \cdots \times T_n$ and, for each agent $i$, a measurable payoff-type function and a measurable belief-type function, denoted 
\[
    \hat{\bm{\th}}_i \colon T_i \to \Th_i^K
    \quad
    \text{and}
    \quad
    \hat{\b}_i \colon T_i \to \D (T_{-i}).
\]
The type space $( T, (\hat{\bm{\th}}_i, \hat{\b}_i)_{i=1}^{n} )$ is common knowledge. Each agent $i$ knows his own type $t_i$, but not the types of others. Agent $i$'s payoff type $\hat{\bm{\th}}_i (t_i) = ( \hat{\th}_i^1 (t_i), \ldots, \hat{\th}_i^K(t_i))$ determines his preferences over decisions in the $K$-composite problem. His belief type $\hat{\b}_i (t_i)$ specifies his subjective belief about the type profile of his opponents. This belief pins down agent $i$'s beliefs of all orders---about other agents' types, about other agents' beliefs about others' types, and so on. In this framework, we perform all analysis at the interim stage; we do not specify a prior over $T$. 


Consider a linking mechanism $(M, g)$. On the type space $(T, (\hat{\bm{\th}}_i, \hat{\b}_i)_{i=1}^{n} )$, a strategy for agent $i$ is a map $\s_{i} \colon T_i \to \D (M_i)$. The solution concept is (interim) Bayes--Nash equilibrium.


We will show that quota mechanisms perform well on type spaces satisfying certain properties, which we now define. A type space $(T, (\hat{\bm{\th}}_i, \hat{\b}_i)_{i=1}^{n} )$ is \emph{payoff-type exchangeable} if each agent believes that each of his opponents' payoff types is exchangeable across the $K$ problems: for each agent $i$ and type $t_i \in T_i$, the measure $\hat{\bm{\th}}_{j} \bigl( \marg_{T_j} \hat{\b}_i ( t_i) \bigr)$ on $\Th_j^K$ is exchangeable for each $j \neq i$.\footnote{Here, $\marg_{T_j} \hat{\b}_i (t_i)$ denotes the marginal distribution of $\hat{\b}_i (t_i)$ over $T_j$. We view $\hat{\bm{\th}}_j \colon T_j \to \Th_j^K$ as a map into $\D ( \Th_j^K)$ whose values are unit masses. Then we extend this map linearly to obtain a map from $\D (T_j)$ to $\D (\Th_j^K)$. Recall that a measure on $\Th_j^K$ is \emph{exchangeable} if it is invariant to permuting the factors.}
A type space $(T, (\hat{\bm{\th}}_i, \hat{\b}_i)_{i=1}^{n} )$ is \emph{payoff-type independent} if each agent believes that his opponents' payoff types are statistically independent: for each agent $i$ and type $t_i \in T_i$, the measure $\hat{\bm{\th}}_{-i} ( \hat{\b}_i ( t_i)) \in \D( \prod_{j \neq i } \Th_j^K)$ is a product measure over the factors $\Th_j^K$ for $j \neq i$.\footnote{We define $\hat{\bm{\th}}_{-i} \colon T_{-i} \to \prod_{j \neq i} \Th_j^K$ by $\hat{\bm{\th}}_{-i} ( t_{-i}) = ( \hat{\bm{\th}}_j (t_j))_{j \neq i}$. We view $\hat{\bm{\th}}_{-i}$ as a map into $\D( \prod_{j \neq i} \Th_j^K)$ whose values are unit masses. Then we extend this map linearly to obtain a map from $\D ( T_{-i})$ to $\D ( \prod_{j \neq i} \Th_j^K)$.} If there are only two agents, then payoff-type independence holds vacuously. 

Our next result generalizes \cref{res:optimal_transport_bound} from independent, common prior type spaces to payoff-type exchangeable, payoff-type independent type spaces. We first discuss the generality of these type spaces. The independent, common prior setting from \cref{sec:setting} can be represented by the following type space. For each agent $i$, let $T_i = \Th_i^K$. Let $\hat{\bm{\th}}_i$ be the identity map and let $\hat{\b}_i$ be the constant map that always equals $\pi_{-i}^{\otimes K}$. Thus, the agents' beliefs are consistent and commonly known. Both of these assumptions can be relaxed on payoff-type exchangeable, payoff-type independent type spaces. First, beliefs can be inconsistent. For example, with at least three agents, it can be common knowledge that each agent $i$ believes that on each problem the payoff type of each opponent $j$ is independently distributed according to $\pi_j^i$, where $\pi_j^i \neq \pi_j^{i'}$ for $i \neq i'$. Second, beliefs need not be common knowledge.  For example, each agent can be uncertain about what his opponents believe about others' payoff types. In fact, each agent can believe that his opponents' beliefs are correlated. 


\begin{thm}[Optimal ex-post error bound on rich type spaces] \label{res:robust}
Assume that there are at least two agents: $n \geq 2$. Fix $q \in \prod_{i=1}^{n} \D ( \Th_i)$. Let $x \colon \Th \to \D (\XX)$ be $q$-cyclically monotone. In the $K$-composite problem, on any payoff-type exchangeable, payoff-type independent type space $(T, (\hat{\bm{\th}}_i, \hat{\b}_i)_{i=1}^{n})$, the $(x,q)$-quota mechanism $(M,g)$ has a Bayes--Nash equilibrium $\s$ satisfying, for all type profiles $t$ in $T$,\footnote{This equilibrium $\s$ is pure, so $\s(t) \coloneqq ( \s_{i} (t_i))_{i=1}^{n} \in M$. Similarly, $\hat{\th}^k(t) \coloneqq ( \hat{\th}_i^k (t_i))_{i=1}^{n} \in \Th$.}
\begin{equation} \label{eq:ineq_t}
    \frac{1}{K}  \sum_{k=1}^{K} \| g^k (\s(t)) - x (\hat{\th}^k (t)) \| 
    \leq \sum_{i=1}^{n} ( |\Th_i| - 1) \| q_i - \marg \hat{\bm{\th}}_i (t_i)  \|.
\end{equation}
Moreover, the constants $|\Th_i| - 1$ cannot be reduced, even using arbitrary linking mechanisms.
\end{thm}

At each realized type profile $t$, the inequality in \eqref{eq:ineq_t} bounds the frequency with which the decision is incorrect (relative to the decision specified by the social choice function $x$). The bound is small if each agent's realized payoff-type vector has an empirical distribution close to that agent's quota. This bound does not depend on the agents' realized belief types. 

In the $(x,q)$-quota mechanism, the quota $q_i$ guarantees that 
agent $i$'s reports on the $K$ problems average to $q_i$. The difficulty is that the quotas do not pin down the reports on any particular problem. To prove \cref{res:robust}, we proceed as follows. On any payoff-type exchangeable, payoff-type independent type space $(T, (\hat{\bm{\th}}_i, \hat{\b}_i)_{i=1}^{n})$, we construct a special Bayes--Nash equilibrium of the $(x,q)$-quota mechanism. 
In this equilibrium, each agent's report vector depends only on his payoff type, not his belief type. More precisely, each agent applies his equilibrium strategy from \cref{res:optimal_transport_bound} to his payoff type. We use payoff-type exchangeability and payoff-type independence to show that under this strategy profile, each type $t_i$ believes that on every problem $k$ the opposing report profile $r_{-i}^k \in \D( \Th_{-i})$ has expectation $\oprod_{j \neq i} q_{j}$. Therefore, we can follow the proof of \cref{res:optimal_transport_bound} to show that this strategy profile is an equilibrium. 

\cref{res:robust} assumes that $x$ is $q$-cyclically monotone. It is well known that every ex-ante Pareto efficient social choice function satisfies the stronger property of ex-post cyclical monotonicity; see, e.g., \citet[p.~254]{jackson2007overcoming}.\footnote{Formally, a social choice function is ex-ante Pareto efficient if it is ex-ante Pareto efficient with respect to \emph{some} full-support prior, or equivalently, with respect to \emph{every} full-support prior; see \cref{ft:PE} in \cref{sec:comparison}.} Even if $x$ is assumed to be ex-post cyclically monotone, the conclusion of \cref{res:robust} still requires restrictions on the type space. Ex-post cyclical monotonicity controls an agent's reporting incentives in the primitive problem, given any fixed belief about his opponents' types. In the composite problem, however, without the exchangeability and independence restrictions, some type may believe that his opponents' payoff-type profile is distributed differently on different problems. In that case, the conclusion of \cref{res:robust} may not hold,  as illustrated in the next example.

\begin{exmp}[Beliefs violating payoff-type exchangeability]
There are two agents. In the primitive problem, there is a single good to be allocated. Each agent $i$'s payoff type is his valuation $\th_i \in \Th_i = \{ \th_L, \th_M, \th_H\}$, where $\th_L < \th_M < \th_H$. Consider the social choice function $x$ that allocates the good to the agent whose valuation is highest, breaking ties uniformly. For each $i$, the allocation probability for agent $i$ is increasing in $\th_i$, so $x$ is ex-post cyclically monotone. 

There are $K = 3$ problem copies. Consider a type space $(T, (\hat{\bm{\th}}_i, \hat{\b}_i)_{i=1}^{n})$. Suppose that there is some type profile $\bar{t} = (\bar{t}_1,\bar{t}_2) \in T$ such that
\begin{equation*}
\begin{aligned}
    \hat{\bm{\th}}_1 ( \bar{t}_1 ) &= (\th_L, \th_M, \th_H), \\
    \hat{\bm{\th}}_2 (\bar{t}_2) &= (\th_M, \th_H, \th_L).
\end{aligned}
\end{equation*}
Suppose further that $\hat{\b}_1 ( \bar{t}_1) = \d_{\bar{t}_2}$. That is, type $\bar{t}_1$ of agent $1$ is certain that agent $2$'s type is $\bar{t}_2$ and hence that agent $2$'s payoff type is $(\th_M, \th_H, \th_L)$. This violates payoff-type exchangeability. 

Consider the uniform quotas $q_1 = q_2 = (1/3,1/3,1/3)$. Under the $(x,q)$-quota mechanism, let $\s$ be a strategy profile that satisfies \eqref{eq:ineq_t} at every type profile $t$. We show that $\s$ is not a Bayes--Nash equilibrium. At the fixed type profile $\bar{t}$, we have $\marg \hat{\bm{\th}}_i (\bar{t}_i) = q_i$ for $i=1,2$, so the decisions induced by $\s$ must match $x$ exactly on each problem. That is, the good is allocated to agent $2$ on the first two problems and to agent $1$ on the third problem. Type $\bar{t}_1$ is certain that agent $2$'s type is $\bar{t}_2$ and hence that agent $2$ will follow $\s_2 ( \bar{t}_2)$. If type $\bar{t}_1$ mimics agent $2$ by deviating from $\s_1 (\bar{t}_1)$ to $\s_{2} (\bar{t}_2)$, then he believes that he will get the good with probability $1/2$ on each problem.  This deviation is strictly profitable if $\th_L + \th_M > \th_H$. 
\end{exmp}

\section{Extensions} \label{sec:discussion} 

In the main model, we assume that the agents have private values and that their information arrives all at once. In this section, we relax these assumptions. 

\subsection{Interdependent values} 

The private values assumption is important for the robustness of quota mechanisms to agents' beliefs about each other (\cref{res:robust}). Quotas control each agent's beliefs about his opponents' \emph{reports}. However, quotas do not affect each agent's beliefs about his opponents' \emph{true types}, and these beliefs are relevant to each agent's best response when values are interdependent.

On the other hand, our results for the independent, common prior setting largely extend to interdependent values. Suppose that in the primitive problem each agent $i$'s utility from decision $x \in \XX$ is given by $u_i (x, \th_{i}, \th_{-i})$ rather than $u_i (x, \th_i)$ as in the main model. As before, types are drawn independently across agents and problems according to the profile $\pi = (\pi_1, \ldots, \pi_n) \in \prod_{i=1}^{n} \D(\Th_i)$. 

Our main definitions can be extended to the setting of interdependent values. Given $p = (p_1, \ldots, p_n) \in \prod_{i=1}^{n} \D (\Th_i)$, a social choice function $x \colon \Th \to \D(\XX)$ is \emph{$p$-cyclically monotone} if for each agent $i$ the following holds: for all integers $J \geq 2$ and all types $\th_{i,1}, \ldots, \th_{i,J} \in \Th_i$, we have
\begin{equation*} 
    \sum_{j=1}^{J}  \E_{\th_{-i} \sim p_{-i}} \Brac{u_i ( x (\th_{i,j}, \th_{-i}), \th_{i,j}, \th_{-i})}
 \geq \sum_{j=1}^{J}  \E_{\th_{-i} \sim p_{-i}} \Brac{ u_i ( x ( \th_{i,j+1}, \th_{-i}), \th_{i,j}, \th_{-i})},
\end{equation*}
where $\th_{i,J+1} = \th_{i,1}$. Similarly, under the profile $\pi$ of priors, a social choice function $x$ is  \emph{one-shot implementable with transfers} if for each agent $i$ there exists a transfer function $T_i \colon \Th_i \to \R$ such that for all $\th_i, \th_i' \in \Th_i$, we have
\begin{equation*}
    \E_{\th_{-i}} \Brac{ u_i (x (\th_i, \th_{-i}), \th_i, \th_{-i})} - T_i (\th_i) \geq \E_{\th_{-i}} \Brac{ u_i ( x (\th_i', \th_{-i}), \th_i, \th_{-i})} - T_i (\th_i'),
\end{equation*}
where the expectations are taken with respect to the priors in $\pi$.

With interdependent values, linking mechanisms and quota mechanisms are defined exactly as in the main model. A version of
the implementation equivalence (\cref{res:equivalence}) goes through, with a weaker notion of asymptotic implementation. Let $(M_K, g_K)_{K \geq 1}$ be a sequence of linking mechanisms. Let $x \colon \Th \to \D (\XX)$ be a social choice function. The sequence $(M_K, g_K)_{K \geq 1}$ \emph{approximately asymptotically implements} $x$ if there exists an associated sequence of strategy profiles $(\s_K)_{K \geq 1}$ and a sequence $(\e_K)_{K \geq 1}$ converging to $0$ such that for each $K$, the profile $\s_K$ is an interim Bayes--Nash $\e_K$-equilibrium of $(M_K, g_K)$,\footnote{That is, under the profile $\s_{K}$, every type of every agent gains at most $\e_K$ in expectation from unilaterally deviating.} and we have
\begin{equation} \label{eq:approx_asymptotic}
 \lim_{K \to \infty} \E_{\bm{\th}} \Brac{  \frac{1}{K} \sum_{k=1}^{K}  \| g_K^k ( \s_K ( \bm{\th})) -  x(\th^k) \|} = 0.
\end{equation}
With this definition,\footnote{This is essentially the notion of implementation used in \cite{MatsushimaEtal2010} for the case of multiple agents, except that they use an ex-ante definition of equilibrium. Their model allows for interdependent values.}  we can state the result.

\begin{thm}[Implementation equivalence with interdependent values] \label{res:interdependent}
Consider the setting of interdependent values. For any social choice function $x \colon \Th \to \D (\XX)$, the following are equivalent:
\begin{enumerate}[label = (\roman*), ref = \roman*]
    \item \label{it:IDP_oneshot} $x$ is one-shot implementable with transfers;
    \item \label{it:IDP_CM} $x$ is $\pi$-cyclically monotone;
    \item \label{it:IDP_quota} $x$ is approximately asymptotically implemented by the $(x, \pi)$-quota mechanisms;
    \item \label{it:IDP_transfers} $x$ is approximately asymptotically implementable by linking mechanisms with transfers.
\end{enumerate}
\end{thm}
    
With interdependent values, $\pi$-cyclical monotonicity is still equivalent to one-shot implementability with transfers (by \cite{Rochet1987}), but $\pi$-cyclical monotonicity is now equivalent to a weaker form of asymptotic implementation. With private values, we prove \cref{res:equivalence} by constructing a sequence $(\s_K)_{K \geq 1}$ of Bayes--Nash equilibria such that on every problem, each agent $i$'s report is \emph{close} to truthful and has expectation \emph{exactly} $\pi_i$. With interdependent values, each agent cares about the joint distribution of his opponents' reports and true types, so the analogous strategy profile $\s_K$ may not be an exact Bayes--Nash equilibrium. Nevertheless, as $K$ grows large, each agent's gain from deviating converges to $0$ because his opponents' strategies converge to truthtelling. 

We caution that in the setting of interdependent values, ex-ante Pareto efficient social choice functions are not necessarily $\pi$-cyclically monotone.\footnote{Indeed, in a setting with transferable utility, \cite{JehielMoldovanu2001} show that with multidimensional (continuous) types, no efficient social choice function is one-shot implementable with transfers, unless a non-generic condition is satisfied.} Next, we give an example of an ex-ante Pareto efficient social choice function that cannot be  approximately asymptotically  implemented by quota mechanisms.\footnote{JS (fn.~8, p.~245) claim that in the setting of (independently distributed) interdependent values, every ex-ante Pareto efficient social choice function is asymptotically implemented by the associated quota mechanisms. \cref{ex:counterexample} is a counterexample to this claim.}


\begin{exmp}[Efficient SCF that is not implementable] \label{ex:counterexample} There are two agents. The principal chooses whether to provide a public good. Agent $1$ has private information. Agent $2$ does not. Agent $1$'s type $\th_1$ is the vector $(\nu_1, \nu_2) \in \{-2, 1, 3\}^2$ specifying each agent's valuation for the public good. Agent $1$'s type is drawn from the uniform prior $\pi$ over the nine possible realizations. Let $x^\ast$ be the ex-ante Pareto efficient social choice function, which provides the public good if and only if  $\nu_1 + \nu_2 \geq 0$. This inequality holds 
with ex-ante probability $2/3$. Let $\tilde{x}$ be agent $1$'s favorite social choice function, which provides the public good if and only if $\nu_1 \geq 0$. This inequality also holds with ex-ante probability $2/3$. Therefore, under the $(x^\ast, \pi)$-quota mechanisms, agent $1$ has a sequence of strategies that asymptotically induce $\tilde{x}$. Since agent $1$ strictly prefers $\tilde{x}$ to $x^\ast$, the $(x^\ast, \pi)$-quota mechanisms cannot approximately asymptotically implement $x^\ast$. 
\end{exmp}

\subsection{Dynamics} \label{sec:dynamics}

The main model is static. Each agent knows his preferences on all problems, and he simultaneously submits a report on every problem. In this section, we assume instead that information arrives over time. We consider an infinite-horizon setting with discounting, and we introduce a dynamic analogue of quota mechanisms. With a single agent, we show that these dynamic quota mechanisms can asymptotically implement any \emph{strictly} cyclically monotone social choice function.

For the dynamic model, we assume that there is a single agent.\footnote{With multiple agents, it is difficult to construct exact equilibria because each agent's best response is determined by the solution of a dynamic optimization problem rather than an optimal transport problem.} For any discount factor $\b \in (0,1)$, define the $\b$-discounted problem as follows. 
The horizon is infinite, with periods indexed by $t = 0, 1, \ldots$. Each period has one problem copy. In each period $t$, the agent learns his type $\th^t \in \Th$. Types are drawn independently across periods from a full-support prior $\pi$ in $\D( \Th)$. The agent's utility from a decision sequence $(x^t)_{t=0}^{\infty}$ is given by the discounted average utility $(1 - \b) \sum_{t = 0}^{\infty} \b^{t} u ( x^t, \th^t)$. 

Fix a social choice function $x \colon \Th \to \D(\XX)$ and a quota $q \in \D(\Th)$. In the $\b$-discounted problem, the \emph{dynamic $(x,q)$-quota mechanism} asks the agent to report, in each period $t$, a distribution $r^t \in \D( \Th)$, subject to the constraint that for every period $t$,
\begin{equation} \label{eq:time_t}
    (1 - \b) \sum_{s = 0}^{t} \b^{s} r^s (\th') \leq q (\th'), \qquad \th' \in \Th.
\end{equation}
In each period $t$, the mechanism selects the decision $x(r^t) \in \D(\XX)$. Note that the inequality \eqref{eq:time_t} holds for every period $t$ if and only if 
\begin{equation*}
    (1 - \b) \sum_{s=0}^{\infty} \b^{s} r^s = q.
\end{equation*}
That is, the $\b$-weighted average of the agent's reports must equal $q$. \cite{Frankel2016discounted} defines a similar discounted quota mechanism in which the agent reports distributions over decisions rather than types.\footnote{In a dynamic sender--receiver game, \cite{RenaultSolanVieille2013} construct a quota-like equilibrium for $\b$ sufficiently large. Time is partitioned into long, finite blocks, and an \emph{undiscounted} quota is applied within each block. If the sender violates the quota within a block, then the receiver punishes the sender until the end of the block. Deviations by the receiver are punished with babbling forever after.} Our formulation reduces to his if $x$ is deterministic and injective; see \cref{rem:decisions}.

Under the dynamic $(x,q)$-quota mechanism, a (pure) strategy for the agent is a sequence $\s = (\s^t)_{t \geq 0}$, specifying for each period $t$ a map $\s^t \colon \Th^{t+1} \to \D (\Th)$, such that the quota is satisfied: for all type sequences $\bm{\th} = (\th^t)_{t \geq 0} \in \Th^{\infty}$, 
\[
    (1 - \b) \sum_{t=0}^{\infty} \b^t \s^t (\th^0, \ldots, \th^t) = q.
\]
Here, $\s^t (\th^0, \ldots, \th^t)$ is the agent's report in period $t$ after the type history $(\th^0, \ldots, \th^t)$.\footnote{This formulation does not allow the agent to condition his report on past decisions. This choice has no effect on the results.} Each strategy $\s = (\s^t)_{t \geq 0}$ induces a distribution, $\rho (\s)$, over sequences
$(\bm{\th}, \bm{r}) \in \Th^{\infty} \times [\D(\Th)]^{\infty}$ in the natural way.

The dynamic $(x,q)$-quota mechanisms \emph{asymptotically implement} $x$ if for each $\b \in (0,1)$ the agent has a pure best response $\s_\b$ to the  $(x,q)$-quota mechanism in the $\b$-discounted problem such that
\begin{equation} \label{eq:asymptotically_implement}
    \lim_{\b \to 1} \E_{ (\bm{\th}, \bm{r}) \sim \rho(\s_\b)} \Brac{ (1 - \b) \sum_{t=0}^{\infty} \b^t  \| x ( r^{t}) -  x(\th^t) \|} = 0.
\end{equation}
That is, the expected $\b$-discounted average decision error converges to $0$ as $\b$ tends to $1$.  

For the next result, recall the definition of strict cyclical monotonicity in \cref{rem:full}. If $\XX$ and $\Th$ are totally ordered, and $u \colon \XX \times \Th \to \R$ is \emph{strictly} supermodular,\footnote{That is, for all $x,x' \in \XX$ and $\th, \th' \in \Th$, if 
$x < x'$ and $\th < \th'$, then $u(x,\th) + u(x', \th') > u ( x, \th') + u(x', \th)$.} then every \emph{weakly} increasing deterministic function $x \colon \Th \to \XX$ is \emph{strictly} cyclically monotone.

\begin{thm}[Implementation with dynamic quota mechanisms] \label{res:dynamic_implementation} Suppose that there is a single agent ($n = 1$). Let $x \colon \Th \to \D(\XX)$ be a social choice function. If $x$ is strictly cyclically monotone, then the dynamic $(x,\pi)$-quota mechanisms asymptotically implement $x$.
\end{thm}

As an application of this dynamic setting, consider the limits imposed by TANF (Temporary Assistance for Needy Families). As described in the introduction, a family can only collect TANF benefits for up to $60$ months over their lifetime. Each month, an eligible family must choose whether to collect benefits that month, without knowing their future needs. On the one hand, families may want to conserve their eligibility in case they face greater hardship in the future.\footnote{\cite{LowMeghirPistaferriVoena2023}
analyze a life-cycle model under these TANF limits. In each period, single women make choices about work, consumption, marriage, and whether to claim TANF benefits.} On the other hand, since this cap is undiscounted, families may prefer to collect benefits earlier. To discourage early collection, some states impose additional moving window caps, e.g., a family can collect benefits for at most 24 months out of any period of 60 consecutive months.\footnote{See \href{https://wrd.urban.org/sites/default/files/documents/2024-02/2022\%20Welfare\%20Rules\%20Databook\%20\%2812\%2022\%202023\%29.pdf}{Welfare Rules Databook: State and Territory TANF Policies as of July 2022} and \href{https://wrd.urban.org/policy-tables}{Table IV.C.1}.} \cref{res:dynamic_implementation} suggests an alternative---capping the \emph{discounted} number of months in which benefits are collected.

The TANF application highlights a distinctive feature of \emph{dynamic} quota mechanisms. Each period, the agent must submit a report without knowing his type realizations in future periods. In the $\b$-discounted problem, suppose that in period $t$, the agent's type history has a discounted empirical distribution that is close to the quota. Nevertheless, the agent may prefer to misreport today in order to conserve certain quotas in case future type realizations are extreme. For this reason, the expected $\b$-discounted average decision error can converge to $0$ arbitrarily slowly as $\b$ tends to $1$, even for type spaces of a fixed size.\footnote{For the same reason, \emph{strict} cyclical monotonicity cannot be relaxed to weak cyclical monotonicity in \cref{res:dynamic_implementation}; see \cref{ex:weak_CM} for a counterexample.} In the static setting, by contrast, our optimal transport techniques bound the rate of convergence uniformly over all type spaces of a fixed size. 

\section{Conclusion} \label{sec:conclusion}

In settings without transfers, quota mechanisms are ubiquitous.  In this paper, we analyze quota mechanisms under more realistic conditions---with finitely many problem copies, and with uncertainty about the population distribution. Using tools from optimal transport theory, we quantify the equilibrium decision error under quota mechanisms when the realized type frequencies differ from the quota, either due to sampling variation or estimation error. Moreover, we show that quota mechanisms satisfy a robust optimality property: the decision error guarantee under quota mechanisms cannot be improved by any other mechanisms without transfers. Together, our results provide a deeper understanding of quota mechanisms and indicate the contexts in which quota mechanisms will perform well.

\newpage
\appendix

\section{Main proofs} \label{sec:proofs}

\subsection{Optimal transport results} \label{sec:OT}

In this section, we state the optimal transport results that we will use in the main proofs. We begin with some measure theory definitions and some useful properties of the total variation norm. All lemmas are proven in \cref{sec:additional}.

\paragraph{Measure theory}
Fix measurable spaces $X$ and $Y$. Given $p \in \D (X)$ and $q \in \D(Y)$, define the product measure $p \otimes q$ to be the unique probability measure on $X \times Y$ (with the product $\s$-algebra) satisfying
\[
   ( p \otimes q )(A \times B) = p(A) q(B),
\]
for all measurable subsets $A$ of $X$ and $B$ of $Y$. 

Let $h \colon X \to \D (Y)$ be measurable.\footnote{That is, for each measurable subset $B$ of $Y$, the map $x \mapsto h(B|x)$ from $X$ to $[0,1]$ is measurable.} Such a map is called a \emph{probability kernel} or a  \emph{Markov transition}. Define the measure $h(p)$ in $\D(Y)$ by
\[
    h ( B | p) = \int_{X} h(B | x) \de p (x),
\]
for all measurable subsets $B$ of $Y$. Define $p \otimes h$ to be the unique probability measure on  $X \times Y$ (with the product $\s$-algebra)
satisfying
\[
    (p \otimes h)(A \times B) = \int_{A} h (B | x) \de p(x),
\]
for all measurable subsets $A$ of $X$ and $B$ of $Y$.\footnote{The notation $\otimes$ is overloaded but consistent: if $h(x) = q$ for all $x$, then $p \otimes h = p \otimes q$, where the product $\otimes$ is between a measure and a Markov transition on the left and between two measures on the right.} These definitions can be interpreted in terms of a compound lottery. Suppose that $x$ in $X$ is drawn from the distribution $p$, and then $y$ in $Y$ is drawn from the distribution $h(x)$. Then $y$ has distribution $h(p)$ and $(x,y)$ has distribution $p \otimes h$.


\paragraph{Total variation distance} The \emph{total variation distance} between two probability measures $p$ and $q$ on a measurable space $X$ can be defined in two equivalent ways:
\[
    \| p  - q \| = \sup_{A} | p(A) - q(A) | = \frac{1}{2} \sup_f | p f - q f |,
\]
where the first supremum is over all measurable subsets $A$ of $X$ and the second supremum is over all measurable functions $f \colon X \to [-1,1]$.\footnote{We write $p f$ to denote the integral of the function $f$ with respect to the measure $p$.} More generally, $\| \cdot \|$ extends to the \emph{total variation norm} on the vector space of finite signed measures.\footnote{For any finite signed measure $\mu$ on a measurable space $X$,  let $\| \mu\| = (1/2) \sup_f | \mu f|$, where the supremum is over all measurable functions $f \colon X \to [-1,1]$. Here, $\mu f$ denotes the integral of the function $f$ with respect to the signed measure $\mu$. This norm is also commonly defined  without the factor $1/2$.} If $X$ is finite, we can view $p$ and $q$ as vectors in $\R^{|X|}$. In this case, $\|p - q \| = (1/2) \| p - q \|_{1}$, where $\| \cdot \|_1$ is the $\ell^1$-norm on $\R^{|X|}$.

The following total variation bounds will be useful in our proofs.

\begin{lem}[Total variation bounds] \label{res:TV_bounds} Let $X$, $Y$, and $X_1, \ldots, X_J$ be measurable spaces. 
\begin{enumerate}[label = (\roman*), ref = \roman*]
    \item \label{it:TV_nonexpansive} For any probability measures $p,q \in \D (X)$ and any measurable map $h \colon X \to \D(Y)$, we have
\begin{equation*}
    \| h(p) - h(q) \| \leq \| p - q \|.
\end{equation*}
    \item \label{it:TV_product} For $j =1 , \ldots, J$, let $p_j$ and $q_j$ be in $\D(X_j)$. We have
    \begin{equation*}
    \| \oprod_{j=1}^{J} p_j - \oprod_{j=1}^{J} q_j \| \leq \sum_{j=1}^{J} \| p_j - q_j \|.
\end{equation*}
\end{enumerate}
\end{lem}

Part~\ref{it:TV_nonexpansive} says that 
the total variation distance between two probability measures cannot increase after the same Markov transition is applied to both measures. Part~\ref{it:TV_product} bounds the total variation distance between two product measures in terms of the total variation distance between the respective component measures.

\paragraph{Optimal transport}  Let $X$ and $Y$ be finite sets. Given probability measures $p$ in $\D(X)$ and $q$ in $\D(Y)$, a \emph{coupling} of $p$ and $q$ is a probability measure $\g$ on the product space $X \times Y$ whose marginal on $X$ is $p$ and whose marginal on $Y$ is $q$. Let $\Pi (p,q)$ denote the set of all couplings of $p$ and $q$. Let $c \colon X \times Y \to \R$ be a cost function. A coupling of $p$ and $q$ is \emph{$c$-optimal} if it minimizes the expected value of $c$ over all couplings in $\Pi (p,q)$. When the cost function $c$ is clear from context, we call such a coupling \emph{optimal}. These definitions extend to arbitrary nonnegative measures $p$ and $q$ with $p(X) = q(Y)$. We use this extension in our proofs of \cref{res:DZ_bound,res:Lip} below.

A \emph{kernel coupling} of $p$ and $q$ is a map $r \colon X \to \D(Y)$ that satisfies $r(p) = q$. If $r$ is a kernel coupling of $p$ and $q$, then $p \otimes r$ is a coupling of $p$ and $q$.\footnote{Conversely, for any coupling $\g$ of $p$ and $q$, there exists a kernel coupling $r \colon X \to \D(Y)$ of $p$ and $q$ such that $p \otimes r = \g$. Namely, for each $x \in \supp p$, define $r(x) \in \D(Y)$ by $r(y | x) = \g (x, y)/ p(x)$. The map $r$ can be defined arbitrarily outside $\supp p$.} 
A kernel coupling $r$ of $p$ and $q$ is \emph{$c$-optimal} if the coupling $p \otimes r$ is $c$-optimal. Below, we state our optimal transport results for couplings, but we will freely apply these results to kernel couplings as well.

A subset $S$ of $X \times Y$ is \emph{$c$-cyclically  monotone} if, for all integers $J \geq 2$ and all $(x_1, y_1), \ldots, (x_J, y_J) \in S$, we have
\begin{equation} \label{eq:CMxy}
    \sum_{j=1}^{J} c(x_j, y_j) \leq \sum_{j=1}^{J} c(x_j, y_{j+1}), 
\end{equation}
where $y_{J+1}$ is defined to equal $y_1$. The set $S$ is \emph{strictly $c$-cyclically  monotone} if \eqref{eq:CMxy} holds strictly whenever $(x_j, y_{j+1}) \not\in S$ for some $j$. Here, we define $c$-cyclical monotonicity as a property of a set, as is standard in optimal transport theory. We now connect this definition to our notion of cyclical monotonicity for social choice functions in the main text. Fix $p = (p_1, \ldots, p_n) \in \prod_{i=1}^{n} \D (\Th_i)$.  Consider a social choice function $x \colon \Th \to \D(\XX)$. For each agent $i$, define the associated cost function $c_i \colon \Th_i \times \Th_i \to \R$ by
\[
    c_i( \th_i, \th_i') = - \E_{\th_{-i} \sim p_{-i}} \Brac{ u_i (x ( \th_i', \th_{-i}), \th_i)}.
\]
It can be verified that the social choice function $x$ is $p$-cyclically monotone in the sense of \eqref{eq:CM} if and only if, for each agent $i$, the diagonal $D_i =  \{ (\th_i, \th_i) : \th_i \in \Th_i \}$ is $c_i$-cyclically monotone. In the single-agent case, define the cost function $c \colon \Th \times \Th \to \R$ by $c( \th, \th') = - u (x(\th'), \th)$. In this case, it can be verified that the social choice function $x$ is strictly cyclically monotone in the sense of \eqref{eq:strict_CM} if and only if the set $D(x) = \{ (\th, \th') \in \Th^2 : x(\th') = x(\th) \}$ is strictly $c$-cyclically monotone.\footnote{\label{ft:strict_CM}To prove this equivalence, observe that in the definition of strict cyclical monotonicity in the main text, we can equivalently impose \eqref{eq:strict_CM} whenever $x(\th_1), \ldots, x(\th_J)$ are not all equal.}



At the heart of our proofs is the following bound on the mass moved in an optimal transport problem on a finite set. This result generalizes Lemma~1 in \citet[p.~o6]{BJK2022}. 

\begin{lem}[Bound on mass moved] \label{res:DZ_bound} Fix a finite set $Z$, a cost function $c \colon Z \times Z \to \R$, and probability measures $p,q \in \D(Z)$. If the diagonal $D = \{ (z,z): z \in Z\}$ is $c$-cyclically monotone, then there exists a $c$-optimal coupling $\g$ of $p$ and $q$ such that
\begin{equation*} 
   1 - \g(D) \leq (| Z| - 1) \| q -  p \|.
\end{equation*}
\end{lem}

To prove \cref{res:DZ_bound}, we use the $c$-cyclical monotonicity of the diagonal $D$ to show that there exists a $c$-optimal coupling $\g$ whose support contains no nontrivial cycles. As illustrated in \cref{fig:cascade}, we show that the coupling $\g$ moves the net probability $\| q - p\|$ along paths of length at most $|Z| - 1$, so that the total probability moved is at most $(|Z| -1 ) \| q - p\|$. 


\cref{res:DZ_bound} is a special case of the following Lipschitz continuity property,\footnote{\cref{res:DZ_bound} can be derived from \cref{res:Lip}, using results about the $c$-cyclical monotonicity of the support of a $c$-optimal coupling. We give an independent proof of \cref{res:DZ_bound}.}  which is used in the proof of \cref{res:error_bound}.

\begin{lem}[Lipschitz continuity of solution set] \label{res:Lip}
Fix finite sets $X$ and $Y$ and a cost function $c \colon X \times Y \to \R$.  Consider probability measures $p,p' \in \D(X)$ and $q,q' \in \D(Y)$. For any $c$-optimal coupling $\g$ of $p$ and $q$, there exists a $c$-optimal coupling $\g'$ of $p'$ and $q'$ such that
\begin{equation} \label{eq:Lip_ineq}
 \| \g' - \g\| \leq \min\{ |X| \wedge |Y|, |X| \vee |Y| - 1\} ( \| p' - p \| + \| q ' -q \|).
\end{equation}
\end{lem}

The constant on the right side of \eqref{eq:Lip_ineq} is sharp, as we show in the proof. If $|X| \neq |Y|$, this constant equals $|X| \wedge |Y|$. If $|X| = |Y|$, this constant equals $|X| - 1$, consistent with \cref{res:DZ_bound}. For a general linear program with constraints $Ax \leq b$ and $Cx = d$, \citet[Theorem 2.4, p.~589]{MangasarianShiau1987} prove that the solution set is Lipschitz continuous in the right-side data $(b,d)$, with respect to the sup-norm on the solution set. Under the assumption that the matrix $C$ has full rank, \citet[Theorem 2.5, p.~24; Theorem 
3.5, p.~29]{Li1993} identifies the sharp Lipschitz constant with respect to any given pair of norms (on the right-side data and on the solution set). Their constant is expressed as the value of a constrained optimization problem involving $A$ and $C$. This result can be used to prove a version of \cref{res:Lip} with a strictly larger constant.\footnote{An optimal transport problem between finite sets can be formulated as a linear program: the inequality $Ax \leq b$ encodes the $|X| \cdot |Y|$ nonnegativity constraints, and the equality $C x = d$ encodes the $|X| + |Y|$ marginal constraints. With this parametrization, the matrix $C$ does not have full rank because one of the marginal constraints is redundant. Therefore, we drop one of the marginal constraints, and then we apply \citet[Theorem 2.5, p.~24]{Li1993} with the $\ell^1$-norm to obtain a Lipschitz constant. Numerical optimization suggests that this constant is $2 (|X| \wedge |Y|)$, which is strictly larger than our Lipschitz constant in \cref{res:Lip}. This does not contradict the sharpness result of \citet[Theorem 
3.5, p.~29]{Li1993} because (i) not all perturbations of the right-side data $(b,d)$ preserve the optimal transport structure, and (ii) dropping the marginal constraint can reduce the norm of the change in the right-side data.} We obtain the sharp constant in \cref{res:Lip} using methods from optimal transport rather than linear algebra.

We caution that the continuity property in \cref{res:Lip} is somewhat special. The solution set of a linear program is not generally continuous in the left-side data $(A,C)$ or the coefficients of the objective function.\footnote{To see this, consider the classical two-good consumer problem of maximizing $u_1 x_1 + u_2 x_2$ subject to the constraints $p_1 x_1 + p_2 x_2 \leq w$ and $x_1, x_2 \geq 0$. For simplicity, suppose that the parameters $u_1$, $u_2$, $p_1$, $p_2$, and $w$ are all strictly positive. At any parameter values satisfying $u_1 / u_2 = p_1/ p_2$, there is a continuum of optimal bundles, and the solution set is not \emph{lower} hemicontinuous in $(p_1, p_2)$ or in $(u_1,u_2)$. In this example, the objective and the feasible set are continuous in the parameters, so the solution set is \emph{upper} hemicontinuous in the parameters, by Berge's theorem. In general, however, the feasible set of a linear program is not necessarily continuous in the left-side data, so Berge's theorem may not apply, and the solution set may violate both upper and lower hemicontinuity. For example, with a constraint of the form $a^T x \leq 1$, the feasible set may fail to be continuous whenever some component of $a$ equals zero.} Moreover, for optimal transport problems on \emph{infinite} sets, the solution set is not necessarily continuous in the marginals with respect to the total variation norm.\footnote{Note that the Lipschitz constant in \cref{res:Lip} grows without bound as $|X|$ and $|Y|$ increase. Here is a counterexample to continuity with infinite sets. For each $\e > 0$, consider the problem of transporting the uniform distribution over $[0,1]$ to the uniform distribution over $[\e, 1 + \e]$, under a squared distance moving cost. It is uniquely optimal to shift each point to the right by $\e$. Therefore, the entire mass of $1$ is moved, but the total variation distance between the marginals is $\e$.} 


\subsection{Proof of Remark~\ref{rem:decisions}} \label{sec:proof_decisions}

Suppose that there is a single agent ($n = 1$). Drop agent subscripts. Under the $(x,q)$-quota mechanism $(M,g)$, we have
\[
    g(M) = \Set{ ( x (r^1), \ldots, x(r^K) ) \in [\D (\XX)]^K : \frac{1}{K} \sum_{k=1}^{K} r^k = q }.
\] 
Consider the set
\[
  A =  \Set{ (\bar{x}^1, \ldots, \bar{x}^K) \in  [\D(\XX)]^K : \frac{1}{K} \sum_{k=1}^{K} \bar{x}^k = x(q) }.
\]

In general, we have $g(M) \subseteq A$. Indeed, if $\frac{1}{K} \sum_{k=1}^{K} r^k = q$, then $
\frac{1}{K} \sum_{k=1}^{K} x(r^k) = x(q)$, by the linearity of the extension of $x$ to $\D(\Th)$. 

Suppose that $x$ is deterministic. In this case, we  claim that $A \subseteq g(M)$. Since $x$ is deterministic, we view $x$ as a map into $\XX$. Enumerate $x(\Th)$ as $\{ x_1, \ldots, x_J \}$. For each $j$, let $\Th_j = \{ \th \in \Th : x(\th) = x_j \}$. Fix $\bar{x} = (\bar{x}^1, \ldots, \bar{x}^K) \in A$.  Thus, $\frac{1}{K} \sum_{k=1}^{K} \bar{x}^k = x(q)$.  For each $k$, we have $\supp \bar{x}^k \subseteq \{ x_1, \ldots, x_J\}$, so we can choose $r^k \in \D(\Th)$ such that (a) $x(r^k) = \bar{x}^k$ and (b) for each $j$, the restriction of $r^k$ to $\Th_j$ is proportional to the restriction of $q$ to $\Th_j$. Let $r= \frac{1}{K} \sum_{k=1}^{K} r^k$. By (a), we have $x(r) = \frac{1}{K} \sum_{k=1}^{K} \bar{x}^k = x(q)$. Since $x$ is deterministic, we have $r(\Th_j) = x(r) (x_j) = x(q)(x_j) =  q (\Th_j)$ for each $j$. By (b), we conclude that $r = q$. Thus, the vector $\bar{x} = (x(r^1), \ldots, x(r^K))$ is in $g(M)$.

\subsection{Proof of Theorem~\ref{res:optimal_transport_bound}} \label{sec:proof_optimal_transport_bound}

We break the proof into parts. 

\paragraph{Upper bound}
First we select a solution of each agent's associated optimal transport problem. For each agent $i$, define the transport cost function $c_i \colon \Th_i \times \Th_i \to \R$ by
\[
c_i(\th_i, \th_i') = - \E_{\th_{-i} \sim q_{-i}} \Brac{ u_i (x (\th_i', \th_{-i}), \th_i)}.
\]
By \cref{res:DZ_bound}, for any $p_i \in \D ( \Th_i)$ there exists a $c_i$-optimal kernel coupling $r_i \colon \Th_i \to \D (\Th_i)$ of $p_i$ and $q_i$ such that 
\begin{equation} \label{eq:truth_bound}
   1 -  \sum_{ \th_i \in \Th_i} p_i (\th_i) r_i ( \th_i | \th_i) \leq ( |\Th_i| - 1) \| q_i - p_i \|.
\end{equation}
To indicate the dependence of $r_i$ on the initial distribution $p_i$, we denote $r_i (\cdot)$ by $r_i ( \cdot ; p_i)$. 

Now we construct the equilibrium strategy profile $\s$. In the $K$-composite problem, let $(M, g)$ denote the $(x,q)$-quota mechanism. For each agent $i$, define the strategy  $\s_{i} \colon \Th_i^K \to M_i$ by
\[
    \s_{i}^k (\bm{\th}_i) = r_i ( \th_i^k ; \marg \bm{\th}_i), \qquad k = 1, \ldots, K.
\]
Note that 
\[
   \frac{1}{K} \sum_{k=1}^{K}  r_i ( \th_i^k ; \marg \bm{\th}_i)  = \sum_{\th_i \in \Th_i} r_i ( \th_i ; \marg \bm{\th}_i) \marg (\th_i | \bm{\th}_i) = q_i.
\]
Write $\s ( \bm{\th}) = (\s_1 ( \bm{\th}_1), \ldots, \s_n ( \bm{\th}_n))$.

First we prove that $\s$ satisfies \eqref{eq:EP_bound}. Fix $\bm{\th} = ( \bm{\th}_1, \ldots, \bm{\th}_n) \in \Th^K$. By \cref{res:TV_bounds}, for each problem $k$ we have
\begin{equation} \label{eq:k_bound_sum}
\begin{aligned}
     \| g^k ( \s ( \bm{\th})) -  x(\th^k) \|  
  &= \| x( \oprod_{i=1}^{n} \s_i^k ( \bm{\th}_i)) -  x(\th^k) \| \\
  & \leq  \| \oprod_{i=1}^{n} \s_i^k ( \bm{\th}_i) -  \d_{\th^k} \| \\
  & \leq \sum_{i=1}^{n} [ 1 - \s_{i}^k ( \th_i^k | \bm{\th}_i) ] \\
  &= \sum_{i=1}^{n} [ 1 - r_{i} ( \th_i^k |  \th_i^k ; \marg \bm{\th}_i) ].
\end{aligned}
\end{equation}
Average the inequality \eqref{eq:k_bound_sum} over problems $k = 1, \ldots, K$ and then apply \eqref{eq:truth_bound}  with $p_i = \marg \bm{\th}_i$ for each agent $i$. We conclude that
\begin{equation*}
\begin{aligned}
\frac{1}{K} \sum_{k=1}^{K} \| g^k ( \s( \bm{\th})) -  x(\th^k) \| 
&\leq \sum_{i=1}^{n} \Brac{1 -  \sum_{k=1}^{K} \frac{r_{i} (\th_i^k | \th_i^k; \marg \bm{\th}_i)}{K}} \\
&\leq \sum_{i=1}^{n} ( |\Th_i| - 1) \| q_i -  \marg \bm{\th}_i  \|.
\end{aligned}
\end{equation*}

Next we prove that $\s$ is a Bayes--Nash equilibrium. 
By construction, for each agent $i$ the strategy $\s_{i}$ is \emph{label-free}: for any permutation $\tau$ on the space of $K$-vectors, we have $\s_{i} ( \tau (\bm{\th}_i)) =  \tau ( \s_{i} (\bm{\th}_i))$ for each $\bm{\th}_i$ in $\Th_i^K$.\label{pg:label_free} Therefore, for each problem $k$, 
\begin{equation} \label{eq:interim_expectations}
    \E_{ \bm{\th}_{-i}} \Brac{ \oprod_{j \neq i} \s_{j}^k ( \bm{\th}_j)}
    =
     \oprod_{j \neq i} \E_{ \bm{\th}_{j}} \Brac{ \s_{j}^k ( \bm{\th}_j)}
     = \oprod_{j \neq i} 
     q_j,
\end{equation}
where the first equality uses the independence of type vectors across agents, and the second uses label-freeness together with the exchangeability of the distribution of each agent's type vector.

By \eqref{eq:interim_expectations}, type $\bm{\th}_i$'s interim expected utility from reporting $\bm{r}_i \in M_i$ when his opponents follow $\s_{-i}$ is given by 
\begin{equation*}
\begin{aligned}
    &\E_{\bm{\th}_{-i}} \Brac{ \frac{1}{K} \sum_{k=1}^{K} u_i \Paren{ x \bigl( r_i^k \otimes (\mathop{\oprod_{j \neq i}}  \s_{j}^k ( \bm{\th}_j)) \bigr) , \th_i^k }}\\
    &=  \frac{1}{K} \sum_{k=1}^{K} u_i \Paren{ x \bigl( r_i^k \otimes (\mathop{\oprod_{j \neq i}}  q_j) \bigr) , \th_i^k } \\
    &= - \frac{1}{K} \sum_{k=1}^{K} \sum_{\th_i' \in \Th_i} c_i ( \th_i^k, \th_i') r_i^k (\th_i') \\
    &= - \sum_{\th_i, \th_i' \in \Th_i}  c_i (\th_i, \th_i') \g_i (\th_i, \th_i'),
\end{aligned}
\end{equation*}
where $\g_i$ is the coupling of $\marg \bm{\th}_i$ and $q_i$ defined by 
\[
\g_i(\th_i, \th_i') = \frac{1}{K} \sum_{k: \th_i^k = \th_i} r_i^k ( \th_i').
\]
By the $c_i$-optimality of the kernel coupling $r_i (\cdot; \marg \bm{\th}_i)$, it follows that $\s_{i}$ is a best response to $\s_{-i}$.  

\paragraph{Upper bound is tight} Suppose $n = 1$. Hereafter, we drop agent subscripts.  Fix a type space $\Th$ with $|\Th| \geq 2$ and an integer $K$ satisfying $K \geq |\Th|$. We construct a decision environment $(\XX, u)$, a cyclically monotone social choice function $x \colon \Th \to \D(\XX)$, and a quota $q \in \D (\Th)$ such that the following holds: For each $\e >0$, no linking mechanism $(M,g)$ in the $K$-composite problem has a Bayes--Nash equilibrium $\s$ that satisfies, for all $\bm{\th}$ in $\Th^K$,
\begin{equation} \label{eq:sharp}
\frac{1}{K} \sum_{k=1}^{K} \| g^k ( \s(\bm{\theta})) - x( \theta^k) \| \leq   ( | \Theta| - 1 - \e) \| q - \marg \bm{\theta} \|.
\end{equation}

Here is the construction. To simplify notation, let $m = |\Th|$. By assumption, $K \geq m \geq 2$. Without loss, we can relabel the types so that $\Th = \{ \th_{1}, \ldots, \th_{m}\}$. Let $\XX = \{x_1, \ldots, x_m\}$. Define the agent's utility function $u \colon \XX \times \Th \to \R$ by
\[
    u( x_p, \th_{\ell}) = 
    \begin{cases} 
    -(m-1) &\text{if}~ p < \ell, \\
    0 &\text{if}~ p = \ell,\\
    1 &\text{if}~ p > \ell.
    \end{cases}
\]
Let $x \colon \Th \to \XX$ be the deterministic social choice function defined by $x( \th_{\ell})  = x_\ell$ for each $\ell \in \{1, \ldots, m \}$. The social choice function $x$ is cyclically monotone. To see this, observe that for each $\ell \in \{1, \ldots, m\}$, we have $\ell \in \argmax_{p = 1, \ldots, m} \Set{ u ( x_p, \th_{\ell}) - p}$. Therefore,  for all integers $J \geq 2$ and all $\ell_1, \ldots, \ell_J \in \{ 1, \ldots, m\}$, we have
\[
    \sum_{j=1}^{J} \Brac{ u(x_{\ell_j}, \th_{\ell_j})  - \ell_j} \geq     \sum_{j=1}^{J} \Brac{ u(x_{\ell_{j+1}}, \th_{\ell_j})  - \ell_{j+1}},
\]
where $\ell_{J+1} = \ell_1$. After canceling the permuted summation over $\ell_j$ from each side, we get
\[
 \sum_{j=1}^{J}  u(x_{\ell_j}, \th_{\ell_j}) \geq \sum_{j=1}^{J} u(x_{\ell_{j+1}}, \th_{\ell_j}).
\]
Finally, let $q$ be the probability distribution that puts probability $1/K$ on types $\th_{1}, \ldots, \th_{m-1}$, and probability $(K - m + 1)/K$ on type $\th_{m}$.

Fix $\e > 0$. Suppose for a contradiction that in the $K$-composite problem there exists a linking mechanism $(M,g)$ with a Bayes--Nash equilibrium $\s$ that satisfies \eqref{eq:sharp} for each $\bm{\th}$ in $\Th^K$. Consider the type vectors $\bar{\bm{\th}}, \hat{\bm{\th}} \in \Th^K$ defined by
\begin{equation}
\begin{aligned}
    \bar{\bm{\th}} &= ( \th_{1}, \th_{1}, \th_{2}, \ldots, \th_{m-1}, \th_{m}, \ldots, \th_{m}), \\
     \hat{\bm{\th}} &= ( \th_{1}, \th_{2}, \th_{3}, \ldots, \th_{m}, \th_{m}, \ldots, \th_{m}).
\end{aligned}
\end{equation}
Note that $\th_{m}$ appears $K - m$ times in $\bar{\bm{\th}}$ and $K- m +1$ times in $\hat{\bm{\th}}$.

To get a contradiction, we prove that type $\bar{\bm{\th}}$ has a profitable deviation from $\s (\bar{\bm{\th}})$ to $\s( \hat{\bm{\th}})$. Apply \eqref{eq:sharp} at $\bm{\th} = \hat{\bm{\th}}$. Since $\marg \hat{\bm{\th}} = q$, we conclude that
\[
    g(\s (\hat{\bm{\th}})) = (x_1, x_2, x_3, \ldots, x_m, x_m, \ldots, x_m).
\]
Therefore, type $\bar{\bm{\th}}$ gets utility $(m-1)/K$ from deviating to $\s (\hat{\bm{\th}})$.

Next, apply \eqref{eq:sharp} at $\bm{\th} = \bar{\bm{\th}}$. Since $\| q - \marg  \bar{\bm{\th}} \| = 1/K$, we get
\begin{equation} \label{eq:eps_bound}
    \frac{1}{K} \sum_{k=1}^{K} \| g^k ( \s( \bar{\bm{\th}})) - x( \bar{\th}^k) \| \leq \frac{ m - 1 - \e}{K}.
\end{equation}
For each $k$, since $x$ is deterministic, we have
\begin{equation}\label{eq:eps_2}
\begin{aligned} 
    &u (g^k (\s( \bar{\bm{\th}})), \bar{\th}^k)  \\
    &\leq u ( x(\bar{\th}^k), \bar{\th}^k) + \| g^k (\s( \bar{\bm{\th}})) - x(\bar{\th}^k) \| \Paren{ \max_{x' \in \XX} u(x', \bar{\th}^k) - u( x(\bar{\th}^k), \bar{\th}^k)} \\
    &\leq \| g^k (\s(\bar{\bm{\th}})) - x(\bar{\th}^k) \|.
\end{aligned}
\end{equation}
Average \eqref{eq:eps_2} over $k = 1, \ldots, K$ and apply \eqref{eq:eps_bound} to conclude that type $\bar{\bm{\th}}$ gets utility at most $(m - 1 - \e)/K$ from $\s(\bar{\bm{\th}})$. Thus, it is strictly profitable for type $\bar{\bm{\th}}$ to deviate from $\s (\bar{\bm{\th}})$ to $\s( \hat{\bm{\th}})$.

\paragraph{Lower bound (\cref{rem:lower_bound})}  In the $K$-composite problem, consider an arbitrary strategy profile $\s$ in the $(x,q)$-quota mechanism $(M,g)$. Since the extension of $x$ to $\D(\Th)$ is linear, every mixed strategy is outcome-equivalent to some pure strategy. Therefore, we may assume without loss that $\s$ is pure. Suppose that $x$ is injective (as defined in \cref{rem:lower_bound}). Fix agent $i$. For each profile $\bm{\th} \in \Th^K$ and each problem $k$,  we have
\begin{equation*}
\begin{aligned}
    \| g^k (\s (\bm{\th})) - x (\th^k) \| 
    &=  \| x ( \oprod_{j=1}^{n} \s_j^k ( \bm{\th}_j) ) - x ( \th^k) \|  \\
    &= \| \oprod_{j=1}^{n} \s_j^k ( \bm{\th}_j) - \d_{\th^k} \| \\
    &\geq \|  \s_i^k ( \bm{\th}_i) -  \d_{\th_i^k} \|,
\end{aligned}
\end{equation*}
where  the second equality uses the injectivity of $x$. Therefore, 
\begin{equation*}
    \begin{aligned}
\frac{1}{K} \sum_{k=1}^{K} \| g^k ( \s( \bm{\th})) -  x(\th^k) \| 
&\geq
\frac{1}{K} \sum_{k=1}^{K} \| \s_i^k( \bm{\th}_i) -  \d_{\th_i^k} \|  \\
&\geq \bigg\| \frac{1}{K} \sum_{k=1}^{K} \bigl( \s_i^k( \bm{\th}_i) - \d_{\th_i^k} \bigr) \bigg \| \\
&= \| q_i - \marg \bm{\th}_i \|.
    \end{aligned}
\end{equation*}
Since $i$ was arbitrary, we can take the maximum of the right side over agents $i = 1, \ldots, n$. This completes the proof. 

\paragraph{Refined bound (\cref{rem:refined_upper_bound})}  We first need a refined version of  \cref{res:DZ_bound}. Given an equivalence relation $\sim$ on a finite set $Z$, let $Z_{\sim}$ denote the space of equivalence classes under $\sim$. For each $z \in Z$, let $[z]_{\sim}$ denote the equivalence class containing $z$. For any probability measure $p \in \D(Z)$, let $p_{\sim} \in \D( Z_{\sim})$ denote the projection of $p$ onto $Z_{\sim}$.  That is, for all $z \in Z$, 
\[
    p_{\sim} ([z]_{\sim}) = \sum_{z' \in [z]_{\sim}} p(z').
\]

\begin{lem}[Bound on mass moved under projection] \label{res:refined_bound} Fix a finite set $Z$, an equivalence relation $\sim$ on $Z$, a cost function $c \colon Z \times Z \to \R$, and probability measures $p,q \in \D(Z)$. Let $D_\sim = \{ (z, z') \in Z^2 : z \sim  z' \}$. If $D_\sim$ is $c$-cyclically  monotone, then there exists a $c$-optimal coupling $\g$ of $p$ and $q$ such that
\begin{equation} 
   1 - \g(D_\sim) \leq (| Z_{\sim}| - 1) \|q_{\sim} - p_{\sim} \|.
\end{equation}
\end{lem}

Using \cref{res:refined_bound}, we now prove \cref{rem:refined_upper_bound}. Suppose $n =1$. Let $x \colon \Th \to \XX$ be a social choice function that is deterministic and cyclically monotone. Define the associated transport cost function $c \colon \Th \times \Th \to \R$ by $c(\th, \th') = -u ( x(\th'), \th)$. Define the equivalence relation $\sim$ on $\Th$ by $\th \sim \th'$ if and only if $x(\th) = x(\th')$. Thus, $D_{\sim} = \{ (\th, \th') \in \Th^2 : x(\th) = x(\th') \}$. First, we check that $D_{\sim}$ is $c$-cyclically monotone. For all integers $J \geq 2$ and all $(\th_1, \th_1'), \ldots, (\th_J, \th_J') \in D_{\sim}$, we have 
\[
    \sum_{j=1}^{J} c( \th_j, \th_j')
    = 
     \sum_{j=1}^{J} c( \th_j, \th_j) 
    \leq \sum_{j=1}^{J} c( \th_j, \th_{j+1})
    = \sum_{j=1}^{J} c( \th_j, \th_{j+1}'),
\]
where the equalities follow from the definitions of $D_{\sim}$ and $c$, and the middle inequality follows from the $c$-cyclical monotonicity of the diagonal $D = \{ (\th, \th): \th \in \Th \}$.

Having shown that $D_{\sim}$ is $c$-cyclically monotone, we can follow the steps in the main proof of the upper bound above, except that instead of applying \cref{res:DZ_bound} to the diagonal $D$, we apply \cref{res:refined_bound} to the set $D_{\sim}$. With this change, 
\eqref{eq:truth_bound} becomes 
\begin{equation*}
\begin{aligned}
    1 - \sum_{(\th, \th') \in D_{\sim}} p (\th) r( \th' | \th) 
    &\leq  ( |\Th_{\sim}| - 1) \|  q_{\sim} - p_{\sim} \| \\
    &= (|x(\Th)| - 1) \| x(q) - x(p) \|,
\end{aligned}
\end{equation*}
where the equality holds because $x$ is deterministic.\footnote{To extend $x$ to $\D(\Th)$, we first view $x$ as a map from $\Th$ into $\D ( \XX)$ whose values are unit masses. Then we extend this map linearly to obtain a map from $\D (\Th)$ to $\D (\XX)$.} Then in \eqref{eq:k_bound_sum}, we observe that
\[
    \| x (\s^k ( \bm{\th})) - x(\th^k) \| \leq 1- \sum_{\th' : \th' \sim \th^k} \s^k ( \th' | \bm{\th}).
\]

\paragraph{Full implementation (\cref{rem:full})} We use the following strict analogue of \cref{res:DZ_bound}.

\begin{lem}[Bound on mass moved under strict cyclical monotonicity] \label{res:S_bound} Fix a finite set $Z$, a cost function $c \colon Z \times Z \to \R$, and probability measures $p,q \in \D(Z)$. Let $S$ be a subset of $Z \times Z$ that includes the diagonal $\{ (z,z): z \in Z \}$. If $S$ is strictly $c$-cyclically monotone, then every $c$-optimal coupling $\g$ of $p$ and $q$ satisfies
\begin{equation*} 
   1 - \g(S) \leq (| Z| - 1) \| q -  p \|.
\end{equation*}
\end{lem}

Using \cref{res:S_bound}, we now prove \cref{rem:full}. Suppose $n =1$. Let $x \colon \Th \to \D(\XX)$ be strictly cyclically monotone. Define the associated transport cost function $c \colon \Th \times \Th \to \R$ by $c(\th, \th') = -u ( x(\th'), \th)$. Let $D(x) = \{ (\th, \th') \in \Th^2 : x( \th ) = x(\th') \}$. The set $D(x)$ is strictly $c$-cyclically monotone; see \cref{ft:strict_CM}. In the $K$-composite problem, let $(M,g)$ denote the $(x,q)$-quota mechanism. Let $\s \colon \Th^K \to M$ be an arbitrary pure best response; considering a pure strategy is without loss since every mixed strategy is outcome-equivalent to some pure strategy. Fix $\bm{\th} \in \Th^K$. Consider the associated coupling $\g$ of $\marg \bm{\th}$ and $q$ defined by 
\[
    \g(\th, \th') = \frac{1}{K} \sum_{k: \th^k = \th} \s^k ( \th' | \bm{\th}).
\]
Since $\s$ is a best response, this coupling $\g$ is $c$-optimal. Apply \cref{res:S_bound} with $S = D(x)$ to conclude that 
\[
    1 -  \g (D(x)) \leq ( |\Th| -1 ) \| q - \marg \bm{\th} \|.
\]
Therefore, we have
\begin{equation*}
\begin{aligned}
        \frac{1}{K} \sum_{k=1}^{K} \| g^k(\s(\bm{\th})) - x ( \th^k) \|
        &= \frac{1}{K} \sum_{k=1}^{K} \| x ( \s^k (\bm{\th})) - x ( \th^k) \| \\
        &\leq 
        \frac{1}{K} \sum_{k=1}^{K} \s^k \Paren{ \{ \th' \in \Th : x(\th') \neq x(\th^k) \} \mid \bm{\th} }  \\
        &= 1 - \g (D(x)) \\ 
        &\leq (|\Th| - 1 )\| q - \marg \bm{\th} \|.
\end{aligned}
\end{equation*}

\subsection{Proof of Theorem~\ref{res:convergence_rate}}
By \eqref{eq:bound_in_q} and the accompanying argument in the main text, it suffices to bound the expectation $\E_{\bm{\th}_i} \|  \pi_i - \marg \bm{\th}_i\|$ for each $i$. From the definition of the total variation norm, we have
\begin{equation} \label{eq:TV_sum}
     \|  \pi_i - \marg \bm{\th}_i \|
    = \frac{1}{2} \sum_{ \th_i \in \Th_i} | \pi_i ( \th_i) -  \marg (\th_i | \bm{\th}_i) |. 
\end{equation}
If we view $\bm{\th}_i$ as a random variable, then for each fixed $\th_i \in \Th_i$, the random variable $\marg (\th_i | \bm{\th}_i)$ has mean $\pi_i (\th_i)$ and variance $\pi_i (\th_i) (1 - \pi_i (\th_i))/K$. Take expectations in \eqref{eq:TV_sum} and apply Jensen's inequality to get
\begin{equation} \label{eq:expectation_bound}
\begin{aligned}
 \E_{\bm{\th}_i} \| \pi_i -  \marg \bm{\th}_i \| 
 &= \frac{1}{2} \sum_{ \th_i \in \Th_i} \E_{\bm{\th}_i} | \pi_i (\th_i) - \marg (\th_i | \bm{\th}_i)  | \\
 &\leq \frac{1}{2} \sum_{\th_i \in \Th_i} \Brac{ \E_{\bm{\th}_i} \Paren{   \pi_i (\th_i) - \marg (\th_i | \bm{\th}_i) }^2}^{1/2} \\
  &=
  \frac{1}{2 \sqrt{K}} \sum_{\th_i \in \Th_i} \Brac{ \pi_i (\th_i) (1 - \pi_i (\th_i))}^{1/2}.
\end{aligned}
\end{equation}
On the other hand, it follows from the central limit theorem (and uniform integrability) that 
\[
\lim_{K \to \infty} 2 \sqrt{K} \E_{\bm{\th}_i} \| \pi_i -  \marg \bm{\th}_i\| =
\sqrt{ \frac{2}{\pi}} \sum_{\th_i \in \Th_i} \Brac{ \pi_i (\th_i) (1 - \pi_i (\th_i))}^{1/2},
\]
where $\pi$ refers to the mathematical constant, not the vector of probability distributions.\footnote{For a standard normal random variable $Z$, recall that $\E |Z| = \sqrt{2/\pi}$.} This shows that the bound in \eqref{eq:expectation_bound} cannot be improved by more than a factor of $\sqrt{\pi/2} \approx 1.25$. 

To complete the proof, we check that
\[
    \sum_{\th_i \in \Th_i} \Brac{ \pi_i (\th_i) (1 - \pi_i (\th_i))}^{1/2} \leq (|\Th_i| - 1 )^{1/2},
\]
with equality if $\pi_i$ is the uniform distribution over $\Th_i$. Observe that for any probability vector $p = (p_1, \ldots, p_d)$, we have
\begin{equation} \label{eq:CS}
    \Paren{ \sum_{j=1}^{d} \Brac{ p_j (1 - p_j)}^{1/2}}^2 \leq d \sum_{j=1}^{d} p_j (1 - p_j) =  d \Paren{ 1-  \sum_{j=1}^{d} p_j^2} \leq d - 1,
\end{equation}
where the first inequality follows from Cauchy--Schwarz (applied with the $d$-vector of ones) and the last inequality holds because the convex map $p \mapsto p_1^2 + \cdots + p_d^2$ is minimized over the simplex (in $\R^d$) at the uniform probability vector. Moreover, if $p$ is uniform, then the leftmost and rightmost expressions in \eqref{eq:CS} agree.

\subsection{Proof of Theorem~\ref{res:equivalence}} \label{sec:proof_equivalence}

The equivalence \eqref{it:oneshot} $\iff$ \eqref{it:CM} follows from \citet[Theorem 1, p.~192]{Rochet1987}.\footnote{\cite{Rochet1987} proves the result for a single agent with an arbitrary type space. The result extends immediately to the multi-agent setting with a common, independent prior. Here $\Th_i$ is finite, so the equivalence can be proven directly from a theorem of the alternative; see \cite{Rahman2011}.}  \cref{res:convergence_rate} implies that \eqref{it:CM} $\implies$ \eqref{it:quota}. It is immediate that {\eqref{it:quota} $\implies$ \eqref{it:transfers}}. Therefore, it suffices to prove that \eqref{it:transfers} $\implies$ \eqref{it:CM}. We prove the contrapositive, following \cite{MatsushimaEtal2010}.

Let $x \colon \Th \to \D (\XX)$ be a social choice function that is not $\pi$-cyclically monotone. Then for some agent $i$, there exists an integer $J \geq 2$ and some \emph{distinct}\footnote{If types in the sequence are repeated, we partition the sequence into cycles.} types $\th_{i,1}, \ldots, \th_{i,J} \in \Th_i$ such that
\begin{equation*}
    G \coloneqq \sum_{j=1}^{J}  \E_{\th_{-i} \sim \pi_{-i}} \Brac{ u_i ( x ( \th_{i,j+1}, \th_{-i}), \th_{i,j})} - \sum_{j=1}^{J}  \E_{\th_{-i} \sim \pi_{-i}} \Brac{u_i ( x (\th_{i,j}, \th_{-i}), \th_{i,j})} > 0,
\end{equation*}
where $\th_{i,J+1} = \th_{i,1}$. Suppose for a contradiction that $x$ is asymptotically implemented by a sequence of linking mechanisms with transfers. By the revelation principle, $x$ is asymptotically implemented by \emph{truthful} equilibria of a sequence $(\Th^K, g_K, t_K)_{K \geq 1}$ of \emph{direct} linking mechanisms with transfers. For each $K$, we extend $g_K$ linearly to $\D(\Th^K)$. In each $K$-composite problem, we construct a  deviation for agent $i$. To get a contradiction, we show that this deviation is ex-ante profitable for all $K$ sufficiently large. 

For each $K$ and each type vector $\bm{\th}_i \in \Th_i^K$, let $N(\bm{\th}_i)$ be the largest nonnegative integer $N$ such that each of the types $\th_{i,1}, \ldots, \th_{i,J}$ appears in the vector $\bm{\th}_i$ at least $N$ times. Define $\s_{K,i} (\bm{\th}_i)$ in $\D ( \Th_i^K)$ to be the distribution of the report vector $\hat{\bm{\th}}_i$ in $\Th_i^K$ selected according to the following random reporting procedure. For each $j = 1, \ldots, J$, let $\mathcal{K}_j (\bm{\th}_i) = \{ k : \th_i^k = \th_{i,j}\}$. By definition, $| \mathcal{K}_j (\bm{\th}_i)| \geq N (\bm{\th}_i)$. For each $j$, independently and uniformly select a $N(\bm{\th}_i)$-element subset $\mathcal{K}_j'$ of $\mathcal{K}_j (\bm{\th}_i)$. On each problem $k \in \mathcal{K}_j'$, report $\th_{i,j+1}$. On each problem $k \not\in \cup_{j=1}^{J} \mathcal{K}_j'$, report truthfully. This construction ensures that agent $i$'s reported vector $\hat{\bm{\th}}_i$ has 
the same ex-ante distribution as agent $i$'s true type vector $\bm{\th}_i$. 
That is, $\s_{K,i} ( \pi_i^{\otimes K}) = \pi_i^{\otimes K}$. 

The type vectors are independent across agents, so agent $i$'s deviation from truthtelling to the strategy $\s_{K,i}$ does not change the ex-ante distribution of the profile of reported type vectors. Therefore, agent $i$'s ex-ante expected transfer payment is unchanged. For each $\bm{\th}_i \in \Th_i^K$, let $\marg^k \s_{K,i}( \bm{\th}_i)$ denote the $k$-th marginal of the probability distribution $\s_{K,i}( \bm{\th}_i) \in \D(\Th_i^K)$. Therefore, we have
\begin{equation} \label{eq:Jensen_id}
\begin{aligned}
    &   \E_{\bm{\th}}  \Brac{ \frac{1}{K} \sum_{k=1}^{K}  
 \| g_K^k ( \s_{K,i} ( \bm{\th}_i), \bm{\th}_{-i}) - x(\marg^k \s_{K,i} (\bm{\th}_i), \th_{-i}^k) \|} \\
 & =  \E_{\bm{\th}}  \Brac{ \frac{1}{K} \sum_{k=1}^{K}  
 \bigl\| \E_{\hat{\bm{\th}}_i \sim \s_{K,i} (\bm{\th}_i)} [ g_K^k (  \hat{\bm{\th}}_i, \bm{\th}_{-i}) - x(\hat{\th}_i^k, \th_{-i}^k) ] \bigr\|} \\
    &\leq      \E_{\bm{\th}} \Brac{ \frac{1}{K} \sum_{k=1}^{K}  
\E_{\hat{\bm{\th}}_i \sim \s_{K,i} (\bm{\th}_i)} \| g_K^k ( \hat{\bm{\th}}_i, \bm{\th}_{-i}) - x(\hat{\th}_i^k, \th_{-i}^k) \| }\\
    &=
    \E_{\bm{\th}}  \Brac{ \frac{1}{K} \sum_{k=1}^{K}    \| g_K^k ( \bm{\th}) - x(\th^k) \|},
\end{aligned}
\end{equation}
where inequality holds because the total variation norm is convex, and the final equality holds because agent $i$'s deviation to the strategy $\s_{K,i}$ does not change the distribution of the profile of reported type vectors.  Since $x$ is asymptotically implemented by the truthful equilibria of the sequence $(\Th^K, g_K, t_K)_{K \geq 1}$, the right side of \eqref{eq:Jensen_id} converges to $0$ as $K \to \infty$. Therefore, for each agent $i,$
\begin{equation} \label{eq:deviation_lim}
\begin{aligned}
&\lim_{K \to \infty} \E_{\bm{\th}} \Brac{ \frac{1}{K} \sum_{k=1}^{K} \Abs{ u_i \bigl( g_K^k ( \s_{K,i} ( \bm{\th}_i), \bm{\th}_{-i}),\th_i^k \bigr) -  u_i \bigl( x(\marg^k \s_{K,i} (\bm{\th}_i), \th_{-i}^k),\th_i^k \bigr)}} \\
&\leq \lim_{K \to \infty} \E_{\bm{\th}} \Brac{ \frac{1}{K} \sum_{k=1}^{K} 2 \| u_i \|_{\infty} \| g_K^k ( \s_{K,i} ( \bm{\th}_i), \bm{\th}_{-i}) - x(\marg^k \s_{K,i} (\bm{\th}_i), \th_{-i}^k) \|} \\
&= 0.
\end{aligned}
\end{equation}
Similarly, 
\begin{equation} \label{eq:equil_lim}
\lim_{K \to \infty} \E_{\bm{\th}} \Brac{ \frac{1}{K} \sum_{k=1}^{K} \Abs{ u_i ( g_K^k ( \bm{\th}),\th_i^k) -  u_i ( x(\th^k),\th_i^k)}}  = 0.
\end{equation}

From \eqref{eq:deviation_lim} and \eqref{eq:equil_lim}, we conclude that
\begin{equation*}
\begin{aligned}
    &\lim_{K \to \infty} \E_{\bm{\th}} \Brac{ \frac{1}{K} \sum_{k=1}^{K}  u_i \bigl( g_K^k ( \s_{K,i} ( \bm{\th}_i), \bm{\th}_{-i}),\th_i^k \bigr)  - \frac{1}{K} \sum_{k=1}^{K} u_i ( g_K^k ( \bm{\th}),\th_i^k)} \\
    &=   \lim_{K \to \infty} \E_{\bm{\th}} \Brac{ \frac{1}{K} \sum_{k=1}^{K}  \Set{ u_i \bigl( x(\marg^k \s_{K,i} (\bm{\th}_i), \th_{-i}^k),\th_i^k \bigr) -  u_i ( x(\th^k),\th_i^k)}} \\
    &=    \lim_{K \to \infty} \E_{\bm{\th}_i} \Brac{ \frac{1}{K} \sum_{k=1}^{K}  \E_{\th_{-i} \sim \pi_{-i}} \Brac{ u_i \bigl( x(\marg^k \s_{K,i} (\bm{\th}_i), \th_{-i}),\th_i^k \bigr) -  u_i ( x(\th_i^k, \th_{-i}),\th_i^k)}} \\
    &=  \lim_{K \to \infty} \E_{\bm{\th}_i} \Brac{ \frac{N(\bm{\th}_i)}{K} G } \\
    &= G \min_{j=1,\ldots, J} \pi_i ( \th_{i,j}) \\
    &> 0,
\end{aligned}
\end{equation*}
where the last equality follows from the law of large numbers (and dominated convergence) and the final inequality holds because $\pi_i$ has full support. Therefore, for all $K$ sufficiently large, it is ex-ante profitable for agent $i$ to deviate from truthtelling to $\s_{K,i}$.


\subsection{Proof of Theorem~\ref{res:error_bound}}

We break the proof into parts. Throughout the proof, we use the following notation. Under any fixed prior profile $\pi$, we denote expectations over $\th \in \Th$ by $\E_{\th}^{\pi}$ and over $\bm{\th} \in \Th^K$ by $\E_{\bm{\th}}^{\pi}$.

\paragraph{Upper bound}
Fix a prior profile $\pi \in \prod_{i=1}^{n} \D(\Th_i)$. First we construct the social choice function $x_{\pi}$ satisfying \eqref{eq:distributional_robustness_bound}. For each agent $i$, define the transport cost function $c_i \colon \Th_i \times \Th_i \to \R$ by 
\[
    c_i(\th_i, \th_i') = - \E_{\th_{-i} \sim q_{-i}} \Brac{ u_i ( x(\th_i', \th_{-i}), \th_i)}.
\]
By \cref{res:DZ_bound}, there exists a $c_i$-optimal kernel coupling $\hat{r}_i \colon \Th_i \to \D(\Th_i)$ of $\pi_i$ and $q_i$ such that
\begin{equation} \label{eq:pi_bar_bound}
    1 - \sum_{\th_i \in \Th_i} \pi_i (\th_i) \hat{r}_i( \th_i | \th_i) \leq ( | \Th_i| - 1) \| q_i - \pi_i \|. 
\end{equation}
Define $x_\pi \colon \Th \to \D(\XX)$ by $x_\pi (\th) = x (\oprod_{i=1}^{n} \hat{r}_i(\th_i))$. For each $\th \in \Th$, \cref{res:TV_bounds} gives
\begin{equation*}
\begin{aligned}
    \| x_\pi (\th) - x(\th) \| 
    &= \| x( \oprod_{i=1}^{n} \hat{r}_i (\th_i)) - x(\th) \| \\
    &\leq \| \oprod_{i=1}^{n} \hat{r}_i(\th_i) - \d_{\th} \| \\
    &\leq \sum_{i=1}^{n} [ 1 - \hat{r}_i ( \th_i | \th_i)].
\end{aligned}
\end{equation*}
Take expectations and apply \eqref{eq:pi_bar_bound} to conclude that
\begin{equation*}
\begin{aligned}
    \E_{\th}^{\pi} \| x_\pi (\th) - x (\th) \|  
    &\leq \sum_{i=1}^{n} \sum_{\th_i \in \Th_i} \pi_i (\th_i) [ 1 - \hat{r}_i ( \th_i | \th_i)] \\
    &\leq \sum_{i =1}^{n} ( | \Th_i| - 1) \| q_i - \pi_i \|. 
\end{aligned}
\end{equation*}

Now we show that under the distribution $\pi$, the $(x,q)$-quota mechanisms asymptotically implement $x_\pi$. First we select solutions of the auxiliary optimal transport problems. For each $i$, recall that $\hat{r}_i$ is a $c_i$-optimal kernel coupling of $\pi_i$ and $q_i$. By \cref{res:Lip}, for each $p_i \in \D ( \Th_i)$, there exists a $c_i$-optimal kernel coupling $r_i \colon \Th_i \to \D (\Th_i)$ of $p_i$ and $q_i$ such that 
\begin{equation} \label{eq:ineq_mixed_first}
     \| p_i \otimes r_i - \pi_i \otimes \hat{r}_i \|  \leq    (|\Th_i| -1) \|\pi_i - p_i\|,
\end{equation}
hence
\begin{equation} \label{eq:ineq_mixed}
\begin{aligned}
    \sum_{\th_i \in \Th_i} p_i (\th_i) \| r_i(\th_i) - \hat{r}_i (\th_i) \| 
    &= \| p_i \otimes r_i - p_i \otimes \hat{r}_i \| \\
    &\leq \| p_i \otimes r_i - \pi_i \otimes \hat{r}_i \| +  \| \pi_i \otimes \hat{r}_i - p_i \otimes \hat{r}_i \| \\
    &\leq (|\Th_i| -1) \|\pi_i - p_i\| + \|\pi_i - p_i\| \\
    &= |\Th_i| \| \pi_i - p_i\|.
\end{aligned}
\end{equation}
To indicate the dependence of $r_i$ on the initial distribution $p_i$, we denote $r_i(\cdot)$ by $r_i( \cdot ; p_i)$.

Next we construct the agents' strategies. For each $K$, let $(M_K, g_K)$ denote the $(x, q)$-quota mechanism in the $K$-composite problem. For each agent $i$, define the strategy $\s_{K,i} \colon \Th_i^K \to M_{K,i}$ by 
\[
    \s_{K,i}^k ( \bm{\th}_i) = r_i ( \th_i^k; \marg \bm{\th}_i), \qquad k  =1, \ldots, K.
\]
By following the proof of \cref{res:optimal_transport_bound} (\cref{sec:proof_optimal_transport_bound}), it can be shown that the strategy profile $\s_{K} = (\s_{K,i})_{i=1}^{n}$ is a Bayes--Nash equilibrium of  $(M_K, g_K)$. 
 
It remains to check that the sequence $(\s_{K})_{K \geq 1}$
asymptotically implements $x_\pi$. For each $K$ and each type-vector profile $\bm{\th} \in \Th^K$, we conclude from \cref{res:TV_bounds} that for each problem $k$, 
\begin{equation} \label{eq:bound_r}
\begin{aligned}
    \| g_K^k (\s_{K} (\bm{\th})) - x_\pi (\th^k) \| 
    &= \| x ( \oprod_{i=1}^{n} \s_{K,i}^k (\bm{\th}_i)) - x ( \oprod_{i=1}^{n} \hat{r}_i(\th_i^k)) \| \\
    &\leq  \| \oprod_{i=1}^{n} \s_{K,i}^k (\bm{\th}_i) - \oprod_{i=1}^{n} \hat{r}_i ( \th_i^k) \| \\
    &\leq \sum_{i=1}^{n} \| \s_{K,i}^k (\bm{\th}_i) - \hat{r}_i ( \th_i^k) \| \\
    &= \sum_{i=1}^{n} \| r_i ( \th_i^k ; \marg \bm{\th}_i) - \hat{r}_i (\th_i^k) \|.
\end{aligned}
\end{equation}
Average the inequality \eqref{eq:bound_r} over problems $k = 1, \ldots, K$, and then for each $i$  apply \eqref{eq:ineq_mixed} with $p_i = \marg \bm{\th}_i$ to get 
\begin{equation*}
\begin{aligned}
    \frac{1}{K} \sum_{k=1}^{K}  \| g_K^k (\s_{K} (\bm{\th})) - x_\pi (\th^k) \| 
    &\leq \sum_{i=1}^{n} \frac{1}{K} \sum_{k=1}^{K} \| r_i (\th_i^k; \marg \bm{\th}_i) - \hat{r}_i (\th_i^k) \| \\
    &\leq \sum_{i=1}^{n} | \Th_i|  \| \pi_i - \marg \bm{\th}_i \|.
\end{aligned}
\end{equation*}
Take expectations over $\bm{\th}$ (with respect to $\pi$) and then pass to the limit as $K \to \infty$.  By the law of large numbers (in $L^1$), $\lim_{K \to \infty} \E_{\bm{\th}_i}^{\pi} \| \pi_i - \marg \bm{\th}_i  \| = 0$ for each agent $i$, so we conclude that
\[
\lim_{K \to \infty} \E_{\bm{\th}}^{\pi} \Brac{ \frac{1}{K} \sum_{k=1}^{K}  \| g_K^k (\s_{K} (\bm{\th})) - x_\pi (\th^k) \| } = 0.
\]

\paragraph{Upper bound is tight} Suppose $n = 1$. Hereafter, we drop agent subscripts. Fix a type space $\Th$ with $|\Th| \geq 2$. We construct a decision environment $(\XX, u)$, a cyclically monotone social choice function $x \colon \Th \to \D(\XX)$, and a quota $q \in  \D (\Th)$ such that, for each $\e >0$, there is no sequence $(M_K, g_K)_{K \geq 1}$ of linking mechanisms satisfying the following \emph{$\e$-strengthened property}: For each $\pi \in \D (\Th)$, the sequence $(M_K, g_K)_{K \geq 1}$ asymptotically implements, under the distribution $\pi$,  some social choice function $x_\pi \colon \Th \to \D (\XX)$ satisfying
\begin{equation} \label{eq:Lepsilon}
        \E_{\th}^{\pi} \| x_\pi (\th) - x(\th) \| \leq ( | \Th| - 1 - \e) \| q - \pi \|.
\end{equation}

The construction is similar to the construction in the proof of \cref{res:optimal_transport_bound} (\cref{sec:proof_optimal_transport_bound}). To simplify notation, let $m = |\Th|$. By assumption, $m \geq 2$. Without loss, we can relabel the types so that $\Th = \{ \th_{1}, \ldots, \th_{m}\}$. Let $\XX = \{x_1, \ldots, x_m\}$. Define the agent's utility function $u \colon \XX \times \Th \to \R$ by
\[
    u( x_p, \th_{\ell}) = 
    \begin{cases} 
    -(m-1) &\text{if}~ p < \ell, \\
    0 &\text{if}~ p = \ell,\\
    1 &\text{if}~ p > \ell.
    \end{cases}
\]
Consider the deterministic social choice function $x \colon \Th \to \XX$  defined by $x( \th_{\ell})  = x_\ell$ for each $\ell \in \{1, \ldots, m \}$. The social choice function $x$ is cyclically monotone, as shown in the proof of \cref{res:optimal_transport_bound}. Let $q$ be the uniform distribution over $\Th$. 

Fix $\e > 0$. Suppose for a contradiction that there exists a sequence $(M_K, g_K)_{K \geq 1}$ of linking mechanisms that satisfies the $\e$-strengthened property. That is, for each $\pi \in \D(\Th)$, there exists a social choice function $x_\pi$ satisfying \eqref{eq:Lepsilon} and a sequence $(\s_{K}^{\pi} )_{K\geq 1}$ of best responses to $(M_K, g_K)_{K \geq 1}$ that asymptotically implement $x_\pi$ under the distribution $\pi$. Consider the distribution $\bar{\pi} = q + (\d_{\th_{1}} - \d_{\th_{m}})/(2m)$.\footnote{Though not strictly necessary for our claimed result, we provide a construction that uses only full-support priors.} Define $r \colon \Th \to \D(\Th)$ as follows. Let $r(\th_{1}) = (2/3) \d_{\th_{1}} + (1/3)\d_{\th_{2}}$. For $\ell = 2, \ldots, m-1$, let $r(\th_{\ell}) = \d_{\th_{\ell}}/2 + \d_{\th_{\ell + 1}}/2$. Let $r(\th_{m}) = \d_{\th_{m}}$. By construction, $r(\bar{\pi}) = q$. For each $K$, define $\bm{r}_{K} \colon \Th^K \to \D(\Th^K)$ by $\bm{r}_{K} (\bm{\th}) = \oprod_{k=1}^{K} r(\th^k)$.

To get a contradiction, we show that under the distribution $\bar{\pi}$, for all $K$ sufficiently large, it is ex-ante profitable for the agent to deviate from $\s_{K}^{\bar{\pi}}$ to $\s_{K}^{q} \circ \bm{r}_{K}$. Here, $\s_{K}^{q}$ is linearly extended to $\D(\Th^K)$, so $\s_{K}^{q} \circ \bm{r}_{K}$ is a function from $\Th^K$ into $\D( M_K)$. For each $K$, by the convexity of the total variation norm, we have
\begin{equation} \label{eq:TV_Jensen_bound}
\begin{aligned}
&\E_{\bm{\th}}^{\bar{\pi}} \Brac{ \frac{1}{K} \sum_{k=1}^{K}  \bigl\| g_K^k \bigl
(\s_K^q (\bm{r}_K(\bm{\th})) \bigr) - x_q ( r (\th^k)) \bigr\|} \\
&= \E_{\bm{\th}}^{\bar{\pi}} \Brac{ \frac{1}{K} \sum_{k=1}^{K}   \bigl\| \E_{\tilde{\bm{\th}} \sim \bm{r}_K ( \bm{\th})} [g_K^k ( \s_K^q (\tilde{\bm{\th}})) - x_q ( \tilde{\th}^k) ] \bigr\| } \\
&\leq \E_{\bm{\th}}^{\bar{\pi}} \Brac{ \frac{1}{K} \sum_{k=1}^{K}  \E_{\tilde{\bm{\th}} \sim \bm{r}_K ( \bm{\th})} \| g_K^k ( \s_K^q (\tilde{\bm{\th}})) - x_q ( \tilde{\th}^k) \|} \\
&= \E_{\bm{\th}}^q \Brac{ \frac{1}{K} \sum_{k=1}^{K}  \| g_K^k ( \s_K^q (\bm{\th})) - x_q ( \th^k) \|},
\end{aligned}
\end{equation}
where the last equality holds because $\bm{r}_K ( \bar{\pi}^{\otimes K}) = q^{\otimes K}$. By assumption, the sequence $(\s_{K}^{q})_{K\geq 1}$ asymptotically implements $x_{q}$ under the distribution $q$, so the right side of \eqref{eq:TV_Jensen_bound} converges to $0$ as $K \to \infty$. Therefore, 
\begin{equation} \label{eq:deviation_payoff}
\begin{aligned}
&\lim_{K \to \infty} \E_{\bm{\th}}^{\bar{\pi}} \Brac{ \frac{1}{K} \sum_{k=1}^{K} \Abs{ u \bigl( g_K^k ( \s_K^q (\bm{r}_K(\bm{\th}))), \th^k \bigr) -  u \bigl( x_q ( r(\th^k)), \th^k \bigr)  } } \\
&\leq \lim_{K \to \infty} \E_{\bm{\th}}^{\bar{\pi}} \Brac{ \frac{1}{K} \sum_{k=1}^{K}  2 \| u\|_{\infty} \| g_K^k ( \s_K^q (\bm{r}_K(\bm{\th}))) - x_q ( r (\th^k)) \| } \\
&= 0. 
\end{aligned}
\end{equation}
Similarly, since the sequence $(\s_{K}^{\bar{\pi}})_{K\geq 1}$ asymptotically implements $x_{\bar{\pi}}$ under the distribution $\bar{\pi}$, we have
\begin{equation} \label{eq:strat_payoff}
\lim_{K \to \infty} \E_{\bm{\th}}^{\bar{\pi}} \Brac{ \frac{1}{K} \sum_{k=1}^{K} \Abs{ u ( g_K^k ( \s_K^{\bar{\pi}} (\bm{\th})), \th^k) -  u (x_{\bar{\pi}} ( \th^k), \th^k)  } }  = 0.
\end{equation}

Finally, we analyze the agent's ex-ante expected payoff from the strategies $\s_{K}^{q} \circ \bm{r}_{K}$ and $\s_{K}^{\bar{\pi}}$, under the distribution $\bar{\pi}$, in the limit as $K \to \infty$. By \eqref{eq:deviation_payoff},
\begin{equation} \label{eq:deviation_final}
\begin{aligned}
    \lim_{K \to \infty} \E_{\bm{\th}}^{\bar{\pi}}
\Brac{\frac{1}{K} \sum_{k=1}^{K} u \bigl( g_K^k ( \s_K^q ( \bm{r}_K (\bm{\th}))), \th^k \bigr) }  
&= \lim_{K \to \infty} \E_{\bm{\th}}^{\bar{\pi}}
\Brac{\frac{1}{K} \sum_{k=1}^{K} u (x_q ( r (\th^k)), \th^k) } \\
&= \E_{\th}^{\bar{\pi}} u (x_q (r(\th)), \th) \\
&=    \E_\th^{\bar{\pi}} u ( x (r(\th)), \th ) \\
&= \frac{m-1}{2m},
\end{aligned}
\end{equation}
where the penultimate equality holds because $x_q= x$, which follows from applying \eqref{eq:Lepsilon} with the prior $\pi = q$ (which has full support). By \eqref{eq:strat_payoff}, 
\begin{equation} \label{eq:on_path_final}
\begin{aligned}
    \lim_{K \to \infty} \E_{\bm{\th}}^{\bar{\pi}} \Brac{ \frac{1}{K} \sum_{k=1}^{K} u \bigl( g_K^k (\s_K^{\bar{\pi}}(\bm{\th})), \th^k\bigr)} 
    &=  \lim_{K \to \infty} \E_{\bm{\th}}^{\bar{\pi}} \Brac{ \frac{1}{K} \sum_{k=1}^{K} u (x_{\bar{\pi}}(\th^k), \th^k)} \\
    &= \E_{\th}^{\bar{\pi}} [u( x_{\bar{\pi}}(\th), \th)]\\
    &\leq \E_{\th}^{\bar{\pi}} [u( x(\th), \th)] + \E_{\th}^{\bar{\pi}}  \| x_{\bar{\pi}}(\th) - x(\th) \| \\
    &\leq \E_\th^{\bar{\pi}} \Brac{ u(x(\th), \th)} + (m-1 -\e) \| q - \bar{\pi} \| \\
    &= \frac{m- 1 - \e}{2m},
\end{aligned}
\end{equation}
where the first inequality holds because the deterministic social choice function $x$ satisfies the inequality $u(x',\th) -u(x(\th), \th) \leq 1$ for all $x' \in \XX$ and $\th \in \Th$; and the second inequality follows from applying \eqref{eq:Lepsilon} with $\pi = \bar{\pi}$. Comparing \eqref{eq:deviation_final} and \eqref{eq:on_path_final}, we conclude that under the distribution $\bar{\pi}$, the deviation from $\s_{K}^{\bar{\pi}}$ to $\s_{K}^{q} \circ \bm{r}_{K}$ is ex-ante profitable for all $K$ sufficiently large. 


\subsection{Proof of Theorem~\ref{res:robust}}

The proof is similar to the proof of \cref{res:optimal_transport_bound} (\cref{sec:proof_optimal_transport_bound}). The tightness of the upper bound is immediate from that proof since the independent, common prior setting is a particular payoff-type exchangeable, payoff-type independent type space; technically, the tightness argument in \cref{res:optimal_transport_bound} considers the case $n = 1$, but we can add additional players with singleton type spaces and the argument still goes through. Here, we construct an equilibrium  satisfying the upper bound. In the $K$-composite problem, let $(M,g)$ denote the $(x,q)$-quota mechanism. For each agent $i$, let $\s_i \colon \Th_i^K \to M_i$ denote the strategy constructed in the proof of \cref{res:optimal_transport_bound} (\cref{sec:proof_optimal_transport_bound}). Let $(T, (\hat{\bm{\th}}_i, \hat{\b}_i)_{i=1}^{n} )$ be a type space that is payoff-type exchangeable and payoff-type independent. For each agent $i$, define the strategy $\hat{\s}_i \colon T_i \to M_i$ by $\hat{\s}_i (t_i) = \s_i ( \hat{\bm{\th}}_i (t_i))$. From the proof of \cref{res:optimal_transport_bound}, it is immediate that $\hat{\s}$ satisfies \eqref{eq:ineq_t}. 

It remains to check that $\hat{\s}$ is a Bayes--Nash equilibrium. For each agent $i$, type $t_i \in T_i$, and problem $k$, we have
\begin{equation*} 
\begin{aligned}
    \E_{t_{-i} \sim \hat{\b}_i(t_i)} \Brac{ \oprod_{j \neq i} \hat{\s}_{j}^k (t_j)} 
    &=    \E_{t_{-i} \sim \hat{\b}_i(t_i)} \Brac{ \oprod_{j \neq i} \s_{j}^k (\hat{\bm{\th}}_j(t_j))} \\
  &=   \E_{\bm{\th}_{-i} \sim \hat{\bm{\th}}_{-i}( \hat{\b}_i (t_i)) } \Brac{  \oprod_{j \neq i} \s_j^k (\bm{\th}_j)} \\
  &= \oprod_{j \neq i} \E_{\bm{\th}_{-i} \sim \hat{\bm{\th}}_{-i}( \hat{\b}_i (t_i)) } \Brac{   \s_j^k (\bm{\th}_j)} \\
  &=\oprod_{j \neq i}\, q_j,
\end{aligned}
\end{equation*}
where the third equality uses payoff-type independence, and the final equality follows from payoff-type exchangeability and the fact that each $\s_j$ is label-free. Therefore, we can complete the proof exactly as in the proof of \cref{res:optimal_transport_bound}.

\subsection{Proof of Theorem~\ref{res:interdependent}}

As in the case of private values, the equivalence \eqref{it:IDP_oneshot} $\iff$ \eqref{it:IDP_CM} follows from \citet[Theorem 1, p.~192]{Rochet1987}. It is immediate that \eqref{it:IDP_quota} $\implies$ \eqref{it:IDP_transfers}. The proof that \eqref{it:IDP_transfers} $\implies$ \eqref{it:IDP_CM}, which uses a direct implementation, is almost identical to the corresponding proof of \cref{res:equivalence} (\cref{sec:proof_equivalence}). Here, we prove that \eqref{it:IDP_CM} $\implies$ \eqref{it:IDP_quota}. 

Let $x \colon \Th \to \D (\XX)$ be $\pi$-cyclically monotone. Let $(M_K, g_K)_{K \geq 1}$ denote the sequence of $(x, \pi)$-quota mechanisms. For each agent $i$, define the cost function $c_i \colon \Th_i \times \Th_i \to \R$ by 
\[
    c_i (\th_i, \th_i')  = -\E_{ \th_{-i} \sim \pi_{-i}} \Brac{ u_i (x(\th_i', \th_{-i}), \th_i, \th_{-i})}.
\]
For each $K$ and each agent $i$, define the strategy $\s_{K,i} \colon \Th_i^K \to M_{K,i}$ as in the 
proof of \cref{res:optimal_transport_bound} (\cref{sec:proof_optimal_transport_bound}) using the quota $q_i = \pi_i$ and this new cost function $c_i$. Take expectations in \eqref{eq:EP_bound} and pass to the limit as $K \to \infty$. By the law of large numbers (in $L^1$), $\lim_{K \to \infty} \E_{\bm{\th}_i} \| \pi_i - \marg \bm{\th}_i \| = 0$ for each $i$, so the sequence $(\s_K)_{K \geq 1}$ satisfies \eqref{eq:approx_asymptotic}. To complete the proof, we check that $(\s_K)_{K \geq 1}$ has the required approximate equilibrium property.

For each $K$, define the cost function $c_{K,i} \colon \Th_i \times \Th_i \to \R$ by
\[
    c_{K,i} (\th_i, \th_i')  = - \E_{ \bm{\th}_{-i}} \Brac{\frac{1}{K} \sum_{k=1}^{K} u_i \bigl(x(\th_i', \oprod_{j \neq i} \s^k_{K,j} (\bm{\th}_j)), \th_i, \th_{-i}^k\bigr)},
\]
where each function $x (\th_i', \cdot)$ is extended linearly from $\Th_{-i}$ to $\D(\Th_{-i})$ and each function $u_i ( \cdot, \th_i, \th_{-i}^k)$ is extended linearly from $\XX$ to $\D(\XX)$. In the $K$-composite problem, type $\bm{\th}_i$'s interim expected utility from reporting $\bm{r}_i \in M_{K,i}$ when his opponents follow $\s_{K,-i}$ is given by 
\[
    - \sum_{\th_i, \th_i' \in \Th_i} c_{K,i} (\th_i, \th_i') \g_i (\th_i, \th_i'),
\]
where $\g_i$ is the coupling of $\marg \bm{\th}_i$ and $\pi_i$ defined by 
\[
    \g_i (\th_i, \th_i') = \frac{1}{K} \sum_{k : \th_i^k = \th_i} r_i^k (\th_i'). 
\]
Since the coupling induced by $\s_{K,i}$ is $c_i$-optimal, it follows that type $\bm{\th}_i$'s interim   expected gain from deviating is at most $2 \| c_{K,i} - c_i \|_{\infty}$.\footnote{To see this, let $\g_i^\ast$ be a $c_i$-optimal coupling of $\marg \bm{\th}_i$ and $\pi_i$, and let $\g_i$ be an arbitrary coupling of  $\marg \bm{\th}_i$ and $\pi_i$. Then $\g_i^\ast c_{K,i} - \g_{i} c_{K,i} =  (\g_i^\ast  c_{K,i} - \g_i^\ast c_i) + (\g_i^\ast  c_i - \g_{i} c_i) + (\g_{i} c_i - \g_{i} c_{K,i})$. Since $\g_i^\ast$ is $c_i$-optimal, the middle term $\g_i^\ast c_i - \g_i c_i$ is nonpositive. Since $\g_i^\ast$ and $\g_i$ are probability measures, we conclude that $\g_i^\ast c_{K,i} - \g_{i} c_{K,i} \leq 2 \| c_{K,i} - c_i \|_{\infty}$.} Let $\e_K = 2 \max_{i=1,\ldots, n} \| c_{K,i} - c_i \|_{\infty}$. By construction, $\s_{K}$ is a Bayes--Nash $\e_K$-equilibrium of $(M_K, g_K)$. 

It remains to check that $\lim_{K \to \infty} \e_K = 0$. Fix agent $i$. Observe that $c_i$ can alternatively be expressed as 
\[
c_i (\th_i, \th_i') =- \E_{ \bm{\th}_{-i}} \Brac{\frac{1}{K} \sum_{k=1}^{K}  u_i (x(\th_i', \th_{-i}^k), \th_i, \th_{-i}^k)}.
\]
Thus, for each $\th_i, \th_i' \in \Th_i$, we have
\begin{equation} \label{eq:cost_bound}
\begin{aligned}
  &| c_{K,i} (\th_i, \th_i') - c_i (\th_i, \th_i') | \\
  &\leq \E_{\bm{\th}_{-i}}  \Brac{ \frac{1}{K}  \sum_{k=1}^{K} \Abs{u_i \bigl( x(\th_i', \oprod_{j \neq i} \s^k_{K,j} (\bm{\th}_j)), \th_i, \th_{-i}^k \bigr) -  u_i (x(\th_i', \th_{-i}^k), \th_i, \th_{-i}^k)}} \\
  &\leq  \E_{\bm{\th}_{-i}} \Brac{ \frac{1}{K} \sum_{k=1}^{K} 2 \| u_i \|_{\infty}  \| x ( \th_i', \oprod_{j \neq i} \s^k_{K,j} (\bm{\th}_j)) -x(\th_i', \th_{-i}^k) \|}.
\end{aligned}
\end{equation}
Following the argument in the proof of \cref{res:optimal_transport_bound} (\cref{sec:proof_optimal_transport_bound}), we see that for each $\th_i' \in \Th_i$ and each $\bm{\th}_{-i} \in \Th_{-i}^K$, we have
\begin{equation} \label{eq:bound_i}
  \frac{1}{K} \sum_{k=1}^{K} \| x ( \th_i', \oprod_{j \neq i} \s^k_{K,j} (\bm{\th}_j)) -x(\th_i', \th_{-i}^k) \| \leq \sum_{j \neq i} (|\Th_j| - 1) \| \pi_j - \marg \bm{\th}_j \|.
\end{equation}
Substitute \eqref{eq:bound_i} into \eqref{eq:cost_bound} to conclude that
\[
 \| c_{K,i} - c_i \|_{\infty}  \leq 2 \| u_i\|_{\infty} \sum_{j \neq i} (|\Th_j| - 1) \E_{\bm{\th}_j} \| \pi_j - \marg \bm{\th}_j \|.
\]
By the law of large numbers (in $L^1$), we have $\lim_{K \to \infty} \E_{\bm{\th}_j} \| \pi_j - \marg \bm{\th}_j \| = 0$ for each $j \neq i$. Thus, $\lim_{K \to \infty} \| c_{K,i} - c_i \|_{\infty} = 0$. Since agent $i$ is arbitrary, we conclude that $\lim_{K \to \infty} \e_K = 0$.

\subsection{Proof of Theorem~\ref{res:dynamic_implementation}} \label{sec:proof_dynamic}

Let $x \colon \Th \to \D(\XX)$ be strictly cyclically monotone. First, we introduce notation. Consider the $\b$-discounted problem, for some fixed $\b \in (0,1)$. Given any strategy $\s$ in the dynamic $(x,\pi)$-quota mechanism, we already defined the associated probability measure $\rho(\s)$ over $\Th^{\infty} \times [\D(\Th)]^{\infty}$. Now define the probability measure $\g_\b(\s) \in \D (\Th^2)$ by 
\[
   \g_\b ( \s)  =  \E_{ (\bm{\th}, \bm{r}) \sim \rho(\s)} \Brac{ (1 - \b) \sum_{t=0}^{\infty} \b^t \sum_{\th' \in \Th} \d_{(\th^t, \th')} r^t (\th')},
\]
where the expectation is taken in the space $\D(\Th^2)$. Note that $\g_\b( \s)$ is a coupling of $\pi$ and $\pi$. The agent's expected utility from any strategy $\s$ in the $(x,\pi)$-quota mechanism is
\[
    \E_{ (\th, \th') \sim \g_\b (\s)} \Brac{ u( x(\th'), \th)}.
\]
Define the cost function $c \colon \Th^2 \to \R$ by $c(\th, \th') = - u ( x(\th'), \th)$. Let
\[
    D(x) = \{ (\th, \th') \in \Th^2: x(\th) = x(\th') \}.
\]
Since $x$ is strictly cyclically monotone, the set $D(x)$ is strictly $c$-cyclically monotone; see \cref{ft:strict_CM}.

With these preliminaries, we turn to the proof.  In fact, we will prove a stronger property, namely that the desired inequality \eqref{eq:asymptotically_implement} holds for every selection of best responses. In each $\b$-discounted problem, the agent has some (pure) best response $\hat{\s}_\b = (\hat{\s}_\b^t)_{t=0}^{\infty}$ to the dynamic $(x, \pi)$-quota mechanism; this follows from  a standard dynamic programming argument.\footnote{To formulate the agent's problem as a stochastic dynamic programming problem, define the state space to be $\D(\Th) \times \Th$. The state $(Q, \th)$ indicates the (normalized) remaining quota $Q \in \D(\Th)$ and the agent's current type $\th \in \Th$. In state $(Q, \th)$, the agent chooses $r \in \D(\Th)$ satisfying $r \leq Q  / (1 - \b)$. The sequence $(\th^t)$ is independent and identically distributed. Initially, $Q^0$ equals the quota $\pi$. For $t \geq 0$, 
\[
    Q^{t+1}  = \b^{-1} \Brac{ Q^t - (1 - \b) r^t}.
\]
The associated Bellman equation is given by
\[
    V(Q,\th) = \sup_{r} \Brac{ (1 - \b) \sum_{\th' \in \Th} r(\th') u (x(\th'), \th) + \b \sum_{\th'' \in \Th} \pi(\th'') V \Paren{ \b^{-1} \Brac{ Q - (1 - \b) r}, \th''} },
\]
where the supremum is over all $r \in \D(\Th)$ satisfying $r \leq Q / (1 - \b)$. Following the argument in \citet[Theorem 4.6, p.~79]{stokey1989recursive}, it can be shown that this Bellman equation is solved by some continuous, bounded value function. The associated optimal policy function $r \colon \D(\Th) \times \Th \to \D(\Th)$ determines the agent's best response.} Let $D$ denote the diagonal in $\Th^2$. We claim (see proof below) that in each $\b$-discounted problem, there exists a feasible strategy $\tilde{\s}_\b$ in the dynamic $(x,\pi)$-quota mechanism such that $\lim_{\b \to 1} \g_\b (\tilde{\s}_\b) (D) = 1$.  For each $\b \in (0,1)$, let $\hat{\g}_\b = \g_\b (\hat{\s}_\b)$ and $\tilde{\g}_\b = \g_\b (\tilde{\s}_\b)$. By construction, $\hat{\g}_\b$ and $\tilde{\g}_\b$ are both couplings of $\pi$ and $\pi$. For each $\b$, we have
\begin{equation} \label{eq:ineq_b}
    \E_{ (\th, \th') \sim \tilde{\g}_\b } \Brac{ u( x(\th'), \th)} \leq 
    \E_{ (\th, \th') \sim \hat{\g}_\b } \Brac{ u( x(\th'), \th)} 
    \leq 
    \E_{ \th \sim \pi } \Brac{ u( x(\th), \th)},
\end{equation}
where the first inequality follows from the optimality of $\hat{\s}_\b$ and the second inequality holds because the coupling of $\pi$ and $\pi$ that keeps all probability fixed is $c$-optimal by \cref{res:DZ_bound} (since $x$ is cyclically monotone and hence the diagonal $D$ is $c$-cyclically monotone). In \eqref{eq:ineq_b}, pass to the limit  as $\b \to 1$. By our claim, $\lim_{\b \to 1} \g_\b (\tilde{\s}_\b) (D) = 1$, so
\[
\lim_{\b \to 1} \E_{ (\th, \th') \sim \tilde{\g}_\b } \Brac{ u( x(\th'), \th)} =  \E_{ \th \sim \pi } \Brac{ u( x(\th), \th)}.
\]
By the sandwich theorem,
\begin{equation} \label{eq:limit_util}
     \lim_{\b \to 1} \E_{ (\th, \th') \sim \hat{\g}_\b } \Brac{ u( x(\th'), \th)} =  \E_{ \th \sim \pi } \Brac{ u( x(\th), \th)}.
\end{equation}

Using \eqref{eq:limit_util}, we now complete the proof. For each $\b \in (0,1)$, we have
\begin{equation*}
\begin{aligned}
    &\E_{ (\bm{\th}, \bm{r}) \sim \rho(\hat{\s}_\b)} \Brac{ (1 - \b) \sum_{t=0}^{\infty} \b^t  \| x ( r^{t}) -  x(\th^t) \| } \\
    &\leq \E_{ (\bm{\th}, \bm{r}) \sim \rho(\hat{\s}_\b)} \Brac{ (1 - \b) \sum_{t=0}^{\infty} \b^t  r^t ( \{ \th' \in \Th:  x(\th') \neq x(\th^t) \} )  } \\
    &= 1 - \hat{\g}_\b (D(x)).
\end{aligned}
\end{equation*}
Therefore, it suffices to check 
that  $\lim_{\b \to 1} \hat{\g}_\b (D(x)) = 1$. Suppose not. Then for some $\e > 0$ there is a sequence $(\b_j)$ converging to $1$ such that $\hat{\g}_{\b_j} ( D(x)) \leq 1 - \e$ for each $j$. Since the space of couplings of $\pi$ and $\pi$ (viewed as a subset of Euclidean space) is compact, the sequence $(\hat{\g}_{\b_j})$ has some limit point $\hat{\g}$. By \eqref{eq:limit_util}, we have
\[
   \E_{ (\th, \th') \sim \hat{\g}} [u(x(\th'), \th)] = \E_{ \th \sim \pi } \Brac{ u( x(\th), \th)}.
\]
Thus, $\hat{\g}$ is a $c$-optimal coupling of $\pi$ and $\pi$ satisfying $\hat{\g} (D(x)) \leq 1 - \e$, contrary to \cref{res:S_bound} (with $S = D(x)$).


\paragraph{Proof of claim}
For each $\b$, let $\tilde{\s}_\b$ be a strategy with the property that the agent reports truthfully (i.e. reports a unit mass on his true type) whenever doing so is feasible. For each $\b \in (0,1)$, time $t$, and (nonrandom) type $\th \in \Th$, define the random variable
\[
    N_t^\b (\th) = (1 - \b) \sum_{s=0}^{t} \b^s [\th^s = \th].
\]
Note that  $N_t^\b (\th)$ is a function of the random sequence $\bm{\th}$. Compute the mean and variance:
\[
    \E  N_t^\b (\th)  
    = (1 - \b^{t +1}) \pi (\th), \qquad
    \var ( N_t^\b (\th)) =
\frac{1 - \b}{1 + \b} \Paren {1 - \b^{2 (t + 1)}} \pi (\th) (1 - \pi(\th)).
\]
In particular, $\var ( N_t^\b (\th)) \leq \frac{1 - \b}{1 + \b}  \pi(\th) (1 - \pi(\th))$, so by Chebyshev's inequality,
\begin{equation} \label{eq:Cheb}
\begin{aligned}
        \P \bigl( N_t^\b (\th) > \pi (\th) \bigr)  
        &\leq \P \bigl( | N_t^\b (\th) - \E   N_t^\b (\th) | \geq \b^{t + 1} \pi (\th) \bigr) \\
    &\leq \frac{ (1 - \b)(1 - \pi(\th))}{(1 + \b) \b^{2 (t +1)}\pi(\th)}.
\end{aligned}
\end{equation}
Fix $\e \in (0,1)$. For each $\b \in (\e,1)$, let $t(\b) = \lfloor  \log_\b \e \rfloor$. By construction, $\b^{2 (t (\b) + 1)} \geq \e^2 \b^2$. Therefore, by \eqref{eq:Cheb}, for each $\th \in \Th$, we have
\begin{equation} \label{eq:prob_limit}
    \lim_{\b \to 1} \P \bigl( N_{t(\b)}^\b (\th) > \pi (\th) \bigr) = 0.
\end{equation}
If $ N_{t(\b)}^\b (\th) \leq \pi(\th)$ for all $\th$, then under the strategy $\tilde{\s}_\b$, the agent is truthful in periods $t = 0, \ldots, t(\b)$. Therefore, applying a union bound gives
\begin{equation} \label{eq:union}
    1 - \g_\b(\tilde{\s}_\b)(D) 
    \leq \b^{t(\b) + 1} + \sum_{\th \in \Th}  \P \bigl( N_{t(\b)}^\b( \th) > \pi (\th) \bigr).
\end{equation}
Note that $\b^{t(\b) + 1} \leq \e$ for each $\b \in (\e, 1)$.
In \eqref{eq:union}, pass to the limit as $\b \to 1$, and apply \eqref{eq:prob_limit} to conclude that $\lim_{\b \to 1} \g_\b(\tilde{\s}_\b) (D) \geq 1 - \e$. Since $\e \in (0,1)$ was arbitrary, the claim follows.

\subsection{Dynamics: Strict cyclical monotonicity is necessary}
 \label{ex:weak_CM} 

In the primitive problem, suppose that there are $m$ decisions and the agent has $m$ types, where $m \geq 3$. Write $\Th = \{ \th_1, \ldots, \th_m\}$ and $\XX = \{x_1, \ldots, x_m\}$. Type $\th_1$ ranks the decisions as $x_1 \succ \cdots \succ x_m$. All other types are indifferent between all decisions. Let $x$ be the deterministic social choice function that assigns decision $x_j$ to type $\th_j$ for each $j$. In particular, $x(\th_1) = x_1$, so $x$ is cyclically monotone. Let $\pi$ be the uniform distribution on $\Th$. Fix $\b \in (0,1)$. Consider the dynamic $(x,\pi)$-quota mechanism in the $\b$-discounted problem. The agent's unique optimal strategy, $\hat{\s}_\b$, is as follows. In period $t$, if $\th^t = \th_1$, make the lowest-indexed feasible report; otherwise, make the highest-indexed feasible report.\footnote{Technically, reports are probability vectors, so ``lowest-indexed'' and ``highest-indexed'' are defined with respect to first-order stochastic dominance. Such reports exist; apply a greedy algorithm.} This strategy ensures that the lowest-indexed reports are conserved for periods in which the agent's type is $\th_1$. This strategy does not distinguish between the realizations $\th_2, \ldots, \th_m$. Since the social choice function $x$ is injective,  we conclude that the dynamic $(x,\pi)$-quota mechanisms do not asymptotically implement $x$.

\newpage
\bibliographystyle{ecta}
\bibliography{lit.bib}

\newpage

\section{Comparing our quota mechanisms with JS's} \label{sec:comparison}

In this section, we formally describe JS's quota mechanisms. Then we establish three results.
\begin{enumerate}[label = \Roman*.]
    \item  Every type of every agent gets weakly higher expected utility under the equilibrium we construct in \cref{res:optimal_transport_bound} than under the equilibrium constructed by JS in their associated quota mechanism.
    \item \cref{res:optimal_transport_bound} fails if we use JS's definition of a quota mechanism.
    \item A weaker version of the bound in \cref{res:optimal_transport_bound} holds for JS's quota mechanisms. This weaker bound is sufficient for the asymptotic results (\cref{res:equivalence} and \cref{res:error_bound}).
\end{enumerate}

\subsection{JS's quota mechanisms} \label{sec:JS_quota}

Consider a quota profile $q \in \prod_{i=1}^{n} \D( \Th_i)$ and a $q$-cyclically monotone social choice function $x \colon \Th \to \D(\XX)$. In the $K$-composite problem, we compare our $(x,q)$-quota mechanism with JS's. 

First, suppose that $q$ is $(1/K)$-divisible, i.e., the components of the probability vectors $q_1, \ldots, q_n$ are all integer multiples of $1/K$. In this special case, our quota mechanism is essentially equivalent to JS's.  Their quota mechanism asks each agent $i$ to report a type vector $(\hat{\th}_i^1, \ldots, \hat{\th}_i^K)$ in which each type $\th_i \in \Th_i$ appears exactly $K q_i (\th_i)$ times. The social choice function $x$ is applied to the profile of reports on each problem. JS consider mixed-strategy equilibria of this mechanism. It can be verified that any profile of mixed strategies in their mechanism is outcome-equivalent to a profile of pure strategies in our mechanism, and conversely, any profile of pure strategies in our mechanism is outcome-equivalent to a profile of mixed strategies in their mechanism.\footnote{Fix agent $i$. Let $M_i$ denote agent $i$'s message set in our mechanism. That is, $M_i$ contains all vectors $\bm{r}_i \in [ \D (\Th_i)]^K$ that average to the quota $q_i$. Let $M_i'$ denote agent $i$'s message set in JS's mechanism.  That is, $M_i'$ contains all vectors $\hat{\bm{\th}}_i \in \Th_i^K$ in which each type $\th_i \in \Th_i$ appears exactly $K q_i (\th_i)$ times. Any mixture $\a_i \in \D ( M_i')$ can be replicated in our mechanism by reporting $\bm{r}_i = (\marg^k \a_i)_{k=1}^{K}$, where $\marg^k \a_i$ is the marginal of $\a_i$ on the $k$-th factor. It is easily verified that $\bm{r}_i$ satisfies the quota $q_i$, hence $\bm{r}_i$ is in $M_i$. For the converse, we check that for each vector $\bm{r}_i \in M_i$, there exists a mixture $\a_i \in \D( M_i')$ such that $\marg^k \a_i = r_i^k$ for all $k$. This follows from \citet[Theorem 1, p.~593]{BCKM2013}. To apply that result, represent each vector $\hat{\bm{\th}}_i \in M_i'$ as the $|\Th_i| \times K$  $[0,1]$-valued matrix that equals $1$ in entries $(\hat{\th}_i^1, 1), \ldots (\hat{\th}_i^K, K)$, and $0$ otherwise.  With this representation, the set $M_i'$ is defined by the constraint that each row $\th_i$ sums to $K q_i(\th_i)$ and each column sums to $1$. The collection of all rows, columns, and singletons forms a \emph{bihierarchy}.}

Next, suppose that $q$ is not $(1/K)$-divisible. JS (p.~247, 252) proceed as follows, under their assumption that each quota $q_i$ has full support. For each agent $i$, let $q_{K,i}$ be the $(1/K)$-divisible quota that is closest to $q_i$ in Euclidean distance. (Ties can be broken arbitrarily.) Let $q_K = (q_{K,i})_{i=1}^{n}$. If there are multiple agents, then a further modification is required to ensure that each agent believes that his opponents' reports on each problem are distributed according to  $\oprod_{j \neq i} q_{j}$.\footnote{\label{ft:PE}JS consider only social choice functions that are ex-ante Pareto efficient with respect to the prior. As long as the prior has full support, these social choice functions are ex-post cyclically monotone, so no further modifications are necessary. The reason for JS's modification can likely be traced to their claim (fn.~14, p.~252) that a social choice function may be Pareto efficient with respect to some prior, but not with respect to some approximation of this prior. But as long as the original prior has full support, then this cannot happen, as we now show. In fact, we prove a slightly stronger result. Let $x \colon \Th \to \D (\XX)$ be a social choice function. Consider priors $p,p' \in \D(\Th)$ such that $\supp p' \subseteq \supp p$.  We show that if $x$ is ex-ante Pareto efficient with respect to $p$, then $x$ is ex-ante Pareto efficient with respect to $p'$. Suppose for a contradiction that $x$ is ex-ante Pareto dominated, under prior $p'$, by some social choice function $y'$. To reach a contradiction, we show that $x$ is ex-ante Pareto dominated, under prior $p$, by the social choice function $y$ defined as follows. Since $\supp p' \subseteq \supp p$, we can choose $\e > 0$ such that $\e p' (\th) \leq p(\th)$ for all types $\th$. For each $\th \in \supp p$, let  
\[
    y ( \th) = \frac{\e p'(\th)}{p(\th)}  y'(\th) + \Paren{ 1 - \frac{\e p'(\th)}{p(\th)}} x(\th). 
\]
(For $\th \not\in \supp p $, we can define $y(\th)$ arbitrarily.)
By construction, under prior $p$, each player's expected gain from $y$ over $x$ equals $\e$ times his expected gain, under prior $p'$, from $y'$ over $x$. Thus, $x$ is not Pareto efficient with respect to $p$, which is a contradiction.} For each agent $i$, choose the smallest probability $\e_{K,i} \in [0,1]$ for which there exists a distribution $p_{K,i} \in \D(\Th_i)$ satisfying
\begin{equation} \label{eq:JS_approx}
    q_i = 
 (1 - \e_{K,i}) q_{K,i} + \e_{K,i} p_{K,i},
\end{equation}
where ties can be broken arbitrarily. Let $\e_K = (\e_{K,i})_{i=1}^{n}$ and $p_K = ( p_{K,i})_{i=1}^{n}$. JS's modified mechanism runs as follows. First, elicit type vector reports as in JS's $(x,q_{K})$-quota mechanism. Then, for each problem $k$ and each agent $i$, independently replace, with probability $\e_{K,i}$, agent $i$'s type report on problem $k$ with an independent draw from $p_{K,i}$. Finally, on each problem, apply the social choice function $x$ to these modified reports. We call this mechanism JS's \emph{$(x,q;q_K, \e_K, p_K)$-quota mechanism}. 

\subsection{Inefficiency of JS's quota mechanisms}

Consider a quota profile $q \in \prod_{i=1}^{n} \D( \Th_i)$ and a $q$-cyclically monotone social choice function $x \colon \Th \to \D(\XX)$. In the $K$-composite problem, we construct a label-free equilibrium of our $(x,q)$-quota mechanism; see the proof of \cref{res:optimal_transport_bound} (\cref{sec:proof_optimal_transport_bound}). Similarly, JS prove their main results by constructing a label-free equilibrium of their $(x,q)$-quota mechanism. We claim that each agent's interim expected equilibrium utility is weakly higher in our equilibrium than in JS's. 

Define $u_i \colon \Th_i^2 \to \R$ by 
\[
    u_i (\th_i' | \th_i) = \E_{\th_{-i} \sim q_{-i}} \Brac{ u_i (x(\th_i', \th_{-i}), \th_i)}.
\]
It follows from our proof of \cref{res:optimal_transport_bound} that in \emph{every} label-free equilibrium of our $(x,q)$-quota mechanism, the interim expected utility of type $\bm{\th}_i$ is given by 
\begin{equation} \label{eq:our_equil}
    \max_{ \bm{r}_i \in M_i} \, \frac{1}{K} \sum_{k=1}^{K} \sum_{\th_i' \in \Th_i} u_i (\th_i' | \th_i^k) r_i^k (\th_i'),
\end{equation}
where $M_i$ contains all vectors $\bm{r}_i \in [\D(\Th_i)]^K$ satisfying $\frac{1}{K} \sum_{k=1}^{K} r_i^k = q_i$. Similarly, following the proof of JS's Theorem 1 (pp.~251--255), it can be shown that in any label-free equilibrium of JS's $(x,q)$-quota mechanism, the interim expected utility of type $\bm{\th}_i$ takes the same form as \eqref{eq:our_equil}, except that the maximum is over all vectors $\bm{r}_i = ((1 - \e_{K,i}) \d_{\hat{\th}_i^k} + \e_{K,i} p_{K,i})_{k =1}^{K}$ in which the reported type vector $\hat{\bm{\th}}_i \in \Th_i^K$ satisfies $\marg \hat{\bm{\th}}_i = q_{K,i}$. Every such vector is in $M_i$, by \eqref{eq:JS_approx}, so the desired utility comparison follows. The next example illustrates the potential magnitude of the inefficiency introduced by JS's approach. 

\begin{exmp}[Inefficiency of JS's approximation]

\begin{table}
    \centering
    \begin{tabular}{l l l}
       $\marg \bm{\th}_i$  &  $q_i = 1/2$ &  $q_{3,i} = 1/3$  \smallskip \\
     \hline 
        $0$ &  $(\cdot, 1/2)$ & $(\cdot, 1/3) \to (\cdot, 1/2)$ \smallskip\\
        $1/3$  & $(1,1/4)$ & $(1,0) \to (1, 1/4)$ \smallskip \\
        $2/3$ & $(3/4,0)$ & $(1/2,0) \to ( 5/8, 1/4)$ \smallskip \\
        $1$ & $(1/2,\cdot)$ &  $(1/3, \cdot ) \to (1/2, \cdot)$
    \end{tabular}
    \caption{Equilibrium reports}
    \label{table:reports}
\end{table}

There are two agents. In the primitive problem, there is a single good to be allocated. Each agent's valuation for the good is equally likely to be high or low, independent of the other agent's valuation. Consider the social choice function $x$ that allocates the good to the agent whose valuation is highest, breaking ties uniformly. 

Consider the $K$-composite problem with $K = 3$. We represent a distribution over the two valuations by the probability of the high valuation. Let $q_1 = q_2 = 1/2$. For JS's modified mechanism, we take $q_{3,1} = q_{3,2} = 1/3$. Thus, $\e_{3,i} = 1/4$ and $p_{3,i} = 1$ for $i = 1,2$. To compare our quota mechanism with JS's, we represent the equilibrium reports of a given type vector $\bm{\th}_i$ in each mechanism as an ordered pair $(r ( \th_H), r(\th_L))$, defined as follows. Suppose that type $\th$ appears in $\bm{\th}_i$. In our mechanism, $r(\th)$ is the average report on valuation-$\th$ problems. In JS's mechanism, $r(\th)$ is the expected share of the valuation-$\th$ problems in which $\th_H$ is reported. \cref{table:reports} lists the equilibrium reports of each type vector $\bm{\th}_i$ in our equilibrium (second column) and in JS's equilibrium, before and after modification (third column). Conditional on the event that the agents have different valuations, it can be shown that the probability that the higher-valuation agent gets the good is $0.75$ in our equilibrium and $0.6875$ in JS's. 
\end{exmp}

\subsection{Theorem~\ref{res:optimal_transport_bound} fails with JS's quota mechanisms}

We construct a counterexample to show that Theorem~\ref{res:optimal_transport_bound} fails if we use JS's quota mechanism in place of ours. Let $n = 2$ and let $|\Th_i| = 2$ for $i = 1, 2$.  Let $x \colon \Th \to \XX$ be a deterministic, injective social choice function. For each $i$, we identify $\D(\Th_i)$ with $[0,1]$. Fix $K \geq 3$. Define the quotas by 
\[
    q_1 = \frac{\lfloor K/2 \rfloor }{K} + \frac{1}{2K}, \qquad q_2 = \frac{1}{K},  
    \]
where $\lfloor K/2 \rfloor$ denotes the greatest integer less than or equal to $K/2$. Consider JS's $(x,q;q_K, \e_K, p_K)$-quota mechanism, as described in \cref{sec:JS_quota}.\footnote{In fact, our counterexample works as long as $q_K$  is $(1/K)$-divisible and \eqref{eq:JS_approx} is satisfied for each agent $i$; the approximations need not be chosen ``optimally.''} Let $\s$ be a Bayes--Nash equilibrium of JS's $(x,q;q_K, \e_K, p_K)$-quota mechanism. We show that the associated inequality \eqref{eq:EP_bound} in \cref{res:optimal_transport_bound} fails at some profile $\bm{\th} \in \Th^K$. We separate into two cases.

First, suppose $q_{K,1} \leq \lfloor K/2 \rfloor /K$. Choose $\bm{\th} \in \Th^K$ with $\marg \bm{\th}_1 =  (\lfloor K/2 \rfloor + 1)/K$ and $\marg \bm{\th}_2 = q_2$. By construction, $\| q_1 - \marg \bm{\th}_1 \| = 1/(2K)$ and $\| q_2 - \marg \bm{\th}_2 \| = 0$, so the right side of \eqref{eq:EP_bound} equals $1/(2K)$. Since $x$ is injective, the average decision error on the left side of \eqref{eq:EP_bound} is at least the expected fraction of problems on which agent $1$'s modified report is different from his true type on that problem. This expected fraction is at least 
\[
    (1 - \e_{K,1}) \| q_{K,1} - \marg \bm{\th}_1 \|  + \e_{K,1} \E_{\th_1 \sim \marg \bm{\th}_1} [1 - p_{K,1} (\th_1)].
\]
Since $K \geq 3$, we have $\| q_{K,1} - \marg \bm{\th}_1 \| \geq 1/K > 1/(2K)$ and 
\begin{equation} \label{eq:p_ineq}
\begin{aligned}
    \E_{\th_1 \sim \marg \bm{\th}_1} [1 - p_{K,1} (\th_1)]
    &\geq 1 - \max_{\th_1 \in \Th_1} \marg (\th_1 |  \bm{\th}_1) \\
    &\geq 1 - (1/2 + 1/K) \\
    &\geq 1/(2K).
\end{aligned}
\end{equation}
Moreover, if $\e_{K,1} = 1$, then $p_{K,1} = q_1$, so the first inequality in \eqref{eq:p_ineq} holds strictly. Therefore, \eqref{eq:EP_bound} is violated at $\bm{\th}$. 

Next, suppose $q_{K,1} \geq (\lfloor K/2 \rfloor + 1)/K$. Choose $\bm{\th} \in \Th^K$ with $\marg \bm{\th}_1 =  \lfloor K/2 \rfloor/K$ and $\marg \bm{\th}_2 = q_2$. The rest of the argument is identical to the first case. 

\subsection{Error bound for JS's quota mechanisms}

We now show that a variant of JS's quota mechanism satisfies a weaker version of the decision error bound \eqref{eq:EP_bound} in \cref{res:optimal_transport_bound}. For any $q_i,q_i' \in \D(\Th_i)$, define $S(q_i | q_i')$ to be the smallest probability $\e_i$ for which there exists a distribution $p_i$ in $\D(\Th_i)$ such that $q_i = (1 - \e_i) q_i' + \e_i p_i$. 

Fix $q \in \prod_{i=1}^{n} \D (\Th_i)$. For each agent $i$, and each 
 $(1/K)$-divisible distribution $q_{K,i} \in \D(\Th_i)$, we may set $\e_{K,i} = S(q_i|q_{K,i})$ and then select $p_{K,i} \in \D(\Th_i)$ such that $q_i = (1 - \e_{K,i}) q_{K,i} + \e_{K,i} p_{K,i}$. Using the correction in \cite{BallKattwinkel2023} and applying a union bound on the probability of the event that some agent's report is modified, it can be shown that JS's $(x,q; q_{K}, \e_K, p_K)$-quota mechanism $(M,g)$ has a Bayes--Nash equilibrium $\s$ satisfying, for all $\bm{\th} \in \Th^K$, 
\begin{equation} \label{eq:JS_bound}
\begin{aligned}
    &\frac{1}{K} \sum_{k=1}^{K} \| g^k ( \s(\bm{\theta})) - x( \theta^k) \|  \\
    &\leq  \sum_{i=1}^{n} ( |\Th_i| - 1) \| q_{K,i} - \marg \bm{\th}_i \|  + \sum_{i=1}^{n} S( q_i | q_{K,i}) \\
    &\leq \sum_{i=1}^{n} ( | \Theta_i| - 1)  \Paren{ \| q_i - \marg \bm{\theta}_i \| + \| q_{K,i} - q_i \|} + \sum_{i=1}^{n} S( q_i | q_{K,i}).
\end{aligned}
\end{equation}

To usefully apply \eqref{eq:JS_bound}, we will choose each $(1/K)$-divisible approximation $q_{K,i}$  to control both $\| q_{K,i} - q_i \|$ and $S( q_i | q_{K,i})$. With a standard approximation, the term $S( q_i | q_{K,i})$ can be large if $q_i$ is near the boundary of $\D(\Th_i)$. We propose an alternative approximation that gives a uniform bound on $S(q_i | q_{K,i})$ over all $q_i \in \D (\Th_i)$.

\begin{lem}[Discrete approximation] \label{res:approximation}
Fix agent $i$. For any probability measure $q_i \in \D(\Th_i)$ and any integer $K \geq 1$, there exists a $(1/K)$-divisible probability measure $q_{K,i} \in \D(\Th_i)$ such that 
\[
    \| q_{K,i} - q_i \| \leq \frac{ |\Th_i| - 1}{ K}
    \quad
    \text{and}
    \quad
 S( q_i | q_{K,i}) \leq (|\Th_i| - 1) \frac{ |\Th_i|}{K}.
\]
\end{lem}

For each agent $i$, we plug this approximation $q_{K,i}$ from \cref{res:approximation} into \eqref{eq:JS_bound} to get
\[
\frac{1}{K} \sum_{k=1}^{K} \| g^k ( \s(\bm{\theta})) - x( \theta^k) \| 
\leq   \sum_{i=1}^{n} ( |\Th_i| - 1) \Paren{ \| q_i - \marg \bm{\th}_i \| + \frac{ 2 |\Th_i| -1}{K} }, 
\]
as claimed in \cref{rem:JS_quota}. Similarly, we can show that JS's quota mechanisms satisfy weaker versions of \cref{res:convergence_rate} and \cref{res:robust}, where the right sides of \eqref{eq:bound_expected} and \eqref{eq:ineq_t} are both increased by $\frac{1}{K} \sum_{i=1}^{n} ( |\Th_i| - 1) (2 |\Th_i| - 1)$. Since the error $1/K$ tends to $0$ as $K$ grows large, the asymptotic results \cref{res:equivalence} and \cref{res:error_bound} hold, without change, for JS's quota mechanisms.

\newpage

\section{Proofs of optimal transport results} \label{sec:additional}

\subsection{Proof of Lemma~\ref{res:TV_bounds}}

(i) For any measurable function $f \colon Y \to [-1,1]$, define the composition $hf \colon X \to \R$ by $(hf) (x) = h(x) f$, where $h(x) f$ denotes the integral of $f$ with respect to the measure $h(x)$. Since $hf$ takes values in $[-1,1]$, we have
\[
    | h(p) f - h (q) f | = | p (hf) - q (hf) | \leq 2 \| p - q \|.
\]
Taking the supremum over all such $f$ gives the desired result. 

(ii) By induction, it suffices to prove the result for $J =2$. For each measurable function $f \colon X_1 \times X_2 \to [-1,1]$, we apply Fubini's theorem to get
\begin{equation*}
\begin{aligned}
     &| (p_1 \otimes p_2) f - (q_1 \otimes q_2) f |  \\
    &\leq |(p_1 \otimes p_2) f - (q_1 \otimes p_2) f | + |(q_1 \otimes p_2) f - (q_1 \otimes q_2) f | \\
     &= | p_1 f (\cdot, p_2) - q_1 f( \cdot, p_2) |
     + |p_2 f( q_1, \cdot) - q_2 f (q_1, \cdot)| \\
     &\leq 2 \| p_1 - q_1 \| + 2 \| p_2 - q_2\|.
     \end{aligned}
\end{equation*}
Taking the supremum over all such $f$ gives the desired result.

\subsection{Proof of Lemma~\ref{res:DZ_bound}} \label{sec:DZ_bound_proof}

We first introduce notation. Given probability measures $p$ and $q$ on a fixed finite set $Z$, define the nonnegative measures $p \vee q$, $p \wedge q$, $(p - q)_+$, and $(q-  p)_+$ by performing the indicated operations on the associated probability mass functions.\footnote{For example, for any subset $Z'$ of $Z$, the probability $(p \vee q) (Z')$ is defined to equal $\sum_{z \in Z'} (p \vee q) (z)$, not $p(Z') \vee q(Z')$.} In particular, $p \vee q = p + (q - p)_+ = q + (p - q)_+$. Note that $\| p - q \| = ( p -q)_+ (Z) = (q - p)_+ (Z)$.  

A \emph{cycle} in $Z$ is a set
\[
    C = \{ (z_1, z_2), (z_2, z_3), \ldots, (z_{m-1}, z_m), (z_m, z_{1}) \},
\]
where $m \geq 1$ and $z_1, \ldots, z_m$ are distinct points in $Z$. We interpret $C$ as a collection of directed edges. Thus, $|C| = m \leq |Z|$. If $|C| = 1$, then $C = \{ (z_1, z_1) \}$, so the cycle $C$ is a self-loop. If $|C| > 1$, then the cycle $C$ is \emph{nontrivial}. The \emph{cycle measure} associated with the cycle $C$ is the discrete measure on $Z \times Z$ given by $\sum_{e \in C} \d_{e}$, where $\d_{e}$ denotes the unit mass (i.e., Dirac measure) on the directed edge $e \in Z^2$.

Now we turn to the proof proper. Among all $c$-optimal couplings of $p$ and $q$, choose a coupling $\g$ that places maximum probability on the diagonal $D$; by compactness, such a coupling exists. We claim that $\supp \g$ contains no nontrivial cycles. Otherwise, moving a small probability ``backwards'' along the cycle preserves the marginals of $\g$, strictly increases the probability on $D$, and weakly reduces the expected cost (by $c$-cyclical monotonicity), contrary to the definition of $\g$.\footnote{\label{ft:perturbation}Formally, suppose for a contradiction that $\supp \g$ contains some nontrivial cycle $C$ through $z_1, \ldots, z_m$. Choose $\e < \min_{e \in \supp \g} \g (e)$ and let $\g' = \g + \e [ \sum_{j=1}^{m} \d_{(z_j, z_j)} - \d_{(z_j, z_{j+1})} ]$, where $z_{m + 1} = z_1$.}

Using the fact that $\supp \g$ contains no nontrivial cycles, we now complete the proof. Let $\hat{\g} = \g - \g\vert_{D}$. By construction, $\hat{\g}(Z^2) = 1 - \g (D)$. Choose an arbitrary measure $\g' \in \Pi ( (q - p)_+, (p- q)_+)$. Thus, $\g' (Z^2) = \|q - p\|$. Let $\z = \hat{\g} + \g'$. By construction, both marginals of $\z$ are 
$p \vee q - \marg_1 (\g\vert_{D})$, where $\marg_1$ denotes the marginal on the first factor. Since $\z$ has equal marginals, we can express $\z$ as a nonnegative combination of cycle measures.\footnote{\label{ft:cycles}To visualize the argument, we represent $\z$ as weighted, directed graph with vertex set $Z$. For each $(z,z') \in \supp \z$, form a directed edge from $z$ to $z'$ with weight $\z (z,z')$. Since $\z$ has equal marginals, this weighted graph is balanced, i.e., $\sum_{z' \in Z} \z(z,z') = \sum_{z' \in Z} \z(z', z)$, for each $z$ in $Z$. Start at an arbitrary vertex with an outgoing edge. Form a path by arbitrarily selecting outgoing edges until the path contains a cycle, $C_1$. Let $\l_1$ be the smallest weight of any edge in this cycle. Repeat with $\z - \l_1 \sum_{e \in C_1} \d_{e}$ in place of $\z$. At each step of this procedure, the set of edges with positive weight decreases by at least $1$, and the graph remains balanced. Therefore, after finitely many steps, the final edge will be removed.} That is, 
\[
    \z = \sum_{j\in J} \l_j \z_j,
\]
where $J$ is a finite set, and for each $j$, the coefficient $\l_j$ is nonnegative and $\z_j$ is the cycle measure associated with a cycle $C_j$ in $Z$. Since $\z (D) = 0$, each cycle $C_j$ is nontrivial, and hence $C_j \not\subseteq \supp \g$.

For each $j$, since $C_j \not\subseteq \supp \g$, we have
\[
   | C_j \cap \supp \g | \leq |C_j| -1 \leq  |Z| - 1 \leq (|Z| - 1) | C_j \setminus \supp \g | 
\]
The outer inequality can be expressed as 
\[
    \z_j (\supp \g) \leq (|Z| - 1) \z_j ( Z^2 \setminus \supp \g).
\]
Multiply this inequality by $\l_j$ and sum over all $j$ in $J$ to get
\[
    \z ( \supp \g) \leq (|Z| - 1) \z ( Z^2 \setminus \supp \g).
\]
Since $ 1 - \g (D) = \hat{\g}(Z^2) \leq  \z ( \supp \g)$ and $\z (Z^2 \setminus \supp \g) \leq \g'(Z^2) = \| q - p \|$, we conclude that
\[
1 - \g (D) \leq (|Z| - 1) \| q - p \|,
\]
as desired.

\subsection{Proof of Lemma~\ref{res:Lip}} 

We use the notation from the proof of \cref{res:DZ_bound} (\cref{sec:DZ_bound_proof}). Let
\[
    M = \min \{ |X| \wedge |Y|, |X| \vee |Y| - 1\}.
\]
We may assume without loss that $X$ and $Y$ are disjoint. Let $Z = X \cup Y$. We view any distribution on $X$ or $Y$ as a distribution on $Z$, and we view any distribution on a subset of $Z^2$ as a distribution on $Z^2$. 


We are given a $c$-optimal coupling $\g$ of $p$ and $q$. Among all $c$-optimal couplings of $p'$ and $q'$, choose a coupling $\g'$ that is closest to $\g$ (in the total variation norm); such a coupling exists by compactness. Let $\hat{\g} = \g - (\g \wedge \g')$ and $\hat{\g}' = \g' - (\g \wedge \g')$. By construction, $\hat{\g}$ and $\hat{\g}'$ have disjoint supports, and
\[
    \hat{\g} (X \times Y) = \hat{\g}' (X \times Y) = \|\g' - \g\|.
\]
Define the inverse coupling $\hat{\g}^{-1}$ by $\hat{\g}^{-1}(y,x) = \hat{\g}( x,y)$ for all $(x,y) \in X \times Y$. Choose arbitrary couplings $\a \in \Pi ( (p - p')_+, (p' - p)_+)$ and $\b \in \Pi ((q' - q)_+, (q - q')_+)$. Define the nonnegative measure $\z$ on $Z^2$ by
\[
 \z = \a + \b + \hat{\g}' + \hat{\g}^{-1}.
\]
By construction, both marginals of $\z$ equal 
\[
    p \vee p' - \marg_1 (\g \wedge \g') + q \vee q' - \marg_2 (\g \wedge \g').
\]
Since $\z$ has equal marginals, we can express $\z$ as a nonnegative combination of cycle measures (see \cref{ft:cycles} in the proof of \cref{res:DZ_bound}). That is, 
\[
   \z = \sum_{j\in J} \l_j \z_j,
\]
where $J$ is a finite set, and for each $j$, the coefficient $\l_j$ is nonnegative and $\z_j$ is the cycle measure associated with a cycle $C_j$ in $Z$. 

For each $j$, we use a perturbation argument to show that 
\begin{equation} \label{eq:Cj_subset}
    C_j \not\subseteq (X \times Y) \cup (Y\times X).
\end{equation}
Suppose not. Then for some $m \geq 1$ and some distinct $x_1, \ldots, x_m \in X$ and distinct $y_1, \ldots, y_m \in Y$, we have
\[
    C_j = \{ (x_1, y_1), (y_1, x_2), \ldots, (x_m, y_m), (y_m, x_1) \}.
\]
Note that $|C_j| = 2 m$. By construction, $\z$ coincides with $\hat{\g}'$ on $X \times Y$ and with $\hat{\g}^{-1}$ on $Y \times X$. Since $\supp \hat{\g}$ and $\supp \hat{\g}'$ are disjoint, we must have $m > 1$.  Moreover, we can choose $\e > 0$ such that $\hat{\g}'(x_\ell, y_\ell) \geq \e$ and $\hat{\g}^{-1} ( y_\ell, x_{\ell+1}) \geq \e$ for all $\ell = 1, \ldots, m$, where $x_{m+1}$ is defined to equal $x_1$. Thus, $\g' ( x_\ell, y_\ell) \geq \e$ and $\g ( x_{\ell + 1}, y_\ell) \geq \e$ for all $\ell =1 , \ldots, m$. Consider the $\e$-perturbed couplings
\[
 \g' + \e \sum_{\ell = 1}^{m} \Brac{ \d_{ ( x_{\ell+1}, y_\ell)} - \d_{(x_\ell, y_\ell)}}
 \quad
 \text{and}
 \quad
 \g - \e \sum_{\ell = 1}^{m} \Brac{ \d_{ ( x_{\ell+1}, y_\ell)} - \d_{(x_\ell, y_\ell)}}.
\]
Since $\g'$ and $\g$ are both $c$-optimal, it follows that
\[
    \sum_{\ell = 1}^{m} c( x_\ell, y_\ell) = \sum_{\ell= 1}^{m} c( x_{\ell+1}, y_\ell).
\]
Thus, the $\e$-perturbation of $\g'$ is another $c$-optimal coupling of $p'$ and $q'$ that is strictly closer (in total variation norm) to $\g$, contrary to the definition of $\g'$.

Having established \eqref{eq:Cj_subset}, we next show that for each $j$, we have
\begin{equation} \label{eq:cycle_bound}
|C_j \cap (X \times Y)|+ |C_j \cap (Y \times X)| 
\leq 
2 M |C_j \cap (X^2 \cup Y^2) |.
\end{equation}
As we traverse the cycle $C_j$, we must switch from $X$ to $Y$ as many times as we switch from $Y$ to $X$. That is, $|C_j \cap (X \times Y)| = |C_j \cap (Y \times X)|$. Denote this common value by $m_j$. By \eqref{eq:Cj_subset}, we know that $|C_j \cap (X^2 \cup Y^2)| \geq 1$. Therefore, to prove \eqref{eq:cycle_bound}, it suffices to show that $m_j \leq M$. Since the cycle $C_j$ cannot pass through any point twice, we must have $m_j \leq |X|$ and $m_j \leq |Y|$. By \eqref{eq:Cj_subset}, we have
\[
    2 m_j < |C_j| \leq |X \cup Y| = |X| + |Y|\leq 2 ( |X| \vee |Y|),
\]
hence $m_j \leq |X| \vee |Y| - 1$. We conclude that $m_j \leq \min \{ |X| \wedge |Y|, |X| \vee |Y| - 1\} = M$, so \eqref{eq:cycle_bound} follows.

Finally, we use \eqref{eq:cycle_bound} to complete the proof. For each $j$, \eqref{eq:cycle_bound} can equivalently be expressed as
\[
    \z_j ( (X \times Y) \cup (Y \times X)) \leq 2 M \z_j ( X^2 \cup Y^2). 
\]
Multiply this inequality by $\l_j$ and sum over all $j$ in $J$ to get
\[
\z ( (X \times Y) \cup (Y \times X)) 
    \leq 2 M \z ( X^2 \cup Y^2).
\]
The proof is complete upon noting that 
\begin{align*}
\z ( (X \times Y) \cup (Y \times X)) &= \hat{\g}' ( X \times Y) + \hat{\g}^{-1} ( Y \times X) =   2 \| \g' - \g \|,  \\
\z ( X^2 \cup Y^2) &= \a(X^2) + \b(Y^2) = \| p' - p \| + \|q' - q\|.
\end{align*}

\paragraph{Tightness} To show that the  constant $M$ is tight, we separately consider the two cases $|X| = |Y|$ and $|X| \neq |Y|$.

Fix $m \geq 2$ and let $X= Y = \{ 1, \ldots, m \}$. Define $c \colon X \times Y \to \R$ by $c(x,y) = -x y$. Let $p$ be uniform over $X$. Let  $q$ and $q'$ be uniform over $Y$. Let $p' = p + (1/m) ( \d_{1} - \d_{m})$. Define the perfectly assortative couplings
\[
    \g = \frac{1}{m} \Brac{ \d_{(1, 1)} + \cdots + \d_{(m, m)}},
    \qquad
    \g' = \frac{1}{m} \Brac{ \d_{(1, 1)} + \d_{(1,2)} + \cdots + \d_{(m-1,m)}}.
\]
Since $-c$ is strictly supermodular, it can be verified that $\g$ is the unique $c$-optimal coupling of $p$ and $q$, and $\g'$ is the unique $c$-optimal coupling of $p'$ and $q'$. We have
\[
    \| \g' - \g \| = \frac{m - 1}{m} = ( m-1) \Paren{ \| p' - p\| + \| q' - q\| }.
\]

Fix $m \geq 1$. Let $X = \{0, \ldots, m\}$ and $Y = \{ 1,\ldots, m \}$. As before, define $c \colon X \times Y \to \R$ by $c(x,y) = -x y$. Let $p(x) = 1/m$ for $x = 1, \ldots, m$. Let $q$ and $q'$ be uniform over $Y$. Let $p' = p + (1/m) (\d_{0}- \d_{m})$. Define the perfectly assortative couplings
\[
    \g = \frac{1}{m}  \Brac{ \d_{(1,1)} + \cdots + \d_{(m,m)} },
    \qquad
    \g' =  \frac{1}{m} \Brac{ \d_{(0,1)} + \cdots + \d_{(m-1,m)}}.
\]
Since $-c$ is strictly supermodular, it can be verified that $\g$ is the unique $c$-optimal coupling of $p$ and $q$, and $\g'$ is the unique $c$-optimal coupling of $p'$ and $q'$. We have
\[
    \|\g' - \g\| = 1 = m \Paren{ \|p' - p\| + \|q' - q \| }.
\]

\subsection{Proof of Lemma~\ref{res:refined_bound}}

We continue using the notation from the proof of \cref{res:DZ_bound} (\cref{sec:DZ_bound_proof}). Recall that $Z_{\sim}$ denotes the space of equivalence classes under $\sim$. Since the equivalence relation $\sim$ is fixed throughout the proof, we write $[z]$ in place of $[z]_{\sim}$ for the equivalence class containing $z$. Among all $c$-optimal couplings of $p$ and $q$, choose a coupling $\g$ that places maximum probability on $D_{\sim}$. Let $\g_{\sim} \in \D ( Z_{\sim} \times Z_{\sim})$ denote the projection of $\g$ onto $Z_{\sim} \times Z_{\sim}$. That is, for all $z,z' \in Z$,
\[
    \g_{\sim} ( [z], [z'] ) = \sum_{\hat{z} \in [z]} \sum_{ \hat{z}' \in [z']} \g (\hat{z}, \hat{z}'). 
\]
By construction, $\g_{\sim}$ is a coupling of $p_{\sim}$ and $q_{\sim}$. Let $D_0$ denote the diagonal in $Z_{\sim} \times Z_{\sim}$. 

It suffices to show that $\supp \g_{\sim}$ contains no nontrivial cycles, for then we can follow the rest of the proof of \cref{res:DZ_bound} in \cref{sec:DZ_bound_proof}, with $(\g_{\sim}, p_{\sim}, q_{\sim}, D_0, Z_{\sim})$ in place of $(\g, p, q, D, Z)$, to conclude that 
\[
    1 - \g_{\sim} (D_0) \leq ( |Z_{\sim}| - 1) \| p_{\sim} - q_{\sim} \|.
\]
The desired inequality follows since $\g_{\sim} (D_0) = \g (D_{\sim})$. 

To complete the proof, we check that $\supp \g_{\sim}$ contains no nontrivial cycles. Suppose for a contradiction that $\supp \g_{\sim}$ contains a nontrivial cycle
\[
    C = \{ ([z_1], [z_2]), \ldots, ([z_{m-1}], [z_m]), ([z_m], [z_1]) \},
\]
for some $m \geq 2$ and some distinct $[z_1],  \ldots, [z_m] \in Z_{\sim}$. Hereafter, we use the convention in indices that $m + 1= 1$. For each $\ell =1, \ldots, m$, we have $([z_{\ell}], [z_{\ell + 1}]) \in \supp \g_{\sim}$, so there exist $\hat{z}_\ell, \hat{z}_{\ell+1}' \in Z$ with $\hat{z}_\ell \sim z_\ell$ and $\hat{z}_{\ell+1}' \sim z_{\ell+1}$ such that $(\hat{z}_\ell, \hat{z}'_{\ell+1}) \in \supp \g$.  Note that for each $\ell = 1, \ldots, m$, we have selected $\hat{z}_\ell ,\hat{z}_\ell' \in [z_\ell]$. Choose $\e > 0$ such that $\g (\hat{z}_\ell, \hat{z}'_{\ell+1}) \geq \e$ for all $\ell = 1, \ldots, m$.  Define the $\e$-perturbed coupling $\tilde{\g}$ by 
\[
    \tilde{\g} = \g + \e \sum_{\ell=1}^{m} \Brac{ \d_{(\hat{z}_\ell, \hat{z}'_{\ell})} - \d_{(\hat{z}_\ell, \hat{z}'_{\ell+1})}}.
\]
By construction, $\tilde{\g}$ is a coupling of $p$ and $q$. Since the set $D_{\sim}$ is $c$-cyclically monotone, we have
\[
    \sum_{\ell =1}^{m} c ( \hat{z}_\ell, \hat{z}_\ell')  \leq
    \sum_{\ell =1}^{m} c ( \hat{z}_\ell, \hat{z}_{\ell+1}').
\]
Therefore, $\tilde{\g}$ is also $c$-optimal. But $\tilde{\g} (D_{\sim}) > \g(D_{\sim})$, contrary to the definition of $\g$.

\subsection{Proof of Lemma~\ref{res:S_bound}} 

We follow the argument from the proof of \cref{res:DZ_bound} (\cref{sec:DZ_bound_proof}). Let $\g$ be an arbitrary $c$-optimal coupling of $p$ and $q$. We claim that for every cycle $C \subseteq \supp \g$, we have $C \subseteq S$. Otherwise, moving a small probability ``backwards'' along the cycle preserves the marginals of $\g$ and strictly reduces the expected cost (since $S$ contains the diagonal and $S$ is strictly $c$-cyclically monotone); see \cref{ft:perturbation} for a formal definition of the perturbation. 

Now we use this claim to complete the proof. Choose an arbitrary measure $\g' \in \Pi ( (q - p)_+, (p- q)_+)$. Thus, $\g' (Z^2) = \|q - p\|$. 
Let $\z = \g + \g'$. By construction, both marginals of $\z$ equal $p \vee q$. Since $\z$ has equal marginals, we can express $\z$ as a nonnegative combination of cycle measures (see \cref{ft:cycles} in the proof of \cref{res:DZ_bound}). That is, 
\[
    \z = \sum_{j\in J} \l_j \z_j,
\]
where $J$ is a finite set, and for each $j$, the coefficient $\l_j$ is nonnegative and $\z_j$ is the cycle measure associated with a cycle $C_j$ in $Z$.

Let $J_0 = \{j \in J : C_j \not\subseteq S \}$. By our claim above, for each $j \in J_0$, we have $C_j \not\subseteq \supp \g$, so
\[
   | C_j \cap \supp \g | \leq |C_j| -1 \leq  |Z| - 1 \leq (|Z| - 1)  | C_j \setminus \supp \g |.
\]
The outer inequality can be expressed as
\[
    \z_j ( \supp \g) \leq (|Z| - 1) \z_j (Z^2 \setminus \supp \g). 
\]
Multiply this inequality by $\l_j$ and sum over all $j$ in $J_0$ to get
\begin{equation} \label{eq:J0_bound}
    \sum_{j \in J_0} \l_j \z_j (\supp \g) 
    \leq (|Z| - 1) \sum_{j \in J_0} \l_j \z_j ( Z^2 \setminus \supp \g). 
\end{equation}
To complete the proof, we bound each side of \eqref{eq:J0_bound}. For the left side, we have
\begin{equation} \label{eq:J0_left}
    \g(Z^2 \setminus S) \leq \z (\supp \g \setminus S) = \sum_{j \in J_0} \l_j \z_j (\supp \g \setminus S) \leq \sum_{j \in J_0} \l_j \z_j (\supp \g),
\end{equation}
where the equality holds because $\z_j (Z^2 \setminus S) = 0$ for all $j \in J \setminus J_0$. The sum on the right side of \eqref{eq:J0_bound} satisfies
\begin{equation} \label{eq:J0_right}
    \sum_{j \in J_0} \l_j \z_j ( Z^2 \setminus \supp \g) \leq \z ( Z^2 \setminus \supp \g) \leq \g'( Z^2) = \| q - p\|.
\end{equation}
Combine \eqref{eq:J0_bound} with \eqref{eq:J0_left} and \eqref{eq:J0_right} to obtain the desired inequality.

\subsection{Proof of Lemma~\ref{res:approximation}}

If $|\Th_i| = 1$ or $K \leq (|\Th_i| - 1)|\Th_i|$, then the result is immediate, so we may assume $|\Th_i| \geq  2$ and $K \geq  (|\Th_i| - 1)|\Th_i| + 1$. We will construct a $(1/K)$-divisible probability measure $q_{K,i} \in \D(\Th_i)$ satisfying:
\begin{enumerate}[label = (\roman*), ref = \roman*]
    \item \label{it:approx_TV} $\| q_{K,i}  -q_i \| \leq (|\Th_i| - 1)/K$;
    \item \label{it:approx_small} for all $\th_i \in \Th_i$, if $q_i(\th_i) < 1/ |\Th_i|$, then 
    $q_{K,i} (\th_i) \leq q_i(\th_i)$.
\end{enumerate}
In words, all sufficiently small probabilities are weakly reduced under the approximation. 


The rest of the proof has two parts. First, we check that properties \eqref{it:approx_TV}--\eqref{it:approx_small} are sufficient for \cref{res:approximation}. Then we construct an approximation satisfying \eqref{it:approx_TV}--\eqref{it:approx_small}.

\paragraph{Sufficiency of \eqref{it:approx_TV}--\eqref{it:approx_small}}
We use properties \eqref{it:approx_TV}--\eqref{it:approx_small} to show that $S( q_i | q_{K,i}) \leq (|\Th_i| - 1) |\Th_i| / K$. Let $\e_{K,i} = (|\Th_i| - 1) |\Th_i| / K$.  Note that $\e_{K,i}$ is in $(0,1)$ by our assumptions on $|\Th_i|$ and $K$. Let
\[
    p_{K,i} = \frac{q_i - (1 - \e_{K,i}) q_{K,i}}{\e_{K,i}}.
\]
By construction, $q_i = (1 - \e_{K,i}) q_{K,i} + \e_{K,i} p_{K,i}$ and $p_{K,i} (\Th_i) = 1$. It remains to check that $p_{K,i}$ is nonnegative. For every type $\th_i \in \Th_i$, we check that
\[
    q_{K,i} (\th_i) - q_i(\th_i) \leq q_{K,i} (\th_i) \e_{K,i}.
\]
If $q_{K,i} (\th_i) \leq q_i(\th_i)$, this is immediate, so suppose that $q_{K,i} (\th_i) > q_i(\th_i)$. By \eqref{it:approx_small}, we have $q_i (\th_i) \geq 1/ |\Th_i|$, so 
\[
    q_{K,i} (\th_i) - q_i(\th_i) \leq \| q_{K,i} - q_i \| \leq (|\Th_i| - 1)/K = \e_{K,i}/ |\Th_i| \leq q_{K,i} (\th_i) \e_{K,i},
\]
where the second inequality follows from \eqref{it:approx_TV}.

\paragraph{Construction of approximation satisfying \eqref{it:approx_TV}--\eqref{it:approx_small}}
We construct a  $(1/K)$-divisible probability measure $q_{K,i} \in \D(\Th_i)$ satisfying \eqref{it:approx_TV}--\eqref{it:approx_small}. For any real $z$, let 
$\lfloor z \rfloor$ denote the greatest integer less than or equal to $z$ and let $\lceil z \rceil$  denote the least integer greater than or equal to $z$. Let $\bar{\Th}_i = \{ \th_i : q_i (\th_i) < 1/ |\Th_i|\}$. Since $q_{i}$ is a probability measure, we have $\bar{\Th}_i \neq \Th_i$. Note that 
$\bar{\Th}_i$ is empty if $q_i$ is uniform. For each $\th_i \in \bar{\Th}_i$, let 
\[
    q_{K,i} (\th_i)  = \lfloor K q_i (\th_i) \rfloor /K.
\]
This ensures that \eqref{it:approx_small} holds. It remains to define $q_{K,i}$ on $\Th_i \setminus \bar{\Th}_i$ so that \eqref{it:approx_TV} holds. Let
\[
    R_0 = \sum_{\th_i \in \bar{\Th}_i} (q_{K,i} (\th_i) - q_i (\th_i)).
\]
Note that $R_0 \leq 0$. Enumerate the types in $\Th_i \setminus \bar{\Th}_i$ as $\th_{i,1}, \ldots, \th_{i,J}$. Recursively define $q_{K,i}$ over $\Th_i \setminus \bar{\Th}_i$ as follows. For each $j \in \{1, \ldots, J\}$, given $q_{K,i} (\th_{i,1}), \ldots, q_{K,i} (\th_{i,j-1})$, let
\[
    R_{j-1} = R_0 + \sum_{\ell = 1}^{j-1} \Paren{q_{K,i} (\th_{i,\ell}) - q_i (\th_{i,\ell})}.
\]
If $j < J$, let
\[
    q_{K,i} (\th_{i,j})  = 
    \begin{cases}
    \lceil K q_i (\th_{i,j}) \rceil /K &\text{if}~  R_{j -1} \leq 0, \\  
    \lfloor K q_i (\th_{i,j}) \rfloor /K &\text{if}~ R_{j-1} > 0.
    \end{cases}
\]
For $j = J$, let
\[
    q_{K,i} ( \th_{i,J}) = q_{i} (\th_{i,J}) - R_{J-1} = 1 - \sum_{\th_i \in \bar{\Th}_i} q_{K,i} (\th_i) - \sum_{\ell=1}^{J-1} q_{K,i} ( \th_{i,\ell}).
\]
By construction, $q_{K,i}$ is $(1/K)$-divisible and $q_{K,i} (\Th_i) = 1$. It is immediate that $q_{K,i} ( \th_i) \geq 0$ for all $\th_i \neq \th_{i,J}$. It remains to prove that $q_{K,i} (\th_{i,J}) \geq 0$ and that \eqref{it:approx_TV} holds. We separate into two cases. 

First, suppose that $R_j \geq 0$ for some $j \in \{1, \ldots, J-1\}$. Let $j^\ast$ be the smallest $j$ for which $R_j \geq 0$. Thus, $0 \leq R_{j^\ast} < 1/K$. It follows that $|R_j| < 1/K$ for all $j \geq j^\ast$. In particular, $R_{J-1} < 1/ K$, so $q_{K,i} (\th_{i,J}) >  - 1/K$. Since $q_{K,i} (\th_{i,J})$ is $(1/K)$-divisible, it follows that $q_{K,i} (\th_{i,J}) \geq 0$. Moreover,  $|q_{K,i} (\th_{i,J}) -q_i ( \th_{i,J})|  = |R_{J-1}| \leq 1/K$, so we conclude that 
\[
     \| q_{K,i} - q_i \|  = \frac{1}{2} \sum_{\th_i \in \Th_i} | q_{K,i} (\th_i) - q_i (\th_i) | \leq \frac{|\Th_i|}{2 K} \leq \frac{ |\Th_i| - 1}{K},
\]
where the final inequality holds by our assumption that $|\Th_i| \geq 2$. 

Next, suppose that $R_j < 0$ for all $j =1, \ldots, J-1$. In this case, $   q_{K,i} (\th_{i,j}) = \lceil K q_i (\th_{i,j}) \rceil /K$ for all $j =1 , \ldots, J-1$, and  $q_{K,i} (\th_{i,J}) \geq q_i ( \th_{i,J})$. Therefore, 
\[
    \| q_{K,i} - q_i \| = \sum_{\th_i \in \Th_i} \max\{ q_i (\th_i) - q_{K,i} (\th_i) ,0\} =   -R_0 \leq \frac{| \bar{\Th}_i|}{K} \leq \frac{ |\Th_i| -1}{K}.
\]

\end{document}